\documentclass{article}
\usepackage{amsbsy}

\begin{document}

%
%
\catcode`\@=11 

%
\global\newcount\secno \global\secno=0
\global\newcount\meqno \global\meqno=1

\def\newsec#1{\global\advance\secno by1
\global\subsecno=0\eqnres@t
\section{ #1}
}
\def\eqnres@t{\xdef\secsym{\the\secno.}\global\meqno=1}
\def\sequentialequations{\def\eqnres@t{\bigbreak}}\xdef\secsym{}
\global\newcount\subsecno \global\subsecno=0
\def\subsec#1{\global\advance\subsecno by1
\subsection{#1}}

\def\draftmode{\message{ DRAFTMODE }
\writelabels
 {\count255=\time\divide\count255 by 60 \xdef\hourmin{\number\count255}
  \multiply\count255 by-60\advance\count255 by\time
  \xdef\hourmin{\hourmin:\ifnum\count255<10 0\fi\the\count255}}}
\def\nolabels{\def\wrlabeL##1{}\def\eqlabeL##1{}\def\reflabeL##1{}}
\def\writelabels{\def\wrlabeL##1{\leavevmode\vadjust{\rlap{\smash%
{\line{{\escapechar=` \hfill\rlap{\tt\hskip.03in\string##1}}}}}}}%
\def\eqlabeL##1{{\escapechar-1\rlap{\tt\hskip.05in\string##1}}}%
\def\reflabeL##1{\noexpand\llap{\noexpand\sevenrm\string\string\string##1}}}
\nolabels

\def\eqn#1#2{
\xdef #1{(\secsym\the\meqno)}
\global\advance\meqno by1
$$#2\eqno#1\eqlabeL#1
$$}

\def\eqalign#1{\null\,\vcenter{\openup\jot\m@th
  \ialign{\strut\hfil$\displaystyle{##}$&$\displaystyle{{}##}$\hfil
      \crcr#1\crcr}}\,}

\def\foot#1{\footnote{#1}} 

%
\global\newcount\refno \global\refno=1
\newwrite\rfile
\def\ref{[\the\refno]\nref}
\def\nref#1{\xdef#1{[\the\refno]}
\ifnum\refno=1\immediate\openout\rfile=refs.tmp\fi
\global\advance\refno by1\chardef\wfile=\rfile\immediate
\write\rfile{\noexpand\bibitem{#1}}\findarg}
\def\findarg#1#{\begingroup\obeylines\newlinechar=`\^^M\pass@rg}
{\obeylines\gdef\pass@rg#1{\writ@line\relax #1^^M\hbox{}^^M}%
\gdef\writ@line#1^^M{\expandafter\toks0\expandafter{\striprel@x #1}%
\edef\next{\the\toks0}\ifx\next\em@rk\let\next=\endgroup\else\ifx\next\empty%
\else\immediate\write\wfile{\the\toks0}\fi\let\next=\writ@line\fi\next\relax}}
\def\striprel@x#1{} \def\em@rk{\hbox{}} 
\def\lref{\begingroup\obeylines\lr@f}
\def\lr@f#1#2{\gdef#1{\ref#1{#2}}\endgroup\unskip}
\def\semi{;\hfil\break}
\def\addref#1{\immediate\write\rfile{\noexpand\item{}#1}} 
\def\listrefs{
}
\def\startrefs#1{\immediate\openout\rfile=refs.tmp\refno=#1}
\def\xref{\expandafter\xr@f}\def\xr@f[#1]{#1}
\def\refs#1{\count255=1[\r@fs #1{\hbox{}}]}
\def\r@fs#1{\ifx\und@fined#1\message{reflabel \string#1 is undefined.}%
\nref#1{need to supply reference \string#1.}\fi%
\vphantom{\hphantom{#1}}\edef\next{#1}\ifx\next\em@rk\def\next{}%
\else\ifx\next#1\ifodd\count255\relax\xref#1\count255=0\fi%
\else#1\count255=1\fi\let\next=\r@fs\fi\next}
\newwrite\lfile
{\escapechar-1\xdef\pctsign{\string\%}\xdef\leftbracket{\string\{}
\xdef\rightbracket{\string\}}\xdef\numbersign{\string\#}}
\def\writedefs{\immediate\openout\lfile=labeldefs.tmp \def\writedef##1{%
\immediate\write\lfile{\string\def\string##1\rightbracket}}}
\def\writestop{\def\writestoppt{\immediate\write\lfile{\string\pageno%
\the\pageno\string\startrefs\leftbracket\the\refno\rightbracket%
\string\def\string\secsym\leftbracket\secsym\rightbracket%
\string\secno\the\secno\string\meqno\the\meqno}\immediate\closeout\lfile}}

\catcode`\@=12 
%

%
\def\noblackbox{\overfullrule=0pt}
\hyphenation{anom-aly anom-alies coun-ter-term coun-ter-terms
}
\def\inv{^{\raise.15ex\hbox{${\scriptscriptstyle -}$}\kern-.05em 1}}
\def\dup{^{\vphantom{1}}}
\def\Dsl{\,\raise.15ex\hbox{/}\mkern-13.5mu D} 
\def\dsl{\raise.15ex\hbox{/}\kern-.57em\partial}
\def\del{\partial}
\def\Psl{\dsl}
\def\tr{{\rm tr}} \def\Tr{{\rm Tr}}
\font\bigit=cmti10 scaled \magstep1
\def\biglie{\hbox{\bigit\$}} 
\def\lspace{\ifx\answ\bigans{}\else\qquad\fi}
\def\lbspace{\ifx\answ\bigans{}\else\hskip-.2in\fi} 
\def\boxeqn#1{\vcenter{\vbox{\hrule\hbox{\vrule\kern3pt\vbox{\kern3pt
	\hbox{${\displaystyle #1}$}\kern3pt}\kern3pt\vrule}\hrule}}}
\def\tilde{\widetilde} \def\bar{\overline} \def\hat{\widehat}
%
\def\CAG{{\cal A/\cal G}} \def\CO{{\cal O}} 
\def\CA{{\cal A}} \def\CC{{\cal C}} \def\CF{{\cal F}} \def\CG{{\cal G}} 
\def\CL{{\cal L}} \def\CH{{\cal H}} \def\CI{{\cal I}} \def\CU{{\cal U}}
\def\CB{{\cal B}} \def\CR{{\cal R}} \def\CD{{\cal D}} \def\CT{{\cal T}}
\def\e#1{{\rm e}^{^{\textstyle#1}}}
\def\grad#1{\,\nabla\!_{{#1}}\,}
\def\gradgrad#1#2{\,\nabla\!_{{#1}}\nabla\!_{{#2}}\,}
\def\ph{\varphi}
\def\psibar{\overline\psi}
\def\om#1#2{\omega^{#1}{}_{#2}}
\def\vev#1{\langle #1 \rangle}
\def\lform{\hbox{$\sqcup$}\llap{\hbox{$\sqcap$}}}
\def\darr#1{\raise1.5ex\hbox{$\leftrightarrow$}\mkern-16.5mu #1}
\def\lie{\hbox{\it\$}} 
\def\ha{{1\over2}}
\def\half{{\textstyle{1\over2}}} 
\def\roughly#1{\raise.3ex\hbox{$#1$\kern-.75em\lower1ex\hbox{$\sim$}}}

%

\font\teneufm=eufm10
\font\seveneufm=eufm7
\font\fiveeufm=eufm5
\newfam\eufmfam
\textfont\eufmfam=\teneufm
\scriptfont\eufmfam=\seveneufm
\scriptscriptfont\eufmfam=\fiveeufm
\def\eufm#1{{\fam\eufmfam\relax#1}}

\font\teneurm=eurm10
\font\seveneurm=eurm7
\font\fiveeurm=eurm5
\newfam\eurmfam
\textfont\eurmfam=\teneurm
\scriptfont\eurmfam=\seveneurm
\scriptscriptfont\eurmfam=\fiveeurm
\def\eurm#1{{\fam\eurmfam\relax#1}}

\font\tenmsx=msam10
\font\sevenmsx=msam7
\font\fivemsx=msam5
\font\tenmsy=msbm10
\font\sevenmsy=msbm7
\font\fivemsy=msbm5
\newfam\msafam
\newfam\msbfam
\textfont\msafam=\tenmsx  \scriptfont\msafam=\sevenmsx
  \scriptscriptfont\msafam=\fivemsx
\textfont\msbfam=\tenmsy  \scriptfont\msbfam=\sevenmsy
  \scriptscriptfont\msbfam=\fivemsy
\def\msam#1{{\fam\msafam\relax#1}}
\def\msbm#1{{\fam\msbfam\relax#1}}


\font\tencmmib=cmmib10  \skewchar\tencmmib='177
\font\sevencmmib=cmmib7 \skewchar\sevencmmib='177
\font\fivecmmib=cmmib5 \skewchar\fivecmmib='177
\newfam\cmmibfam
\textfont\cmmibfam=\tencmmib
\scriptfont\cmmibfam=\sevencmmib
\scriptscriptfont\cmmibfam=\fivecmmib
\def\cmmib#1{{\fam\cmmibfam\relax#1}}

\def\a{\alpha}    \def\b{\beta}       
\def\c{\chi}       
\def\d{\delta}       \def\D{\Delta}    
\def\e{\varepsilon} \def\f{\phi}       
\def\F{\Phi}
\def\g{\gamma}    \def\G{\Gamma}      \def\k{\kappa}     
\def\l{\lambda}
\def\L{\Lambda}   \def\m{\mu}         \def\n{\nu}        
\def\r{\rho}
\def\vr{\varrho}  \def\o{\omega}      \def\O{\Omega}     
\def\p{\psi}
\def\P{\Psi}      \def\s{\sigma}      \def\S{\Sigma}     
\def\th{\theta}
\def\vt{\vartheta}\def\t{\tau}        \def\w{\varphi}    
\def\x{\xi}
\def\z{\zeta}
\def\CA{{\cal A}}
\def\CB{{\cal B}}
\def\CC{{\cal C}}
\def\CG{{\cal G}}
\def\CH{{\cal H}}
\def\CK{{\cal K}}
\def\CM{{\cal M}}
\def\CN{{\cal N}}
\def\CE{{\cal E}}
\def\CL{{\cal L}}\def\CJ{{\cal J}}
\def\CD{{\cal D}}
\def\CW{{\cal W}}
\def\CQ{{\cal Q}} \def\CV{{\cal V}}
\def\ep{\epsilon}
\def\BK{\cmmib{K}}
\def\V{\msbm{V}}
\def\E{\msbm{E}}
\def\CP{\msbm{CP}}
\def\R{\msbm{R}}
\def\C{\msbm{C}}
\def\Da{d_{\! A}}

\def\rd{\partial}
\def\grad#1{\,\nabla\!_{{#1}}\,}
\def\gradd#1#2{\,\nabla\!_{{#1}}\nabla\!_{{#2}}\,}
\def\om#1#2{\omega^{#1}{}_{#2}}
\def\vev#1{\langle #1 \rangle}
\def\darr#1{\raise1.5ex\hbox{$\leftrightarrow$}
\mkern-16.5mu #1}
\def\Ha{{1\over2}}
\def\ha{{\textstyle{1\over2}}}
\def\fr#1#2{{\textstyle{#1\over#2}}}
\def\Fr#1#2{{#1\over#2}}
\def\rf#1{\fr{\rd}{\rd #1}}
\def\rF#1{\Fr{\rd}{\rd #1}}
\def\df#1{\fr{\d}{\d #1}}
\def\dF#1{\Fr{\d}{\d #1}}
\def\DDF#1#2#3{\Fr{\d^2 #1}{\d #2\d #3}}
\def\DDDF#1#2#3#4{\Fr{\d^3 #1}{\d #2\d #3\d #4}}
\def\ddF#1#2#3{\Fr{\d^n#1}{\d#2\cdots\d#3}}
\def\fs#1{#1\!\!\!/\,}   
\def\Fs#1{#1\!\!\!\!/\,} 
\def\roughly#1{\raise.3ex\hbox{$#1$\kern-.75em
\lower1ex\hbox{$\sim$}}}
\def\ato#1{{\buildrel #1\over\longrightarrow}}
\def\up#1#2{{\buildrel #1\over #2}}
\def\opname#1{\mathop{\kern0pt{\rm #1}}\nolimits}
\def\tr{\opname{Tr}}
\def\ch{\opname{ch}}
\def\Re{\opname{Re}}
\def\Im{\opname{Im}}
\def\End{\opname{End}}
\def\pr{\prime}
\def\ppr{{\prime\prime}}
\def\bs{\cmmib{s}}
\def\bbs{\bar\cmmib{s}}
\def\Dp{\rd_{\!A}}
\def\Dpp{\bar\rd_{\!A}}
\def\mapr{\!\smash{
	    \mathop{\longrightarrow}\limits^{\bs_+}}\!}
\def\mapl{\!\smash{
	    \mathop{\longleftarrow}\limits^{\bs_-}}\!}
\def\mapbr{\!\smash{
	    \mathop{\longrightarrow}\limits^{\bbs_+}}\!}
\def\mapbl{\!\smash{
	    \mathop{\longleftarrow}\limits^{\bbs_-}}\!}
\def\mapd{\Big\downarrow
 	 \rlap{$\vcenter{\hbox{$\scriptstyle \bbs_-$}}$}}
\def\mapu{\Big\uparrow
	  \rlap{$\vcenter{\hbox{$\scriptstyle\bbs_+$}}$}}
\def\ne{\nearrow}
\def\se{\searrow}
\def\nw{\nwarrow}
\def\sw{\swarrow}

\def\etal{et al.}
\def\git{/\kern-.25em/}

\def\cmp#1#2#3{Comm.\ Math.\ Phys.\ {{\bf #1}} {(#2)} {#3}}
\def\pl#1#2#3{Phys.\ Lett.\ {{\bf #1}} {(#2)} {#3}}
\def\np#1#2#3{Nucl.\ Phys.\ {{\bf #1}} {(#2)} {#3}}
\def\prd#1#2#3{Phys.\ Rev.\ {{\bf #1}} {(#2)} {#3}}
\def\prl#1#2#3{Phys.\ Rev.\ Lett.\ {{\bf #1}} {(#2)} {#3}}
\def\ijmp#1#2#3{Int.\ J.\ Mod.\ Phys.\ {{\bf #1}} 
{(#2)} {#3}}
\def\jmp#1#2#3{J.\ Math.\ Phys.\ {{\bf #1}} {(#2)} {#3}}
\def\jdg#1#2#3{J.\ Differ.\ Geom.\ {{\bf #1}} {(#2)} {#3}}
\def\pnas#1#2#3{Proc.\ Nat.\ Acad.\ Sci.\ USA.\ {{\bf #1}} 
{(#2)} {#3}}
\def\top#1#2#3{Topology {{\bf #1}} {(#2)} {#3}}
\def\zp#1#2#3{Z.\ Phys.\ {{\bf #1}} {(#2)} {#3}}
\def\prp#1#2#3{Phys.\ Rep.\ {{\bf #1}} {(#2)} {#3}}
\def\ap#1#2#3{Ann.\ Phys.\ {{\bf #1}} {(#2)} {#3}}
\def\ptrsls#1#2#3{Philos.\ Trans.\  Roy.\ Soc.\ London
{{\bf #1}} {(#2)} {#3}}
\def\prsls#1#2#3{Proc.\ Roy.\ Soc.\ London Ser.\
{{\bf #1}} {(#2)} {#3}}
\def\am#1#2#3{Ann.\ Math.\ {{\bf #1}} {(#2)} {#3}}
\def\mm#1#2#3{Manuscripta \ Math.\ {{\bf #1}} {(#2)} {#3}}
\def\ma#1#2#3{Math.\ Ann.\ {{\bf #1}} {(#2)} {#3}}
\def\cqg#1#2#3{Class. Quantum Grav.\ {{\bf #1}} {(#2)} {#3}}
\def\ivm#1#2#3{Invent.\ Math.\ {{\bf #1}} {(#2)} {#3}}
\def\plms#1#2#3{Proc.\ London Math.\ Soc.\ {{\bf #1}} 
{(#2)} {#3}}
\def\dmj#1#2#3{Duke Math.\  J.\ {{\bf #1}} {(#2)} {#3}}
\def\bams#1#2#3{Bull.\ Am.\ Math.\ Soc.\ {{\bf #1}} {(#2)} 
{#3}}
\def\jams#1#2#3{Bull.\ Am.\ Math.\ Soc.\ {{\bf #1}} {(#2)} 
{#3}}

\def\jgp#1#2#3{J.\ Geom.\ Phys.\ {{\bf #1}} {(#2)} {#3}}
\def\ihes#1#2#3{Publ.\ Math.\ I.H.E.S. \ {{\bf #1}} {(#2)} {#3}}

\def\ack{\bigbreak\bigskip\centerline{
{\bf Acknowledgements}}\nobreak}
\def\subsubsec#1{\ifnum\lastpenalty>9000\else\bigbreak\fi
  \noindent{\it #1}\par\nobreak\medskip\nobreak}

\def\submit{\baselineskip=20pt plus 2pt minus 2pt}
\def\lin#1{\medskip\noindent {${\hbox{\it #1}}$}\medskip }
\def\linn#1{\noindent $\bullet$ {\it #1} \par}


\lref\Wittendb{
E. Witten,
{\it Dynamical breaking of supersymmetry},
\np{B 188}{1981}{513}
}

\lref\Wittencsb{
E. Witten,
{\it Constraints on supersymmetry breaking},
\np{B 202}{253}{1982}.
}

\lref\Wittenmorse{
E.~Witten,
{\it Supersymmetry and Morse theory},
\jdg{17}{1982}{692}.
}

\lref\TFT{
E. Witten,
{\it Topological quantum field theory},
Commun. Math. Phys. {\bf 117}  (1988) 353.
}

\lref\tdYM{
E. Witten,
{\it Two dimensional gauge theories revisited},
{\tt hep-th/9204084}.
}

\lref\TSM{
E. Witten, 
{\it Topological sigma models},
Commun. Math. Phys. {\bf 118}  (1988) 411. 
}

\lref\Wittenmirror{
E.~Witten, 
{\it Mirror manifolds and topological field theory},
{\tt  hep-th/9112056}.
}

\lref\Wittenkaehler{
E.~Witten, 
{\it Supersymmetric Yang-Mills theory on a four-manifold},
J. Math. Phys. {\bf 35} (1994) 5101.
{\tt hep-th/9303195}.
}

\lref\Wittencss{
E. Witten,
{\it Chern-Simons gauge theory as a string theory},
{\tt hep-th/9207094}.
}

\lref\Wittenbg{
E. Witten,
{\it Quantum background independence in string theory},
{\tt hep-th/9306122}.
}

\lref\GLSM{
E. Witten
{\it Phases of $N=2$ theories in two dimensions},
Nucl. Phys. {\bf B403} (1993) 159-222,
{\tt hep-th/9301042}.
}

\lref\Wittengwzw{
E. Witten,
{\it The $N$ matrix model and gauged WZW models},
\np{B 371}{1992}{191}.
}

\lref\Wittengr{
E.~Witten, 
{\it The Verlinde algebra and the cohomology of the Grassmannian},
{\tt hep-th/9312104}.
}

\lref\SWinv{
E.~Witten,
{\it Monoples and four manifolds},
Math.~Research Lett.~{\bf 1} (1994) 769.
{\tt hep-th/9411102}.
}

\lref\Wittenadhm{
E.~Witten,
{\it Sigma Models and the ADHM construction of instantons},
{\tt hep-th/9410052}.
}

\lref\sdstring{
E. Witten, 
{\it Some Comments on String Dynamics}, 
{\tt hep-th/9510135}.
}

\lref\Wittensmall{
E.~Witten,
{\it Small instantons in string theory},
{\tt hep-th/9511030}.
}

\lref\withiggs{ 
E. Witten, 
{\it On the Conformal Field Theory of the Higgs Branch},
{\tt hep-th/9707093}.
}

\lref\Wittenabel{
E.~Witten,
{\it On $S$-duality in abelian gauge theory},
{\tt hep-th/9505186}.
}

\lref\DW{
R. Donagi and E. Witten,
{\it Supersymmetric Yang-Mills theory and Integrable systems},
{\tt hep-th/9510101}.
}

\lref\MW{
G.~Moore and E.~Witten,
{\it Integration over the $u-$ plane in Donaldson theory},
{\tt hep-th/9709193}.
}

\lref\SWa{
N. Seiberg and E. Witten,
{\it Electric-magenectic duality, monopole condensation, and confinement
in $N=2$ supersymmetric Yang-Mills theory},
\np{B 426}{1994}{19}.
}

\lref\SWb{
N. Seiberg and E. Witten,
{\it Monople, duality and chiral symmetry breaking in $N=2$
supersymmetric QCD},
{\tt hep-th/9408099}.
}

\lref\VW{C. Vafa and E. Witten,
{\it A strong coupling test of S-Duality},
Nucl. Phys. {\bf B431} (1994) 3-77,
{\tt  hep-th/9408074}.
}


\lref\DM{
R. Dijkgraaf and G. Moore,
{\it Balanced topological field theories},
Commun. Math. Phys. {\bf 185} (1997) 411-440,
{\tt hep-th/9608169}.
}

\lref\DPS{
R. Dijkgraaf, B.J. Schroers and J.-S. Park,
{\it N=4 supersymmetric Yang-Mills theory on a K\"{a}hler surface}, 
{\tt hep-th/9801066}.
}

\lref\NMVW{
J.A.~Minahan, D, Nemeschansky, C.~Vafa and N.P.~Warner,
{\it E-strings and $N=4$ topological Yang-Mills theory}, 
{\tt hep-th/9802168}.
}

\lref\BJSV{
M. Bershadsky, A. Johansen, V. Sadov and C. Vafa,
{\it Topological reduction of 4D SYM to 2D $\sigma$--models},
Nucl. Phys. {\bf B448} (1995) 166
{\tt hep-th/9501096}.
}

\lref\MNS{
G. Moore, N. Nekrasov and S. Shatashvili,
{\it Integrating Over Higgs Branches},
{\tt hep-th/9712241}.
}

\lref\MGM{
M. Mari\~no and G. Moore,
{\it The Donaldson-Witten function for gauge groups
of rank larger than one},
{\tt hep-th/9802185}..
}

\lref\LNS{
A. Losev, N. Nekrasov and S. Shatashvili,
{\it Issues in Topological Gauge Theory},
\np{B 534}{1998}{549}, 
{\tt hep-th/9711108}.
}

\lref\MMGP{    
Marcos Marino, Gregory Moore, Grigor Peradze,
{\it     
Superconformal invariance and the geography of four-manifolds},
{\tt hep-th/9812055}.
}

\lref\Kanno{
H. Kanno,
{\it Weil algebraic structure and geometrical meaning of 
the BRST transformation in
topological quantum field theory},
Z.~Phys.~{\bf C 43} (1989) 477.
}


\lref\TaubesB{
C.H.~Taubes,
{\it Casson's invariant and gauge theory},
J.~Differ.~Geom.~{\bf 31} (1990) {547}.
}

\lref\AtiyahW{
M.F.~Atiyah,
{\it New invariants of three and four dimensional manifolds},
Proc.~Symp.~Pure Math. {\bf 48} (1988) 285.
}

\lref\FloerB{
A.~Floer,
{\it An instanton invariants for $3$-manifolds},
\cmp{118}{1988}{215}.
}

\lref\Kobayashi{
S.~Kobayashi,
{\it 
Differential geometry of complex vector bundles},
Iwanami Shoten, Publishers and Princeton Univ. Press,
1987.
}

\lref\Mukai{
S.~Mukai, 
{\it Symplectic structure of the moduli spaces of sheaves on
an abelian or K3 surface},
\im{77}{1984}{101}.
}

\lref\AHS{
M.F. Atiyah, N.J. Hitchin and I.M. Singer,
{\it Self-duality in four dimensional Riemannian
geometry},
Proc.~Roy.~Soc.~London {\bf 362} (1978) {425}.
}

\lref\Itoh{
M. Itoh,
{\it Geometry of Yang-Mills connections over K\"{a}hler
surface},
Proc.~Japan Acad. {\bf 59}{1983}{431}.
}

\lref\Kim{
H.J. Kim,
{\it Curvatures and holomorphic bundles},
Ph. D. thesis, Berkely, 1995.
}

\lref\DK{
S.K.~Donaldson and P.B.~Kronheimer,
{\it The geometry of four-manifolds},
Clarendon Press, Oxford 1990.
}

\lref\Donaldson{
S.K.~Donaldson,
{\it Polynomial invariants for smooth $4$-manifolds},
\top{29}{1990}{257}.
}

\lref\KM{
P. Kronheimer and T.  Mrowka,
{\it Recurrence relations and asymptotics for
four manifold invariants},
\bams{30}{1994}{215}.
}

\lref\Taubes{
C.~H.~Taubes,
{\it SW $\rightarrow$ GR. From the Seiberg-Witten equations to
pseudo-holomorphic curves},
\jams{\bf 9}{1996}{845}.
}

\lref\DonaldsonC{
S.K.~Donaldson,
{\it Anti-self-dual Yang-Mills connections on complex
algebraic surfaces and stable vector bundles},
\plms{3}{1985}{1}\semi
{\it Infinite determinants, stable bundles and curvature},
\dmj{54}{1987}{231}.
}

\lref\UY{
K.K.~Uhlenbeck and S.T.~Yau,
{\it The existence of Hermitian Yang-Mills connections on stable
bundles over K\"{a}hler manifolds}.
}

\lref\DT{
S.K. Donaldson and R.P. Thomas,
{\it Gauge theory in higher dimensions}. Preprint.
}

\lref\ThomasA{
R.P. Thomas,
{\it A Holomorphic Casson invariant for Calabi-Yau $3$-folds, and
bundles on $K3$ fibrations},
{\tt math.AG/9806111}.
}

\lref\ThomasB{
R.P. Thomas,
{\it Gauge Theory on Calabi-Yau manifolds},
PhD Thesis.
}


\lref\HarveyMoore{
J. Harvey and G. Moore,
{\it One the algebras of BPS states},
Commun. Math. Phys. {\bf 197} (1998) 489
{\tt hep-th/9609017}.
}

\lref\Inflow{
M. Green, J. harvey and G. Moore,
{\it I-brane inflow and anomalous couplings on D-branes}, 
{\tt hep-th/9605033}.
}


\lref\Atiyah{
M. Atiyah,
{\it Convexity and commuting Hamiltonians},
Bull.~London Math.~Soc. {\bf 14} (1982) 1.
}

\lref\AB{
M.~Atiyah and R.~Bott,
{\it The momentum map and equivariant cohomology},
Topology {\bf 23} (1984) 1.
}

\lref\BGV{
N.~Berline, E.~Getzler and M.~Vergne,
{\it Heat kernels and Dirac operators},
Springer-Verlag,  1992.
}

\lref\MQ{
V. Mathai and D. Quillen,
{\it Superconnections, Thom classes and equivariant differential
forms},
\top{25}{1986}{85}.
}

\lref\MarW{
J. Marsden and A.D. Weinstein,
{\it Reductiion of symplectic manifolds with
symmetry},
Reports on Math. Physics {\bf 5} (1974) 121.
}

\lref\DH{
J.J.~Duistermaat and G.J.~Heckman,
{\it On the variation in the cohomology of the 
symplectic form of the reduced phase space},
\ivm{69}{1982}{259}; Addendum, \ivm{72}{1983}{153}.
}

\lref\Wu{
S.~Wu,
{\it An integration formula for the square of momentum
maps of circle actions},
Lett.~Math.~Phys. {\bf 29} (1993) {311}.
}

\lref\HKLR{
N.J. Hitchin, A. Karlhede, U. Lindstr\"om and M. Rocek,
{\it Hyper-K\"{a}hler metrics and supersymmetry},
\cmp{108}{1987}{535}.
}

\lref\Kirwan{
F.~Kirwan,
{\it 
Cohomology of Quotients in Symplectic and Algebraic
Geometry}, 
Math. Notes {\bf 31}. Princeton Univ. Press,
(Princeton, 1984).
}

\lref\Newstead{
P. Newstead,
{\it Introduction to moduli problems and orbit spaces},
(Tata Institute, 1978).
}

\lref\Mumford{
D. Mumford and J. Fogarty,
{\it Geometrical Invariant Theory},
(Springer, 1982).
}

\lref\AJ{
M. Atiyah and L. Jeffreys,
{\it Topological Lagrangians and cohomology},
\jgp{\bf 7}{1990}{119}.
}


\lref\HitchinA{
N.J. Hitchin,
{\it The self-duality equations on a Riemann surface},
\plms{3, 55}{1987}{59}.
}

\lref\HitchinB{
N.J. Hitchin,
{\it Stable bundles and integrable systems},
\dmj{54}{1987}{91}.
}

\lref\Simpson{
C.T.~Simpson,
{\it Constructing variations of Hodge structure using Yang-Mills theory
and applications to uniformization},
J.~Amer..~Math.~Soc. {\bf 1} (1988) 867.
}

\lref\SimpsonA{
C.T.~Simpson,
{\it Higgs bundles and local systems},
\ihes{75}{1992}{5}.
}

\lref\SimpsonB{
C.T.~Simpson,
{\it Moduli of representations of the fundamental group of
a smooth projective variety}, I:\ihes{79}{1994}{47};
II:\ihes{80}{1994}{5}.
}

\lref\SimpsonC{
C.T.~Simpson,
{\it The Hodge filtration on non-Abelian cohomology},
{\tt alg-geom/9604005}.
}

\lref\SimpsonD{
C.T.~Simpson,
{\it Mixed twistor structure},
{\tt alg-geom/9705006}.
}

\lref\Fujiki{
A.~Fujiki,
{\it HyperK\"{a}hler structure on the moduli space of flat bundles},
in "Prospects in complex geometry",
Lecture Notes in Math.~p. 1-83, Springer-Verlag, 1991.
}

\lref\Kaledin{
D.~Kaledin,
{\it Hyperk\"ahler structures on total spaces of
holomorphic cotangent bundles},
{\tt alg-geom/9710026}.
}

\lref\Hausel{
T.~Hausel,
{\it Compactification of moduli of Higgs bundles},
{\tt math.AG/9804083}.
}


\lref\CVG{
N.~Chriss and V.~Ginzburg,
{\it Representation theory and complex geometry},
Birkh\"auser.
}

\lref\KN{
P.B. Kronheimer and H. Nakajima,
{\it Yang-Mills instantons on ALE gravitational instantons},
\ma{288}{1990}{263}.
}

\lref\NJale{
H. Nakajima,
{\it Homology of moduli spaces of instantons on ALE spaces I},
\jdg{40}{1994}{105}.
}

\lref\NJresol{
H. Nakajima,
{\it Heisenberg algebra and Hilbert schemes of points on
projective surfaces},
{\tt alg-geom/9507012}.
}

\lref\NJquiver{
H. Nakajima,
{\it Instantons on ALE spaces, quiver varieties and Kac-Moody
algebras},
Duke. Math. Jour. {\bf 76} (1994) 365.
}

\lref\NJintro{
H. Nakajima,
{\it Gauge theory on resolutions of simple singularities and
simple Lie algebras},
Int. Math. Res. Notices. {\bf  2} (1994) 61.
}

\lref\IGIT{
S.K. Donaldson,
{\it Instantons and geometrical invariant theory},
\cmp{93}{1984}{453}.
}

\lref\Macio{
A.~Maciocia,
{\it Metrics on the moduli spaces of instantons over
Euclidean 4-space},
\cmp{135}{1991}{467}.
}

\lref\NJhs{
H.~Nakajima,
{\it Lectures on Hilbert schemes of points on surface},
Preprint.
}

\lref\ADHM{
M.F. Atiyah, V. Drinfeld, N.J. Hitchin and
Yu.I. Mannin,
{\it Construction of instantons},
Phys.~Lett.~{\bf A 65} (1978) 185.
}


\lref\chiralrings{
W. Lerche, C. Vafa and N.P. Warner,
{\it Chiral rings in N=2 superconformal theories},
Nucl. Phys. {\bf B324} (1989) 427.
}

\lref\CeVa{
S. Cecotti and C. Vafa, 
{\it Topological anti-topological fusion,} 
Nucl. Phys. {\bf B367} (1991) 359.
}

\lref\kstheory{
M. Bershadsky, S. Cecotti, H. Ooguri  and C. Vafa
{\it Kodaira-Spencer theory of gravity and exact results for 
quantum string amplitudes},
Commun. Math. Phys. {\bf165} (1994) 311-428,
{\tt hep-th/9309140}.
}

\lref\Holo{
M. Bershadsky, S. Cecotti, H. Ooguri and C. Vafa (with 
an appendix by S.Katz)
{\it Holomorphic anomalies in topological field theories},
\np{B405}{1993}{279-304}, 
{\tt  hep-th/9302103}.
}

\lref\Dixon{
L.J.~Dixon,
{\it Some world-sheet results on the superpotential 
form Calabi-Yau compactifications}, 
Superstrings, Unified Theories, and Cosmology
1987, G.~Furlan et al., eds., World  Scientific, 
Singapore, 1988. pp. 67.
}
\lref\Nfour{
N. Berkovits and C. Vafa,
{\it N=4 Topological Strings},
Nucl. Phys. B433 (1995) 123-180,
{\tt hep-th/9407190}.
}

\lref\Ntwo{
H. Ooguri and C. Vafa,
{\it All Loop N=2 String Amplitudes},
 Nucl. Phys. B451 (1995) 121,
{\tt  hep-th/9505183}.
}


\lref\MirrorA{
{\it Essays on mirror manifolds}, 
S.-T.~Yau ed., International Press, 
Hong Kong,1992.
}

\lref\MirrorB{
{\it Mirror symmetry II}, B. Greene and S.-T.~Yau eds, 
AMS/IP Studies in Adv. Math. 1., AMS, Providence, RI;
International Press, Cambridge, MA (1997).
} 

\lref\Morrison{
D.R. Morrison,
{\it Mathematical aspects of mirror symmetry},
{\tt alg-geom/9609021}.
}

\lref\AGM{
P.S.~Aspinwall and B.R.~Greene, and D.R.~Morrison,
{\it Calabi-Yau moduli space, mirror manifolds and spacetime topology
change in string theory},
\np{B 416}{1994}{414}.
}

\lref\MorP{
D.R. Morrison, M.R. Plesser
{\it Summing the instantons: Quantum cohomology and 
mirror symmetry in toric varieties},
Nucl. Phys. {\bf B440} (1995) 279-354,
{\tt hep-th/9412236}.
}

\lref\GivK{
A. Givental and B. Kim,
{\it Quantum cohomology of flag manifolds and
Toda lattices}, 
{\tt hep-th/9312096}.
}

\lref\GMP{
B. Greene, D. Morrison, M. Plesser,
{\it Mirror manifolds in higher dimension},
\cmp{173}{1995}{559},
{\tt  hep-th/9402119}.
}


\lref\Kontsevich{
M.~Kontsevich,
{\it Homological algebra of mirror symmetry},
Proceedings of the International Congress of Mathematicians,
pp. 120, Birkh\"auser, 1995, 
{\tt alg-geom/9411018}.
}

\lref\BarKon{
S. Barannikov and M. Kontsevich,
{\it Frobenius manifolds and formality of Lie algebras of polyvector
fields},
{\tt alg-geom/970032}.
}

\lref\Kontformal{
M.~Kontsevich,
{\it Deformation quantization of Poisson Manifolds, I},
{\tt q-alg/9709040}.
}

\lref\ManninF{
Yu.I. Mannin,
{\it Three contructions of Frobenius manifolds: a comparative
study}, 
{\tt math.QA/9801006}.
}

\lref\Koszul{
J.-L. Koszul,
{\it Crochet de Schouten-Nijenhuis et cohomologie},
in "Elie Cartan t les mathe\'ematique d'aujourd'huis',
Asterisque (1985) 251.
}

\lref\Merk{
S.A. Merkulov,
{\it Formality of canonical symplectic complexes and Frobenius
manifolds,}
{\tt math.SG/9805072}.
}

\lref\FukayaA{
K. Fukaya,
{\it Morst homotopy, $A^\infty$ category, and Floer homologies},
in {\it GARC Workshop on Geometry and Topology}. 1993.
Seoul Nat. Univ.
}

\lref\FukayaB{
K. Fukaya,
{\it Mirror symmetry of Abelian variety and multi Theta functions},
Preprint, Kyoto Univ.
}

\lref\FukayaC{
K. Fukaya,
{\it Lagrangian submanifolds and mirror symmetry},
in {Symplectic geometry, mirror symmetry and string theory}.
1999. KIAS Lecture Notes.
}

\lref\FKO{
K.~Fukaya, M.~Kontsevich, Y.~Oh, H.~Ohta and K.~Ono,
{\it Anomaly in Lagrangian intersection Floer homology},
to appear.
}

\lref\Tian{
G. Tian,
{\it  Geometry of Calabi-Yau manifolds},
in {Symplectic geometry, mirror symmetry and string theory}.
1999. KIAS Lecture Notes.
}

\lref\PZ{
A.~Polishchuk and E.~Zaslow,
{\it Categorial mirror symmetry: the elliptic curve,}\
{\tt math.ag/9801119}.
}

\lref\SYZ{
A. Strominger, S.-T. Yau and E. Zaslow,
{\it Miror symmetry is a $T$-duality},
\np{B 479}{1996}{243},
{\tt hep-th/9612121}.
}

\lref\LV{
N.C. Leung and C. Vafa,
Adv. Theor. Math. Phys. {\bf 2} (1998) 91.
}

\lref\Vafa{
C. Vafa, {\it Extending mirror conjecture to Calabi-Yau with bundles},
{\tt hep-th/9804131}.
}

\lref\Tyurin{
A. Tyurin, 
{\it Geometric quantization and mirror symmetry},
{\tt math/9902027}.
}

\lref\Vafaicm{
C. Vafa,
{\it Geometric physics},
{\tt hep-th/9810149}.
}

\lref\BBS{
K.~Becker, M.~Becker and A. Strominger,
{\it Fivebranes, membranes and non-perturbative string theory},
{\tt hep-th/9507158}.
}

\lref\McLean{
R. McLean,
{\it Deformation of Calibrated submanifold},
preprint, 1996.
}


\lref\Zumino{
B. Zumino,
{\it Supersymmetry and K\"{a}hler Manifolds},
\pl{B 87}{1979}{203}.
}

\lref\AGF{
L.~Alvarez-Gaume, D.Z. Freedman,
{\it Geometrical structure and ultraviolet
finiteness
in the supersymmetric sigma model},
\cmp{80}{1981}{443}.
}

\lref\GHR{
S.J. Gates, Jr, C.M. Hull and M. Rocek,
{\it Twisted multiplets and new supersymmetric non-linear
$\s$-models},
\np{B 248}{1984}{157}.
}


\lref\Park{
J.-S.~Park,
{\it N=2 topological Yang-Mills theory
on compact K\"{a}hler surfaces},
\cmp{163}{1994}{113},
{\tt hep-th/9304060}.
}

\lref\HYM{
J.-S. Park,
{\it  Holomorphic Yang-Mills theory on compact K\"{a}hler manifolds},
Nucl. Phys. {\bf B423} (1994) 559, 
{\tt hep-th/9305095}.
}

\lref\HyunPa{
S.~Hyun and J.-S.~Park,
{\it N=2 topological Yang-Mills theories
and Donaldson's polynomials},
J. Geom. Phys. {\bf 20} (1996) 31-53, 
{\tt hep-th/9404009}.
}

\lref\HYMDP{
S.~Hyun and J.S.~Park,
{\it Holomorphic Yang-Mills theory and the
variation of the Donaldson polynomials},
{\tt hep-th/9503092}.
}

\lref\HPA{
C.M.~Hofman and J.-S.~Park,
{\it Sigma Models for Quiver Variety},
unpublished.
}

\lref\HPB{
C.~Hofman and J.-S.~Park,
{\it Sigma Models for bundles on Calabi-Yau: a proposal for
matrix string compactifications},
Nucl.~Phys.~B in press.
{\tt hep-th/9904150}.
}

\lref\HPC{
C.~Hofman and J.-S.~Park,
{\it Some comments on Holomorphic Yang-Mills theory},
in preparation.
}

\lref\Monads{
J.-S.~Park,
{\it Monads and D-instantons},
\np{B 493}{1997}{198}.
}


\lref\Mtheory{
E. Witten, 
{\it String theory dynamics in various dimensions}, 
Nucl. Phys. {\bf B443} (1995) 85,
{\tt hep-th/9503124}.
}

\lref\Polchinski{ J.~Polchinski, 
{\it Dirichlet branes and Ramond Ramond charges},  
{\tt hep-th/951001}\semi
J. Polchinski, S. Chaudhuri, and C. Johnson, 
{\it Notes on D-branes}, 
{\tt hep-th/9602052}\semi
J. Polchinski, {\it TASI Lectures on D-branes}, 
{\tt hep-th/9611050}.
}

\lref\Wittenbound{ E.~Witten, 
{\it Bound states of strings and $p$-branes},
Nucl. Phys.  {\bf B460} (1996) 335--350,
{\tt hep-th/9510135}.
}


\lref\BFSS{ T. Banks, W. Fischler, S.H. Schenker, and L. Susskind,
{\it  Theory As A Matrix Model: A Conjecture}, 
Phys. Rev. {\bf D55} (1997) 5112-5128,
{\tt hep-th/9610043}.
}

\lref\Wati{
W.~Taylor, 
{\it D-brane field theory on compact spaces},
\pl{B394}{1997}{283},
{\tt hep-th/9611042}.
}

\lref\motl{
L.~Motl, 
{\it Proposals on non-perturbative superstring interactions},
{\tt hep-th/9701025}.
}

\lref\tomnati{
T.~Banks and N.~Seiberg, 
{\it Strings from matrices},
{\tt hep-th/9702187}.
}

\lref\matrixstring{ 
R. Dijkgraaf, E. Verlinde, and H. Verlinde, 
{\it Matrix string theory},
{\tt hep-th/9703030}.
}

\lref\Why{
N. Seiberg,
{\it Why is the matrix model correct?},
Phys. Rev. Lett. {\bf 79} (1997) 3577-3580,
{\tt hep-th/9710009}.
}

\lref\Sen{
A. Sen,
{\it D0 branes on $T^n$ and matrix theory}, 
Adv. Theor. Math. Phys. {\bf  2} (1998) 51,
{\tt hep-th/9709220}.
}


\lref\SV{
A. Strominger and C. Vafa,
{\it Microscopic origin of the Bekenstein-Hawking entropy},
\pl{B379}{1996}{99}, 
{\tt hep-th/9601029}.
}

\lref\fivebrane{ 
R.~Dijkgraaf, E.~Verlinde and H.~Verlinde, 
{\it BPS spectrum of the five-brane and black hole entropy}, 
Nucl. Phys.  {\bf B486} (1997) 77, 
{\tt hep-th/9603126}; 
{\it BPS quantization of the five-brane},
Nucl. Phys. {\bf B486} (1997) 89, 
{\tt hep-th/9604055}.
}

\lref\dyon{
R. Dijkgraaf, E. Verlinde, and H. Verlinde, 
{\it Counting Dyons in N=4 String Theory,}'
Nucl.Phys. {\bf B 484} (1997) 543, 
{\tt hep-th/9607026}.
}

\lref\kawai{T. Kawai, ``$N=2$ Heterotic String Threshold
{\it Correction, $K3$ Surface and Generalized Kac-Moody Superalgebra},
{\tt hep-th/9512046}.
}

\lref\orbifolds{
 R. Dijkgraaf, G, Moore, E. Verlinde, and H. Verlinde,
{\it Elliptic genera of symmetric products and second quantized strings},
Commun. Math. Phys. {\bf 185} (1997) 197,
{\tt hep-th/9608096}.
}


\lref\berkoozdouglas{
M. Berkooz and M. Douglas, 
{\it Fivebranes in M(atrix) Theory},
{\tt hep-th/9610236}.
}

\lref\mbh{
R. Dijkgraaf, E. Verlinde, and H. Verlinde, 
{\it 5D black holes and matrix strings},
{\tt hep-th/9704018}.
}

\lref\DVVrev{
R. Dijkgraaf, E. Verlinde, H. Verlinde,
{\it Notes on matrix and micro strings},
{\tt hep-th/9709107}.
}

\lref\JuanLN{
J.M. Maldacena
{\it The large N limit of superconformal field theories and 
supergravity},
{\tt hep-th/9711200}.
}

\lref\Robbert{
R. Dijkgraaf,
{\it Instanton strings and HyperK\"{a}hler geometry},
{\tt hep-th/9810210}.
}

\lref\ABKSS{
O. Aharony, M. Berkooz, S. Kachru, N. Seiberg and E. Silverstein,
{\it Matrix description of interacting theories in six dimensions},
Adv. Theor. Math. Phys.{\bf 1} (1998) 148-157,
{\tt hep-th/9707079}.
}

\lref\ABS{
O. Aharony, M. Berkooz and N. Seiberg, 
{\it Light-cone description of (2,0) superconformal theories 
in six dimensions}, 
Adv. Theor. Math. Phys. {\bf 2} (1998) 119-153,
{\tt hep-th/9707079}.
}

\lref\deBor{
J. de Boer,
{\it Six-dimensional supergravity on $S^3 X AdS_3$ and 
2d conformal field theory},
{\tt hep-th/9806104}\semi
{\it Large N elliptic genus and AdS/CFT correspondence},
{\tt hep-th/9812240}.
}


\lref\AMK{ 
P.S. Aspinwall and D.R. Morrison, 
{\it String theory on $K3$ surfaces,} 
{\tt hep-th/9404151}, in {\sl Essays on mirror manifolds
2}.\\ 
P.S. Aspinwall, 
{\it $K3$ Surfaces and string duality}, 
{\tt hep-th/9611137}.
}

\lref\Vafac{
C.~Vafa,
{\it c-Theorem and the topology of 2-d QFTs},
\pl{B212}{1988}{28}.
}

\lref\Douglas{
M.R. Douglas,
{\it D-branes in curved space},
{\tt hep-th/9702048}\semi
{\it D-branes and matrix theory in curved space},
lecture given at  Strings '97.
{\tt hep-th/9707228}.
}

\lref\DKO{
M.R. Douglas, A. Kato and H. Ooguri,
{\it D-brane actions on K\"{a}hler manifolds},
{\tt hep-th/9708012}.
}

\lref\Wynter{
Ph. Brax, T. Wynter,
{\it Limits of matrix theory in curved space},
{\tt hep-th/9806176}.
}


\lref\Yamron{
J.~Yamron, 
{\it Topological actions in twisted supersymmetric theories},
\pl{B213}{1988}{353}.
}


\lref\GT{
D.J.~ Gross,
{\it Two dimensional QCD as a string
theory}
\np{B400}{1993}{161}
{\tt hep-th/9212149}.
\semi
D.J.~Gross and W.~Taylor,
{\it Two dimensional QCD is a string theory},
\np{B400}{1993}{181},
{\tt hep-th/9301068}.
}

\lref\Horava{ 
P.~Horava,
{\it
Topological rigid string theory and two dimensional QCD},
{\tt hep-th/9507060};
{\it
Topological strings and QCD in two dimensions},
{\tt hep-th/931115}.
}

\lref\CMR{
S.~Cordes, G.~Moore, S.~Ramgoolam,
{\it Large N 2D Yang-Mills theory and topological string 
theory},
{\tt hep-th/9402107};
{\it Lectures on 2D Yang-Mills theory, equivariant 
cohomology and topological field theories },
{\tt hep-th/9411210}.
}

\lref\LMNS{
A.~Losev, G.~Moore, N.~Nekrasov and S.~Shatashvili,
{\it Four-dimensional avatars of two-dimensional RCFT},
{\tt hep-th/9509151};
{\it Chiral Lagrangians, anomalies, supersymmetry, 
and holomorphy},
{\tt hep-th/9606082}.
}

\lref\NS{ 
V.P. Nair, J.~Schiff,
{\it K\"{a}hler Chern-Simons theory and quantization of 
instanton moduli spaces},
\pl{B246}{1990}{423};
{\it K\"{a}hler Chern-Simons theory
and symmetries of antiselfdual gauge fields},
\np{B371}{1992}{329}.
}

\lref\Gott{
L.~G\"{o}ttsche,
{\it Modular forms and Donaldson invariants for 4-manifolds 
with $b_+=1$},
{\tt alg-geom/9506018}.
}


\lref\twozero{J. Distler and B. Greene,
{\it Aspects of (2,0) string compactifications},
Nucl. Phys. {\bf B304} (1988) 1.
}

\lref\AspMor{P. Aspinwall and D. Morrison, 
{\it Topological Field Theory and
Rational Curves}, 
Comm. Math. Phys. {\bf 151} (1993) 245.
}

\lref\CandMir{P. Candelas, X. de la Ossa, P. Green and L. Parkes,
{\it A Pair of Calabi-Yau Manifolds as an Exactly Soluble Superconformal
Theory},
Nucl. Phys. {\bf B359} (1991) 21.
}

\lref\DKone{J. Distler and S. Kachru, 
{\it (0,2) Landau-Ginzburg Theory},
Nucl. Phys. {\bf B413} (1994) 213, 
{\tt hep-th/9309110}.
}

\lref\DKtwo{J. Distler and S. Kachru, 
{\it Singlet Couplings and (0,2)
Models}, 
Nucl. Phys. {\bf B430} (1994) 13, 
{\tt hep-th/9406090}.
}

\lref\DKthree{J. Distler and S. Kachru, 
{\it Quantum Symmetries and
Stringy Instantons}, 
Phys. Lett. {\bf 336B} (1994) 368, 
{\tthep-th/9406091}..
}

\lref\DKfour{J.Distler and S. Kachru, 
{\it Duality of (0,2) String vacua},
{\tt hep-th/9501111}.
}

\lref\TKawai{T. Kawai and K. Mohri, 
{\it Geometry of (0,2) Landau-Ginzburg
Orbifolds}, 
Nucl. Phys. {\bf B425} (1994) 191, 
{\tt hep-th/9402148}.
}

\lref\Distler{
J. Distler,
{\it  Notes on (0,2) superconformal field theories},
{\tt hep-th/9502012}.
}


\lref\NekS{
N. Nekrasov and A. Schwarz,
{\it 
Instantons on noncommutative $R^4$, and (2,0) superconformal
six dimensional theory},
\cmp{198}{1998}{689},
{\tt hep-th/9802068}..
}

\lref\IKKT{N. Ishibashi, H. Kawai, Y. Kitazawa and A. Tsuchiya
{\it A large-N reduced model as superstring},
\np{B 498}{1997}{467},
{\tt hep-th/9612115}..
}

\lref\Givental{
A. Givental, {\it Equivariant Gromov-Witten invariants},
Int.~Math.~Res.~Notices {\bf 13} (1990) 337.
}

\lref\Dubrovin{
B. Dubrovin,
{\it Geometry of $2d$ topological field theory},
LNM 1620, Springer, 1996, pp.120.
}

\lref\Roberticm{
R.Dijkgraff,
{\it The matematics of five branes},
Lectures given at International Congress of Mathematicians (ICM 98),
Berlin, Germany, 18-27 Aug 1998, 
{\tt hep-th/9810157}.
}

\lref\MO{ 
C. Montonen and D. Olive, 
{\it Magnetic monopoles as gauge particles?}, 
Phys. Lett. {\bf 72B} (1977) 117.
}

\lref\Osborn{
H. Osborn, 
{\it Topological charges for $N=4$ supersymmetric gauge
theories and monopoles of spin 1,} 
Phys. Lett. {\bf 83B} (1979) 321..
}

\lref\Udual{
C. Hull and P. Townsend, 
{\it Unity of superstring dualities,} Nucl. Phys. {\bf B 438} (1995) 109, 
{\tt hep-th/9410167}.
}

\lref\Wittenduality{ E. Witten, 
{\it String theory in various dimensions,} 
Nucl. Phys. {\bf B 443} (1995) 85, 
{\tt hep-th/9503124}.
}

\lref\Schwarzsl{J.H. Schwarz, 
{\it An $SL(2,sZ)$ multiplet of type IIB superstrings,} 
Phys. Lett. {\bf B360} (1995) 13--18, 
{\tt hep-th/9508143}..
}

\lref\BBRT{ 
D. Birmingham, M. Blau, M. Rakowski and G. Thompson, 
{\it Topological field theory}, 
Phys. Rep. {\bf 209} 4 \& 5 (1991)..
}

\lref\DVV{R. Dijkgraaf, E. Verlinde, and H. Verlinde, 
{\it Notes on topological string theory and 2d quantum gravity,} 
in {\sl String Theory and Quantum Gravity}, 
Proceedings of the Trieste Spring School1990, 
M. Green {\it et al.} Eds. (World-Scientific, 1991).
}

\lref\Kontmanin{ 
M. Kontsevich and Yu. Manin, 
{\it Gromov-Witten classes, quantum cohomology, and enumerative geometry,}
Commun. Math. Phys. {\bf 164} (1994) 525--562, 
{\tt hep-th/9402147}.
}

\lref\Gromov{
M. Gromov, 
{\it Pseudo-holomorphic curves on almost
complex manifolds,} 
Inv. Math. {\bf 82}(1985) 307.
}

\lref\CHSW{P. Candelas, G. Horowitz, A. Strominger, and E. Witten,
{\it Vacuum configurations for superstrings,} 
Nucl. Phys. {\bf B258} (1985) 46..
}

\lref\Wittencoft{E. Witten, 
{\it Introduction to cohomological field
theories,} 
Int. J. Mod. Phys. {\bf A6} (1991) 2775.
}

\lref\CMR{ S.\ Cordes, G.\ Moore, and S.\ Rangoolam, 
{\it Lectures on 2-D Yang-Mills theory, equivariant cohomology 
and topological field theories,} 
Proceedings of the Trieste Spring School and the Les Houches
Summer School 1994, 
{\tt hep-th/9411210}.
}

\lref\BT{ M.~Blau and G.~Thompson, 
{\it $N=2$ topological gauge theory, the Euler characteristic of 
moduli spaces, and the Casson invariant,} 
Commun. Math. Phys. {\bf 152} (1993) 41-72. 
}.

\lref\WO{
E. Witten and D. Olive, 
{\it Supersymmetry algebras that include topological charges,} 
Phys. Lett.{\bf 78B} (1978) 97.
}

\lref\Strominger{
A. Strominger, {\it Massless black holes and
conifolds in string theory,} 
Nucl. Phys. {\bf B451} (1995) 96--108, 
{\tt hep-th/9504090}.
}

\lref\BSV{ 
M. Bershadsky, V. Sadov, and C. Vafa, 
{\it D-Branes and topological field theories,} 
{\tt hep-th/9511222}.
}

\lref\LaLD{J. M. F. Labastida and C. Lozano,
{\itThe Vafa-Witten theory for gauge group SU(N)},
{\tt hep-th/9903172}.
}

\lref\LaLC{J. M. F. Labastida and C. Lozano,
{\it 
Duality in twisted N=4 supersymmetric gauge theories in four
dimensions},
Nucl.Phys. {\bf B 537} (1999) 203,
{\tt hep-th/9806032}.
}

\lref\LaLB{
J. M. F. Labastida, Carlos Lozano,
{\it 
Mass perturbations in 
twisted N=4 supersymmetric gauge
theories},
Nucl.Phys. {\bf B 518} (1998) 37,
{\tt hep-th/9711132}.
}

\lref\BKS{
L. Baulieu, H. Kanno, I. M. Singer
{\it Special quantum field theories In eight and other dimensions}
 \cmp{194}{1998}{149},,
 {\tt hep-th/9704167}.
}

\lref\QMWA{
B. S. Acharya, M. O'Loughlin, B. Spence
{\it Higher Dimensional Analogues of Donaldson-Witten Theory}
\np{B503 }{1997}{657},
{\tt hep-th/9705138}..
}

\lref\BTE{
M.~Blau and G.~Thompson,
{\it Euclidean SYM theories by time reduction and
special holonomy manifolds},
{\tt hep-th/9706225}.
}

\lref\QMWB{
BS Acharya, JM Figueroa-O'Farrill, M O'Loughlin
and B Spence,
{\it Euclidean D-branes and higher-dimensional gauge
theory},
\np{B 514}{1998}{583},
{\tt hep-th/9707118}.
}

\lref\LLN{ 
L. Baulieu, A. Losev, N. Nekrasov
{\it Chern-Simons and twisted supersymmetry in higher
    dimensions},
\np{B 522}{1998}{82},
{\tt hep-th/9707174}.
}

\lref\FIM{
JM.~ Figueroa-O'Farrill, A.~Imaanpur, and J.~McCarthy,
{\it Supersymmetry and gauge theory in Calabi-Yau $3$-folds},
{\tt hep-th/9709178}.
}

\lref\Fulton{
W.~Fulton,
{\it Intersection Theory},
Springer-Verlag 1984.
}


\rightline{ITFA-99/29, CU-TP-956}
\bigskip
\bigskip
\centerline{\LARGE \sc
Cohomological Field Theories}
\medskip
\centerline{\LARGE\sc  with K\"{a}hler Structure}
\bigskip
\centerline{Jae-Suk Park}
\medskip
\centerline{\small Institute for Theoretical Physics, 
University of Amsterdam}
\centerline{\small Valckenierstraat 65, 1018 XE Amsterdam}
\centerline{\small and}
\centerline{\small Department of Physics, Columbia University}
\centerline{\small 538 West 120th Street, New York, N.Y.~10027}

\vskip 1cm

{\small
This paper is devoted to a general and self-contained
approach to any cohomological 
field theory with K\"{a}hler structure. 
}

\newsec{Introduction}

Historically  the cohomological field
theory first has been introduced as  a {\it twisted} version
of global space-time supersymmetric quantum field theory,
specifically the $N=2$ supersymmetric Yang-Mills theory
in four dimensional space-time \TFT.
The global space-time supersymmetry, by definition,
requires the existence of  a spinor which is constant  everywhere on
the space-time manifold $M$. A spinor does exist on a spin
manifold. A spin manifold, however,
rarely admits a constant spinor.
The canonical way  overcoming the above difficulty 
is localizing the supersymmetry,
which  procedure almost magically introduces (super-)gravity into
the picture.

There is a  second option called twisting, meaning
that  one defines a new Lorentz symmetry group by a suitable
combination of the original Lorentz symmetry with
an internal global symmetry of the theory. As a result,
the supercharges transform differently under the new Lorentz 
symmetry, which typically includes some components which
transform as scalars. 
Such a scalar component $Q$,  which is nilpotent
$Q^2=0$, is regarded as a supercharge of
the twisted theory.
The resulting theory is well-defined on an arbitrary space-time
since there are no global obstruction for a scalar
and enjoys general covariance without gravity.
The path integral of the theory depends only on the
global cohomology of $Q$, provided that one uses $Q$-invariant
observables, which property coined
the adjective cohomological \Wittencoft.

A twisted theory is closely related to the underlying
space-time supersymmetric theory.
Namely the path integral of the twisted theory computes
a certain chiral (or BPS) sector of physical 
amplitudes \Wittenmirror\chiralrings.
This is due to the trivial holonomy of flat space-time
where the physical theory is usually defined.
Then twisting is a physically invisible operation.
The typical physical application of a twisted theory
is a non-perturbative test of certain duality utilizing
the semi-classical exactness of the path integral. 
Two famous examples
are given by mirror symmetry \MirrorA\MirrorB\
and the $S$-duality of $N=4$ supersymmetric Yang-Mills theory 
in four-dimensions \VW.
The twisted version of four dimensional $N=2$ supersymmetric
Yang-Mills theory - the Donaldson-Witten
theory \TFT\Donaldson,  also provided crucial hints 
\KM\Wittenkaehler\ on the celebrated 
Seiberg-Witten solutions of the original $N=2$ theory \SWa.\foot{
We must stress here that solutions of the underlying physical
theory provides us with invaluable insights in the mathematical
problem defined by the twisted theory.  Perhaps one of the most
beautiful properties of quantum field theory is that the theory
depends on a scale. The equally beautiful property of cohomological
field theory is that the theory does not depends on a scale.
Thus the mathematical problem defined by the latter theory
can be solved in terms of the former theory in different
scale where its relevant degrees of freedom is, often, completely 
different from the original microscopic ones. The historical example is,
of course, the Donaldson versus Seiberg-Witten invariant \SWinv.
}

In general  we may forget about the underlying physical
origin of a cohomological field theory and define
the theory as  a quantum field theory  with global
fermionic symmetry.  Such a theory may not be directly obtainable
as a twisted version of underlying space-time 
supersymmetric theory.
The most fundamental property of a quantum field
theory with global fermionic symmetry is the fixed point
theorem of Witten \Wittenmirror\Wittengwzw. Almost
all the other properties of cohomological field theory
can be obtained as certain a lemma of the theorem.

In this paper
we  develop a general approach
which identifies any cohomological field theory
with $0+0$-dimensional supersymmetric sigma
model. Being in zero-dimensions the (space-time) 
supersymmetry simply means global fermionic symmetry. 
The target space of our sigma-model may be some
function space $X$ in the theorem quoted above.
Such a space may be any (non-linear or linear
and finite or infinite dimensional) endowed with
any of
\eqn\vvv{
\hbox{
Riemannian  $\supset$
K\"{a}hler $\supset$  hyper-K\"{a}hler
}
}
structures. Actually the above structures may not be regarded 
as {\it a priori} notions. The cohomological field theory
can be classified by the number  $N_c=(\CN^+_c, N_c^-)$ 
of global supercharges, where we have $N_c^+ + N_c^-$
independent {\it mutually nilpotent}  fermionic charges
and $N_c^\pm$ denote
the number of charges carrying fermionic (or ghost) numbers $\pm 1$.
Then we have the following sequence of fermionic
symmetries
\eqn\vvvv{
N_c^+=1 \supset N_c^+=2 \supset N_c^+=4,
}
which determine the sequence of geometrical structures \vvv.\foot{
The above correspondence
is originally due to supersymmetric 
sigma models in two-dimensions \Zumino\AGF\HKLR.
In certain respects, such a correspondence in zero-dimensional
models is more striking since we do not need any underlying
geometrical objects like the two-dimensional space-time.
Actually the sequence \vvvv\ leads to more general geometrical
structures including torsion \GHR. However, the author is not
aware of any examples of traditional cohomological
field theory with torsion in the space of fields.
}  

In this paper we specialize to   models with a K\"{a}hler structure.
Those models are quite general and allow us to have very compact
formulations.  The initial data will be some function space $X$ endowed
with a complex structure compatible with the supersymmetry.
Then most of the other structures of the models can be fixed.
We will introduce three types of
models, two with $N_c=(2,0)$ and one with $N_c=(2,2)$
symmetry, and establish general interrelations. 
For each type we will consider non-linear
$X$  and linear or non-linear $X$ with a group $\CG$ acting on $X$. 
In due course the relation between our construction 
and two-dimensional supersymmetric field theory  
will become obvious.    
This opens up possibilities of stringy
generalizations of those differential-topological invariants 
defined by cohomological field theory. 
This chapter may be also viewed, after slight modifications, 
as an unorthodox introduction to  two-dimensional supersymmetric
field theories.  Our presentation for models with a group action
will  parallel  the original literature on $N_{ws}=(2,2)$
and $N_{ws}=(2,0)$ 
gauged linear sigma-models in two dimensions 
\GLSM\Wittengr. We should also mention the influential paper
of Witten on supersymmetry and Morse theory \Wittenmorse\
dealing with $(0+1)$-dimensional supersymmetric sigma models,
which can be regarded as the origin of cohomological field
theory.\foot{It is ironical since his construction can be regarded
as, in our viewpoints, a generalized cohomological field theory.
} 
 
Perhaps our definition for cohomological field theory
as a zero dimensional sigma model might be confusing.  
If the target space $X$ is the function space
of certain fields  on a manifold $M$  we have a traditional 
cohomological field theory on $M$.
As some  general literature for cohomological field
theory we refer to   \Wittencoft\ and \AJ\ for short
but lucid introductions. We refer to \CMR\
for a general reference for the Riemannian version
of $\CN_c=(2,0)$ models.
For the Riemannian version
of $\CN_c=(2,2)$ model, called balanced cohomological
field theory, we refer to \DM.
For a mathematician the path integral of a cohomological
field theory is Mathai-Quillen formalism of integral representation
of Thom class \MQ\AJ. Though we will never refer to  Mathai
and Quillen, our (path) integral formula can be viewed as 
K\"{a}hler version of Mathai-Quillen formalism. More precisely 
our formua should be viewed as a certain equivariant generalization 
of Fulton and MacPerson's intersection theory \Fulton.
For a physicist
a cohomological field theory a supersymmetric gauged sigma model in
$(0+0)$-dimensions. Though we will never use the superspace
formalism our construction is equivalent to $N=2$ superspace
formalism.

The most fundamental property of a quantum field
theory with a global fermionic symmetry is the fixed point
theorem of Witten.  Almost
all the other properties of cohomological field theory
can be obtained as a certain lemma of the theorem.
We refer to the original references \Wittenmirror\Wittengwzw\ 
for this crucial theorem.

\newsec{Standard Models of Cohomological Field Theory}

This and the next chapters are devoted to an elementary and self-contained 
introduction to cohomological field theory.  
Though elementary,
we will develop the most general construction of cohomological
field theory involving K\"{a}hler geometry.

\def\E{\msbm{E}}
In this chapter we consider supersymmetric sigma models
in $(0+0)$ dimensions, whose target space is a compact complex
K\"{a}hler manifold $X$.  Those models may be
regarded as the quantum theory of single point-like "instanton" -
the point-like event  of $X$ or point-like instanton probes
of the classical geometry of $X$ by means of the
path integral.
The space of all bosonic field will be the configuration
space of the instanton, which is a copy of the manifold
$X$. We will start from the simplest $N_c=(2,0)$ model
as a toy model.  A slightly more complicated $N_c=(2,2)$
model follows. Then we generalize it to another $N_c=(2,0)$
model. We will survey how those supersymmetric theories
probe or give rise to the classical geometry of K\"{a}hler 
manifolds $X$, its tangent
bundle $TX$ and holomorphic Hermitian vector bundle $\E$
over $X$. The models to be covered here will be used 
as the prototypes of all the other more elaborated models 
to be introduced later. 
We refer  to the models in this chapter as standard models
since any cohomological field theory will reproduce to one of
those models if it is "generic".

We follow a typical
procedure of defining supersymmetric field theory,
namely introducing bosonic fields, supercharges 
with their algebra, fermionic superpartners,
supersymmetric action functional, and studying path integrals. 
Due to the triviality of the model
everything can be made completely rigorous.
Assuming existence of nil-potent supercharges, a simple application
of Poincar\'e lemma leads to an appropriate supersymmetric
action functional. All the other geometrical structures then
naturally follow.
We will also clarify the geometrical meaning of the supercharges.

\subsec{A Toy Model}

In this section we design perhaps the simplest path
integral, which has many of the basic properties of
cohomological field theory.

Consider a compact complex $n$-dimensional  space $X$ .
We pick local coordinates $x^I$, $I=1,\ldots, 2n$  on $X$.
The local complex coordinates on $X$
will be denoted as $z^i$, $i=1,\ldots,n$; their complex conjugates
are $z^{\bar i}=\bar{z^i}$.  
Let $X^I$ be local coordinates fields describing
the position of an instanton  on $X$. 
More precisely, the
$X^I$ parameterize a map 
\eqn\aaa{
X^I :point \rightarrow X.
}
We denote by $X^i$  local complex coordinates fields
and $X^{\bar i}$ be their complex conjugates.
We call $X^i$ and 
$X^{\bar i}$ bosonic fields. 
We introduce anti-commuting operators $\bs$
and $\bbs$ called supercharges satisfying the following 
anti-commutation relations,
\eqn\aab{
\bs^2=0,\qquad \{\bs,\bbs\}=0,\qquad
\bbs^2=0.
}
We define a pair of graded quantum number (ghost numbers)
$(p,q)$ such that $\bs$ and $\bbs$ carry the following
ghost numbers
\eqn\aac{
\bs:(1,0), \qquad\bbs:(0,1).
}
We call the supersymmetry \aab\ of type $N_c=(2,0)$,
meaning that we have two supercharges both 
carrying positive ghost numbers.

We assume that the $X^i$ are holomorphic 
fields, meaning that  $\bbs X^i=0$, 
and their complex conjugate  $X^{\bar i}$ are anti-holomorphic,  
$\bs X^{\bar i}=0$. Then we can postulate the following
supersymmetry transformation laws 
\eqn\aad{
\eqalign{
\bs X^i &= i\p^i,\cr
\bbs X^i&=0,\cr
\bs X^{\bar i}&=0,\cr
\bbs X^{\bar i}&=i\p^{\bar i},\cr
}\qquad
\eqalign{
\bs \p^i &= 0,\cr
\bbs \p^i&=0,\cr
\bs \p^{\bar i}&=0,\cr
\bbs \p^{\bar i}&=0.\cr
}
}
{}From the above we may write $\bs$ and $\bbs$ as follows
\eqn\aadd{
\bs =i\p^i\Fr{\rd}{\rd X^i},\qquad
\bbs =i\p^{\bar i}\Fr{\rd}{\rd X^{\bar i}}.
}
We call the anti-commuting superpartners  
$\p^i$ and $\p^{\bar i}$ of
$X^i$ and $X^{\bar i}$, respectively,  fermionic fields.
They carry the ghost numbers $(1,0)$ and $(0,1)$, respectively.
In general, a field with ghost number $(p,q)$ is fermionic
if $p+q$ is odd while, otherwise, it is bosonic.

Now we consider an action functional 
$S(X^i,X^{\bar i}, \p^i,\p^{\bar i})$ which is 
invariant under  both of  thesupersymmetries with supercharges 
$\bs$ and $\bbs$. 
The conditions for supersymmetry  $\bs S=\bbs S=0$
together with the anti-commutation relations \aab\
imply, due to the Poincar\'{e} lemma, that $S$ may be written as
\eqn\aae{
S =i \bs\bbs \CK(X^i, X^{\bar i}),
}
where $\CK$ is a  locally defined real functional of $X^i$ and $X^{\bar i}$.
Applying the transformation laws \aad\ we have
\eqn\aaf{
S=i\left(\Fr{ \rd^2 \CK }{\rd X^i\rd X^{\bar j}}\right)\p^i \p^{\bar j}
:= -i \CK_{i\bar j} \p^i\p^{\bar j}.
}

Now we consider the Feynman path integral of our model.
The partition function 
is defined as  integration over the space of all fields weighted by $e^{-S}$,
\eqn\aae{
Z = \int [\CD X \CD\bar X\CD\p\CD\bar\p] \;e^{-S}.
}
In everyday quantum field theory, we usually do not  have a 
well-defined path integral measure though 
we have  well-established rules of doing the path integral
at least for the perturbative regime.
For our trivial quantum field theory  the path integral measure is 
perfectly well-defined.
The space of all bosonic fields is 
a copy of $X$.  Thus the path integral is an integral
over $X$. We have
\eqn\aag{
Z = \int_X\prod_{k,\bar k=1}^n dX^k dX^{\bar k} 
d\p^k d\p^{\bar k}
 \exp\left(i\CK_{i\bar j}\p^i\p^{\bar j}\right).
}
Remark that the path integral measure carries ghost number
$(n,n)$, i.e., the ghost number anomaly.  
In the above evaluation we used the basic fact of integration over Grassmann numbers
that the integrand should also carry the net ghost number 
$(n,n)$ to have a non-vanishing integral. 
Performing the integral over $\p^i$ and $\p^{\bar i}$,
using the law of integral over Grassmannian number,
we have
\eqn\aah{
Z =  
\int_X \prod_{k,\bar k=1}^n dX^k dX^{\bar k} det(i \CK_{i\bar j}).
}

Now we compare the properties of our model with the differential
geometry of the K\"{a}hler manifold $X$.
We denote the space of $r$-forms on $X$ by $\O^{r}(X)$.
We have the exterior derivative 
$$d:\O^r(X)\rightarrow \O^{r+1}(X)$$
satisfying $d^2=0$. For any complex manifold  we have
decompositions 
$$\O^r(X)=\bigoplus_{r=p+q}\O^{p,q}(X)$$
of $r$-forms into type $(p,q)$-forms with $p+q=r$.
Similarly we have a decomposition $d =\rd + \bar\rd$
such that 
\eqn\aai{
\rd:\O^{p,q}(X)\rightarrow \O^{p+1,q}(X),
\qquad
\bar\rd:\O^{p,q}(X)\rightarrow \O^{p+1,q}(X),
}
and
\eqn\aaj{
\rd^2 =0,\qquad \{\rd,\bar\rd\}=0,\qquad \bar\rd^2=0.
}
In terms of the local complex coordinates $z^i$ and $z^{\bar i}$
we have
\eqn\aak{
\rd= dz^i\Fr{\rd}{\rd z^i},\qquad
\bar\rd =  dz^{\bar i}\Fr{\rd}{\rd z^{\bar i}}.
}
A complex manifold is K\"{a}hler iff there exists a non-degenerated
type $(1,1)$-form $\varpi$
satisfying
$d\varpi=0$. 
A basic fact of the K\"{a}hler geometry
is that the K\"{a}hler metric tensor $g_{i\bar j}$
can be written as
\eqn\aam{
g_{i\bar j} =\Fr{\rd^2 f}{\rd z^i \rd z^{\bar j}},
}
where $f$ is a K\"{a}hler potential.  The K\"{a}hler form
$\varpi$ is given by
\eqn\aal{
\varpi = \varpi_{i\bar j} dz^i\wedge dz^{\bar j}
= ig_{i\bar j}dz^i\wedge dz^{\bar j},
} 
where $\varpi_{i\bar j}=-\varpi_{\bar j i}$ while
$g_{i\bar j} = g_{\bar j i}$.

A comparison with our supersymmetric theory leads
to the following obvious dictionary
\eqn\aao{
\eqalign{
z^i&\rightarrow X^i,\cr
z^{\bar i} & \rightarrow X^{\bar i},\cr
}\qquad
\eqalign{
dz^i & \rightarrow i\p^i,\cr
dz^{\bar i} &\rightarrow i\p^{\bar i}.\cr
}
}
Under the above isomorphism  the relations \aaj\ and \aak\ become
\aab\ and \aadd, respectively, 
such that
\eqn\aap{
\rd \rightarrow \bs,\qquad
\bar\rd  \rightarrow \bbs.
}
Also the K\"{a}hler form $\varpi$ in \aal, after
identifying $\CK$ with a K\"{a}hler potential $f$ of $X$,
i.e., $\CK_{i\bar j}=g_{i\bar j}$,
becomes (minus)  our  action functional $S$ in \aaf.
Now we examine the partition function $Z$ defined by \aag.
It is obvious that, compare with \aah
\eqn\aaq{
\eqalign{
Z 
=  \int_X e^{\varpi} 
= \int_X \Fr{\varpi^n}{n!}
= \int_X \prod_{k,\bar k=1}^n dz^k dz^{\bar k} det(i g_{i\bar j}),\cr
} 
}
where the second identity follows from the fact that
the integrand should be a top form and the third identity
follows from the definition of $\varpi$. Thus the partition function
of our first supersymmetric field theory
is the symplectic volume of $X$.
We remark that the second identity is 
equivalent to the condition of the ghost number anomaly
cancellation. 

One may formalize the above correspondence
as follows. For the tangent bundle $TX$
we define an associated superspace $\widehat{TX}$
where the hat symbol denotes the parity change of the
fiber as in \aao.  
Then the supercharges $\bs$ and $\bbs$ are odd vectors and
the action $S$ is a function on $\widehat{TX}$. 

Now we move on to observables and correlation functions.
A supersymmetric observable $\hat\a$ is  a quantity 
invariant under the symmetry of the theory and annihilated
by supercharges.
We consider the following polynomial function on $\widehat {TX}$,
\eqn\aar{
\hat\a^{p,q} = \a_{i_1\ldots i_p\bar j_1\ldots \bar j_q} 
\p^{i_1}\ldots \p^{i_p}\p^{\bar j_1}\ldots \p^{\bar j_p},
}
carrying the ghost numbers $(p,q)$. 
Due to the isomorphism \aao\
$\bbs \hat\a^{p,q}=0$ iff $\bar\rd \a^{p,q}=0$
where $\a^{p,q}\in \O^{p,q}(X)$ is the $(p,q)$-form on $X$ defined by
\eqn\aat{
\a^{p,q} = \a_{i_1\ldots i_p\bar j_1\ldots \bar j_q} 
dz^{i_1}\wedge\ldots\wedge dz^{i_p}\wedge dz^{\bar j_1}\wedge
\ldots\wedge dz^{\bar j_p}.
}
Note that $\bbs$ defines a Dolbeault cohomology on the space of 
observables graded by the ghost numbers which correspond
to the form degrees.
In the above we showed
that the $\bbs$ cohomology is isomorphic to the Dolbeault
cohomology $(\bar\rd,  \O^{*,*}(X))$    on $X$. 

The correlation function of observables or the expectation
value is defined by
\eqn\aasm{
\left<\prod_{m=1}^r\hat\a^{p_m, q_m}\right>
= \int [\CD X \CD\bar X\CD\p \CD\bar\p] 
\prod_{m=1}^r\hat\a^{p_m, q_m}\cdot 
e^{-S}.
}
For the present model
we see that
\eqn\aasn{
\left<\prod_{m=1}^r\hat\a^{p_m, q_m}\right>
= \int_X \a^{p_1,q_2}\wedge \ldots\wedge \a^{p_r,q_r} \wedge e^{\varpi}.
}
Obviously we have non-vanishing correlation function
if the observables satisfy the ghost number anomaly cancellation condition
\eqn\aaso{
\sum_{m=1}^r (p_m, q_m) = (\ell,\ell),\qquad \ell \leq n
}
Then
\eqn\aasp{
\left<\prod_{m=1}^r\hat\a^{p_m, q_m}\right>
=\Fr{1}{(n-\ell)!} \int_X \a^{p_1,q_2}\wedge \ldots\wedge \a^{p_r,q_r} 
\wedge \varpi\wedge\ldots\wedge \varpi.
}
It follows that correlation functions of supersymmetric  observables
depend only on the cohomology classes of observables and
the K\"{a}hler form $\varpi$.\foot{Consider the integral
$\int_X \b e^\varpi$ where $\b$ is a closed $(\ell,\ell)$-form.
Let $\g$ be homology cycle Poincar\'{e} dual to $\varpi^\ell$.
Then the integral reduces to $\Fr{1}{(n-\ell)!}\int_\g \b$.
Let $\b^\pr$ belongs to the same cohomology class as $\b$,
i.e., $\b^\pr = \b + d\a$. We have, using Stokes'
theorem, $\int_\g (\b^\pr -\b)
=\int_\g d\a = \int_{\rd \g}\a =0$.
}
Thus the correlation function computes the classical
cohomology ring of the target space $X$.
Equivalently the correlation function computes intersection
numbers of homology cycles dual to $\a^{p,q}\in H^{p,q}(X)$.

Using our toy model we illustrated many of the basic properties
of cohomological field theory. In general, however, life
is never as simple as in the idealized world.
Typically we encounter an infinite dimensional space of
certain set of fields on a manifold $M$ as our target space $X$.
Furthermore there usually exists
an infinite dimensional group action on the target space.
Nonetheless one is eventually interested in the
subspace defined as the solution space of certain first order 
differential equations, modulo the gauge symmetry.  
Thus we will need a machinery to reduce the path integral 
to such a subspace and to take care of the group action,
as we will do later.

For the time being we ignore those things and assume
that the path integral is eventually reduced to some
finite dimensional moduli space. Then it may be equivalent
to our toy model. We may call a quantum field theory on $M$
with such a property a cohomological field theory. Usually the
differential geometrical structures of the moduli space are induced
from those of $M$. 
Such a field theory on $M$ has global supersymmetry
equivalent to $(0+0)$-dimensional supersymmetry. The cohomology
of such a global supersymmetry is isomorphic to a  certain cohomology
of $M$. Consequently the correlation functions of supersymmetric
observables are differential topological invariant of $M$.
We refer to the original paper \TFT\ of Witten for a lucid exposition
of general properties of such a cohomological field theory. 
Here we repeated many
of his arguments, perhaps in a slightly different context.

\subsec{$N_c=(2,2)$ Model}

In this section we consider a somewhat more interesting model
by generalizing the toy model of the previous section.
We introduce two copies $(\bs_\pm,\bbs_\pm)$ 
of the fermionic charges $(\bs,\bbs)$. We regard
the above doubling as a $Z_2$-grading in the sense
that supercharges carry the following ghost numbers $(p,q)$ introduced 
for the toy model,
\eqn\abaa{
\eqalign{
\bs_+:(+1,0),\cr
\bs_-:(-1,0),\cr
}\qquad
\eqalign{
\bbs_+:(0, +1),\cr
\bbs_-:(0,-1).\cr
}
}
Thus the supercharges $\bs_+$ and $\bbs_+$
can be identified with the original supercharges $\bs$
and $\bbs$ of the toy model.
We want to define a supersymmetric theory
invariant under all four supercharges.
Obviously we will have a $Z_2$-symmetry
exchanging the $+$ and $-$ indices.
We will say that  the resulting theory is of type $N_c=(2,2)$.
We will see that such a model is related with the geometry
of tangent bundle $TX$ of a K\"{a}hler manifold $X$.
The partition function of this model can be identified
with the Euler characteristic of $X$.

\subsubsection{Basic Structures}

We postulate that the supercharges satisfy the following anti-commutation
relations,
\eqn\aba{
\bs_\pm^2=0,\qquad 
\{\bs_\pm,\bbs_\pm\}=0,\qquad
\bbs_\pm^2=0,
}
and
\eqn\abac{
\{\bs_+, \bs_-\}=0,\qquad
\{\bs_\pm,\bbs_\mp\}=0,\qquad
\{\bbs_+,\bbs_-\}=0,
}
which is an obvious generalization of \aab.
We will consider the same bosonic fields $X^i$ and $X^{\bar i}$ 
as in our toy model.
We demand $X^i$ to be bi-holomorphic or {\it chiral}, meaning that
$\bbs_\pm X^i=0$.\foot{Note that
this choice is arbitrary. We may also demand twisted bi-holomorphicity
or {\it twisted chirality}
by imposing $\bbs_+ X^i =\bs_- X^i=0$. A model with
both chiral and twisted chiral 
multiplets has very interesting
properties.} We call the complex conjugates $X^{\bar i}$
{\it anti-chiral}, meaning that $\bs_+ X^{\bar i}=\bs_- X^{\bar i}=0$.
Now the anti-commutation relations among supercharges
suggest that we have the following chiral multiplets 
\eqn\abb{
\def\normalbaselines{\baselineskip20pt
\lineskip3pt \lineskiplimit3pt}
\matrix{
\p^i_- &\mapl & X^i &\mapr & \p^i_+\cr
             & 
 	 \rlap{\lower.3ex\hbox{$\scriptstyle s_{\!+}$}}\searrow
 & &\swarrow\!\!\!\rlap{\lower.3ex\hbox{$\scriptstyle s_{\!-}$}} 
& \cr
&&H^i&&
}.
}
In the above $H^i$ are called auxiliary fields, which are introduced
due to the
conditions
\eqn\abbc{
\{\bs_+,\bs_-\} X^i = i\bs_+\p^i_- + i \bs_-\p^i_+=0,
}
can be solved  as $\bs_\pm \p^i =\pm H^i$ while they 
are indeterminate.\foot{
The equation might also be solved as $\bs_\pm\p^i=0$ without
introducing $H^i$. However, the auxiliary fields are indispensable.
The moral is that  we better keep it whenever we encounter redundancy.} 
Denoting 
$
\d =\bs_+\bar\ep_- +\bs_-\bar\ep_+ +\bbs_+\ep_- +\bbs_-\ep_+
$
we have the following transformation laws for
chiral and anti-chiral multiplets,
\eqn\abc{
\eqalign{
\d X^i=& i\bar\ep_-\p^i_+ +i\bar\ep_+\p^i_-,\cr
\d\p^i_+ =& +\bar\ep_+ H^i ,\cr
\d\p^i_- =& -\bar\ep_- H^i ,\cr
\d H^i =& 0
}\qquad
\eqalign{
\d X^{\bar i} =& i\ep_-\p^{\bar i}_+ +i\ep_+\p^{\bar i}_-,\cr
\d\p^{\bar i}_+ =&+ \ep_+  H^i ,\cr
\d\p^{\bar i}_- =& -\ep_-  H^{\bar i} ,\cr
\d H^{\bar i} =&0.
}
}

Now we define a natural supersymmetric action functional.
The requirements $\bs_\pm S=\bbs_\pm S=0$ for $S$
to have $N_c=(2,2)$ supersymmetry and the anti-commutation
relations \aba\ and \abac\ imply, by repeatedly applying
the Poincar\'{e} lemma,  that we can write $S$ as
follows,
\eqn\abd{
S=\bs_+\bbs_+\bs_-\bbs_-\CK(X^i,X^{\bar i}),
}
where $\CK(X^i, X^{\bar i})$ is a locally defined real functional.
Expanding the above we have
\eqn\abe{
S
=
g_{i\bar j}H^i H^{\bar j}
+i\rd^{}_k g_{i\bar j}\; \p^k_+  \p^i_- H^{\bar j}
+i\rd_{\bar k}g_{i\bar j}\;\p^{\bar k}_+H^i\p^{\bar j}_-
+\rd^{}_{\ell}\rd_{\bar k}g_{i\bar j}\;
\p^\ell_+\p^{\bar k}_+\p^i_-\p^{\bar j}_-,
\phantom{\biggr)}
}
where 
$\rd^{}_i = \rd/\rd X^i$ and $\rd^{}_{\bar j} = \rd/\rd X^{\bar j}$
and we set $g_{i\bar j} :=\rd^{}_i\rd_{\bar j}\CK$. 
We can integrate out the auxiliary fields $H^i$ 
and $H^{\bar i}$ by a Gaussian integral,  or, equivalently, 
eliminate them by
pluging in   the algebraic equations of
motions for $H^i$ and $H^{\bar i}$;
\eqn\abaux{
\eqalign{
H^i &=  - i g^{i\bar j} \rd^{}_k g_{\ell\bar j}\;\p^k_+\p^\ell_-,\cr
H^{\bar j} &
= +i g^{i\bar j}\rd_{\bar k} g_{i\bar\ell}\;\p^{\bar k}_+\p^{\bar \ell}_-,
}
}
where $g^{i\bar j}$ is the inverse of $g_{i\bar j}$.
Then we obtain the new action functional $S^\pr$,
\eqn\abaction{
S^\pr= -
 R_{\ell\bar k i\bar j}\;\p^\ell_+\p^{\bar k}_+\p^i_-\p^{\bar j}_-,
}
where 
\eqn\abf{
R_{\ell \bar k i\bar j}
= -\rd^{}_{\ell}\rd_{\bar k}g_{i\bar j}
+ g^{p\bar q}\rd^{}_i g_{p\bar k}\rd_{\bar j}g_{\ell \bar q},
}
which can be identified with the 
the Riemann curvature tensor of $TX$ if $\CK$ is a K\"{a}hler potential
of $X$. Remark that the non-vanishing components of the Christoffel
symbols in the K\"{a}hler geometry are
\eqn\christ{
\G^{i}_{k\ell} = g^{i\bar j}\rd_k g_{\ell\bar j},\qquad
\G^{\bar i}_{\bar k\bar\ell} = g^{i\bar j}\rd_{\bar k} g_{i\bar \ell}.
}
The new action $S^\pr$ is invariant under the supersymmetry
after modifying the transformation laws \abc\ by replacing
$H^i$ and $H^{\bar i}$ by their on-shell expressions \abaux.

Now we examine the path integral.
The partition function is defined as usual,
\eqn\abg{
\eqalign{
Z &= \int 
[\CD X \CD \bar X \CD \p_\pm \CD \bar\p_\pm]
e^{-S^\pr},\cr
&=\left(\Fr{1}{2\pi}\right)^{n}\int \prod^n_{k,\bar k=1}
d X^k d X^{\bar k} d\p^k_+ d\p^{\bar k}_+ d\p^k_- d\p^{\bar k}_-
\exp\left(R_{\ell\bar k i\bar j}\;\p^\ell_+\p^{\bar k}_+
\p^i_-\p^{\bar j}_-\right),
}
}
where the integration is over the space of all fields. 
The bosonic part of the path integral is
an integration over a copy of $X$.
We first perform the integral over $\p^i_+$ and $\p^{\bar i}_+$
which, as we saw earlier, is equivalent to replacing
$R_{\ell\bar k i\bar j}\;\p^\ell_+\p^{\bar k}_+\p^i_-\p^{\bar j}_-$
by the $(1,1)$-form $\cmmib{R}_{i\bar j}:= R_{k\bar\ell i\bar j} 
dz^k\wedge d\bar z^{\bar \ell}$
on $X$,
\eqn\abg{
Z =\left(\Fr{1}{2\pi}\right)^{n}\int_X \prod^n_{k,\bar k=1}
d\p^k_- d\p^{\bar k}_-
\exp\left(\cmmib{R}_{i\bar j}\;
\p^i_-\p^{\bar j}_-\right).
}
Integration over $\p^i_-$ and $\p^{\bar i}$ leads
to 
\eqn\abbh{
Z = \Fr{1}{(2\pi)^n}\int_X det(\cmmib{R}_{i\bar j}) = \int_X e(TX)
:=\chi(TX).
}
The last identity is due to the Gauss-Bonnet theorem.
Thus the partition function computes the Euler characteristic
$\chi(X)=\chi(TX)$ of the manifold $X$.

\subsubsection{Geometrical Interpretation of Supercharges}

Now we examine the geometrical meaning of our supercharges.
In Sect. $2.1.1$ we already saw that the supercharges
$\bs_+$ and $\bbs_+$ are associated with the $\rd$ and
$\bar\rd$ differential on the target space $X$. 
Our task is to understand the geometrical meaning of
the remaining supercharges $\bs_-$ and $\bbs_-$.

We begin with discarding the obvious candidates
for $\bs_-$ and $\bbs_-$, namely the
operators $\rd^*$ and $\bar\rd^*$ defined by
\eqn\aca{
\eqalign{
\rd^* = - * \bar\rd*:\O^{p,q}(X)\rightarrow \O^{p-1,q}(X),\cr
\bar\rd^* = - * \rd*:\O^{p,q}(X)\rightarrow \O^{p,q-1}(X),\cr
}
}
where $*$ denote the Hodge star.
They satisfy the
following relations
\eqn\acb{
\rd^{*2} =0,\qquad \{\rd^*,\bar\rd^*\}=0,\qquad \bar\rd^{*2}=0,
}
and decrease the form degree by  $(-1,0)$
and $(0,-1)$ , respectively.  We have, however, well-known
relations in K\"{a}hler geometry
\eqn\acc{
\{\rd ,\rd^*\}= \{\bar\rd,\bar\rd^*\} 
= \Fr{1}{2}\{d, d^*\} = \Fr{1}{2}\nabla,
}
where $\nabla$ is the Laplacian.
On the other hand we have $\{\bs_+,\bs_-\}=\{\bbs_+,\bbs_-\}=0$.
We also have more obvious problem from
$\rd^* X^i=0$, while $\bs_- X^i=i\p^i_-\neq 0$.
Thus we have to seek an alternative set of operators.

We first consider the real symplectic case and then specialize
to the K\"{a}hler case.
Consider a symplectic manifold with symplectic form
$\varpi =\varpi_{IJ} dx ^I\wedge dx^J$. Since the matrix
$\varpi_{IJ}=-\varpi_{JI}$ is non-degenerated we have a well-defined
inverse matrix $\varpi^{JI}$. Using $\varpi^{JI}$ we have a canonical
map from a cotangent vector to a tangent vector.\foot{We may
also consider a Poisson manifold with a bi-vector $\varpi^{IJ}$.}
Denoting $\a =\a_I d  x^I$ and $\tilde \a =\tilde\a^I \Fr{\rd}{\rd x^I}$
for a cotangent vector and its dual tangent vector, respectively,
we have
\eqn\acd{
\tilde \a^{I} = \varpi^{IJ}\a_J.
}
One may define the corresponding operator $\sqcap$ as
follows
\eqn\ace{
\sqcap := \Fr{\varpi^{IJ}}{2}\left(
\left(\otimes\Fr{\rd}{\rd x^I}\right)\Fr{\rd}{\rd(d x^J)}
-\left(\otimes \Fr{\rd}{\rd x^J}\right)\Fr{\rd}{\rd(d x^I)}\right),
}
where the symbol $\otimes \Fr{\rd}{\rd x^I}$ means
taking tensor product. For instance we have
\eqn\acf{
\sqcap \a = \varpi^{IJ}\a_L \left(\otimes \Fr{\rd}{\rd x^I}\right)
\Fr{\rd(d x^L)}{\rd(d x^J)} = \varpi^{IJ}\a_J \Fr{\rd}{\rd x^I}
= \tilde \a^I\Fr{\rd}{\rd x^I} = \tilde \a.
}
Similarly $\sqcap$ induce an isomorphism
\eqn\acg{
\sqcap:\G(\wedge^p T^*\!X\otimes \wedge^q T\!X)
\rightarrow 
\G(\wedge^{p-1} T^*\!X\otimes \wedge^{q+1} T\!X),
}
where $TX$ and $T^*X$ are the tangent and cotangent
vector, respectively, and $\G$ denotes the space of sections.
Note that 
$\G(\wedge^p T^*\!X\otimes \wedge^q T\!X)=\O^p(X,\wedge^q T\!X)$. 

Now we can define a first order differential operator 
by taking the composition of
$\sqcap$ and the exterior derivative $d$,
\eqn\ach{
d:\O(\wedge^p  T^*\!X\otimes \wedge^q T\!X)
\rightarrow 
\O(\wedge^{p+1} T^*\!X\otimes \wedge^{q} T\!X),
}
as follows,
\eqn\aci{
\tilde d:=(\sqcap d - d\sqcap ):\O(\wedge^p T^*\!X\otimes \wedge^q T\!X)
\rightarrow 
\O(\wedge^{p} T^*\!X\otimes \wedge^{q+1} T\!X).
}
We will conveniently assign the form degree $-1$ to the operator
$\tilde d$. One can check
\eqn\acj{
\tilde d^2=0,\qquad \{d,\tilde d\}=0,
}
after a direct computations.
We also have the following obvious but important relation
\eqn\ack{
\eqalign{
d: x^I & \rightarrow d x^I,\cr
\tilde d: x^I &\rightarrow \varpi^{IJ}\Fr{\rd}{\rd x^I}.
}
}
Thus for a symplectic manifold $X$ with the symplectic form
$\varpi$ we have 
$$(x^I, dx^I, \rd/\rd x^I; d,\tilde d),$$ where $dx^I$
and $\rd/dx^I$
denote local coordinates in the fiber of $TX$ and the fiber of $T^*X$,
respectively. To relate with supersymmetry we perform the parity changes 
for both the fibers of $TX$ and $T^*X$, i.e.,
$\widehat{TX}$ and $\widehat{T^*X}$. Then we have
a map
\eqn\acl{
 (x^I, dx^I, \rd/\rd x^I; d,\tilde d) \rightarrow 
(X^I, i\p^I_+, i\chi_{I}; Q_+, Q_-),
}
where everything is in real coordinates, 
$\p^I_- := \varpi^{IJ}\chi_{J}$ and
$Q_\pm =\bs_\pm +\bbs_\pm$.

One may compare our operator $\tilde d$ with the (different) operator
$\D$ defined by Koszul \Koszul. The operator $\D$ is
define as
\eqn\kosza{
\D := \sqcap_{k} d - d\sqcap_k,
}
where $\sqcap_{k}$ in the notation of \ace\ is given by
\eqn\koszb{
\sqcap_k := \Fr{\varpi^{IJ}}{2}\left(
\Fr{\rd^2}{\rd (d x^I)\rd(d x^J)}
- \Fr{\rd^2}{\rd (dx^J)\rd(d x^I)}\right).
}
Thus $\D$ is a second order differential operator
with degree $-1$ on $\G(\wedge^*T^*\!X)=\O^*(X)$
and we have $\D x^I=0$.\foot{ Koszul proved $\D^2=\{d,\D\}=0$
and defined a covariant Schouten-Nijenhuis bracket, $\a,\b \in \O^*(X)$,
$$
\{\a,\b\}_{SN} = (\D\a)\wedge \b + (-1)^{|a|}\a\wedge \D\a 
-\D(\a\wedge\b).
$$
}

Now we return to  a K\"{a}hler manifold $X$ with K\"{a}hler form
$\varpi=\varpi_{i\bar j} dz^i\wedge dz^{\bar j}$ and show that the above 
interpretation is indeed the correct one.
It is suffice to consider the holomorphic half, say $\bs_+$ and $\bs_-$.
The operator $\sqcap$ is decomposed as
$\sqcap =\sqcap^\pr +\sqcap^\ppr$ where
\eqn\acm{
\eqalign{
\widehat\sqcap^\pr & = -\Fr{1}{2}\varpi^{i\bar j}(X^\ell, X^{\bar \ell})
\chi_{\bar j}\Fr{\rd}{\rd \p^i_+},\cr
\widehat\sqcap^\ppr & = +\Fr{1}{2}\varpi^{i\bar j}(X^\ell, X^{\bar \ell})
\chi_{i}\Fr{\rd}{\rd \p^{\bar j}_+},\cr
}
}
where we did parity change $\sqcap\rightarrow \widehat\sqcap$
by
\eqn\acn{
\eqalign{
dz^i \rightarrow i\p^i_+,\cr
dz^{\bar i} \rightarrow i\p^{\bar i}_+,\cr
}\qquad
\eqalign{
{\rd}/{\rd z^i }\rightarrow i\c_{i},\cr
{\rd}/{\rd z^{\bar i}} \rightarrow i\c_{i}.\cr
}
}
Now we define
\eqn\aco{
\bs_- = \widehat\sqcap^\pr \bs_+ -\bs_+\widehat\sqcap^\pr.
}
{}From
\eqn\acp{
\bs_+ = i\p^i_+\Fr{\rd}{\rd X^i},
}
we have 
\eqn\acq{
\bs_- = -\Fr{i}{2}\varpi^{i\bar j}\chi_{\bar j}\Fr{\rd}{\rd X^i}
+\Fr{i}{2}\Fr{\rd \varpi^{i\bar j}}{\rd X^k}\p^k_+
\chi_{\bar j}\Fr{\rd}{\rd \p^i_+}.
}

Now we can check if the above identification of the
supercharge $\bs_-$ is the correct one.
After direct computations we find the following relations
\eqn\acr{
\eqalign{
\bs_+ X^i &= i\p^i_+,\cr
\bs_+ X^{\bar i} &= 0,\cr
\bs_- X^{\bar i} &= 0,\cr
\bs_- X^{ i} &= i\p^i_-,\cr
}\qquad
\eqalign{
\bs_+\p^i_+ &=0,\cr
\bs_+\p^i_- & = 
 +i\varpi_{\bar j\ell}\Fr{\rd\varpi^{i\bar j}}{\rd X^k}\p^k_+\p^\ell_-,\cr
\bs_-\p^i_+ & = 
 -i\varpi_{\bar j\ell}\Fr{\rd\varpi^{i\bar j}}{\rd X^k}\p^k_+\p^\ell_-,\cr
\bs_-\p^i_-&=0,
}
}
where we defined
\eqn\acs{
\p^i_- = -\Fr{1}{2}\varpi^{i\bar j}\chi_{\bar j}.
}
In checking $\bs_-\p^i_-=0$  we used the torsion-free condition
of the Hermitian connection of $TX$, equivalent to  
the condition $d\varpi=0$.
Using the relation
\eqn\act{
\varpi_{i\bar j} = ig_{i\bar j} = -\varpi_{\bar j i}
}
we see that the above is exactly the supersymmetry
algebra of $\bs_+$ and $\bs_-$ in \abc\ after replacing
the auxiliary fields $H^i$ by their on-shell values given by \abaux.

Now we summarize. We have the following operators
\eqn\acv{
\eqalign{
\bs_+ &= i\p^i_+\Fr{\rd}{\rd X^i},\cr
\bbs_+ &= i\p^{\bar i}_+\Fr{\rd}{\rd X^{\bar i}},\cr
}\qquad
\eqalign{
\bs_- &= \widehat\sqcap^\pr\bs_+ -\bs_+\widehat\sqcap^\pr,\cr
\bbs_- &= \widehat\sqcap^\ppr \bbs_+ -\bbs_+\widehat\sqcap^\ppr,\cr
}
}
such that
\eqn\acx{
\eqalign{
\bs_+ :&\widehat\O^{p,q}\left( 
\wedge^{r} \widehat{ \CT X} \otimes \wedge^{s}
\widehat{\bar \CT X}\right)
\rightarrow
\widehat \O^{p+1,q}\left( 
\wedge^{r} \widehat {\CT X}\otimes \wedge^{s} 
\widehat {\bar \CT X}\right),\cr
\bbs_+ :&\widehat\O^{p,q}\left( 
\wedge^{r} \widehat {\CT X}\otimes \wedge^{s} \widehat{\bar \CT X}\right)
\rightarrow
\widehat \O^{p,q+1}\left( 
\wedge^{r} \widehat{ \CT X} \otimes \wedge^{s} \widehat{ \bar \CT X}\right),\cr
\bs_- :& \widehat \O^{p,q}\left( 
\wedge^{r} \widehat{ \CT  X}\otimes \wedge^{s} \widehat {\bar \CT X}\right)
\rightarrow
\widehat \O^{p,q}\left( 
 \wedge^{r+1} \widehat{\CT X} \otimes \wedge^{s} \widehat{ \bar \CT X}\right),\cr
\bbs_- :&\O^{p,q}\left( 
\wedge^{r} \widehat {\CT X}\otimes \wedge^{s} \widehat{\bar \CT X}\right)
\rightarrow
\widehat \O^{p,q}\left( 
\wedge^{r} \widehat{\CT X}\otimes \wedge^{s+1} \widehat {\bar \CT X}\right),\cr
}
}
where $\CT X$ denotes the holomorphic parts of the
tangent bundle $TX = \CT X \oplus \bar\CT X$ of $X$.

\subsubsection{Introducing a Holomorphic Potential}

Now we consider a more general action functional.
We pick a holomorphic function $\CW(X^i)$ of the chiral
fields $X^i$. Since $\bbs_\pm X^i=0$ we have $\bbs_\pm \CW(X^i)=0$.
It follows that we have the following more general  $N=(2,2)$ 
supersymmetric action functional,
\eqn\ada{
\eqalign{
S(\l)=&
\bs_+\bbs_+\bs_-\bbs_- \CK(X^i, X^{\bar i})
 + {\l}\bs_+\bs_- \CW(X^i)
 +{\l}\bbs_+\bbs_- \bar\CW(X^i),
}
}
where $\l$ is certain coupling constant introduced
for convenience.
Expanding $S(\l)$ we find
\eqn\adb{
\eqalign{
S(\l)
=&g_{i\bar j}H^i H^{\bar j}
 +i\left(\rd^{}_{k}g_{i\bar j} \p^k_+  \p^i_- 
- {\l}V_{\bar j}\right)H^{\bar j}
  +iH^i\left(\rd_{\bar k}g_{i\bar j}\p^{\bar k}_+\p^{\bar j}_-
  -{\l}V_i\right) 
\cr
&
  -\l\Fr{\rd V_i}{\rd X^j} \p^i_+\p^j_-
  -\l
 \Fr{\rd V_{\bar i}}{\rd X^{\bar j}} \p^{\bar i}_+\p^{\bar  j}_-
  +\rd^{}_\ell\rd_{\bar k} g_{i\bar j}
\p^\ell_+\p^{\bar k}_+\p^i_-\p^{\bar j}_-.
}
}
where we set $V_i := \rd\CW/\rd X^i$.
Now we integrate out the auxiliary fields by their algebraic
equations of motions
\eqn\adc{
\eqalign{
H^i &= -i \G^i_{k\ell}\p^k_+\p^\ell_- +i\l g^{i\bar j}V_{\bar j},\cr
H^{\bar i} &= +i\G^{\bar i}_{\bar k\bar\ell}\p^{\bar k}_+\p^{\bar \ell}_-
-i\l g^{j\bar i} V_j,
}
}
where we used the notations in \christ.  We have
\eqn\ade{
\eqalign{
S^\pr(\l)=&  
{\l^2} g^{i\bar j} V_i V_{\bar j}
-{\l}\Fr{D V_i}{D X^j} \p^i_+\p^j_-
-{\l}\Fr{D V_{\bar i}}{D X^{\bar j}} \p^{\bar i}_+\p^{\bar j}_-
-R_{\ell\bar k i\bar j}\p^\ell_+\p^{\bar k}_+\p^i_-\p^{\bar j}_-.
}
}
where 
\eqn\adf{
\eqalign{
\Fr{D V_i}{D X^j}  &:=\Fr{\rd V_i}{\rd X^j} + \G^\ell_{ij} V_{\ell}.
}
}

\subsubsection{The Partition Function}

The partition function  is independent of $\l$ since $\l$ dependent
term is $\bs_\pm$-exact deformation of $S$. 
In the limit $\l\rightarrow \infty$
the dominant contributions to the path integral are from the vanishing
locus of holomorphic vector fields $V_i$. Or we may simply
apply the fixed point theorem of Witten to reach  the same
conclusion; from the supersymmetry transformation laws \abc\
we see that the fixed point equations are $\p^{\bar j}_\pm=H^{\bar i}=0$.
{}From the relations \acb\ the above implies $V_i=0$.

For generic choices the vanishing locus will be zero dimensional
and consists of isolated points. Then there are no fermionic zero-modes
and the action functional evaluated at such a point is simply $0$.
Thus the partition function
is just the sum of contributions of each point weighted by the one
loop determinants of the transverse degrees of freedom. Due
to the Bose-Fermi symmetry such a determinant is $\pm 1$, depending
on a certain orientation, 
due to supersymmetry and due to the ambiguity
in taking the square root of the determinant. In our case
they always can be set $+1$ since the ambiguities from
holomorphic and anti-holomorphic contribution cancel 
each other. Thus the partition function is the number
of zeros. If we turn off the potential 
 we recover the original model. This gives rise to the
Poincar\'e-Hopf theorem. We should mention that the usual derivation
of Poincar\'e-Hopf theorem uses supersymmetric quantum
mechanics, i.e., the $(0+1)$ 
dimensional sigma model \Wittendb\Wittencsb\Wittenmorse,
but with essentially the same arguments.

For a non-generic vector field $V_i$ the vanishing locus
can be a positive dimensional submanifold.
One may try to perturb the vector field $V_i$, thus $\CW(X^i)$,
to a generic one or just evaluate the path integral.
We will give a detailed analysis for this case in
the next section in a more general context.

\subsec{Generalization to $N_c=(2,0)$ Model}

The model in the previous section enjoys a perfect symmetry
between things with $+$ and $-$ indices.
Now we want to relax such a symmetry.
We shall see that such symmetry is due to the restriction
of considering a very special Hermitian 
holomorphic vector bundle, namely the
tangent bundle $TX$, over $X$. 
By maintaining only the $N_c=(2,0)$ supersymmetry generated
by $\bs_+$ and $\bbs_+$ we arrive at a more general model,
which is related with a Hermitian holomorphic bundle 
$\E$ over $X$.

\subsubsection{The Basic Structures}

First we write our action functional $S(\l)$ \ada\  in  form
such that only the $\bs_+$ and $\bbs_+$ are manifest,
\eqn\aea{
\eqalign{
S(\l)&= -\bs_+\bbs_+
\!\left(g_{i\bar j}(X^i, X^{\bar i})\p^i_-\p^{\bar j}_- \right)
+ {i}{\l}\bs_+\!\left(\p^i_- V_i(X^j)\right) 
 +{i}{\l}\bbs_+\!\left(\p^{\bar i}_- V_{\bar i}(X^{\bar j})\right). 
}
}
Similarly we disconnect the diagram \abb\ by removing the
link $\bs_-$
\eqn\aeb{
\def\normalbaselines{\baselineskip20pt
\lineskip3pt \lineskiplimit3pt}
\matrix{
\p^i_- & & X^i &\mapr & \p^i_+\cr
             & 
 	 \rlap{\lower.3ex\hbox{$\scriptstyle s_{\!+}$}}\searrow
 & & 
& \cr
&&H^i&&
}.
}
Now we can regard the above as two independent sets
of multiplets. Then we  rename various fields as
follows
\eqn\aec{
\eqalign{
\p^i_- \rightarrow \chi^\a_-,\cr
\p^{\bar i}_- \rightarrow \chi^{\bar \a}_-,\cr
}\qquad
\eqalign{
H^i \rightarrow H^\a,\cr
H^{\bar i} \rightarrow H^{\bar \a},\cr
}\qquad
\eqalign{
V_i \rightarrow \eufm{S}_\a(X^j),\cr
V_{\bar i} \rightarrow \eufm{S}_{\bar \a}(X^{\bar j}),\cr
}\qquad
 g_{i\bar j} \rightarrow h_{\a\bar\b}(X^i, X^{\bar i}),
}
where the new indices run as $\a,\b=1,\ldots,r$ and we
maintain the Hermiticity of $h_{\a\bar\b}$.
The $\bs_+$ and $\bbs_+$ transformation laws are
\eqn\aed{
\eqalign{
\d X^i&= i\bar\ep_-\p^i_+  ,\cr
\d X^{\bar i} &=i\ep_-\p^{\bar i}_+,\cr
}\qquad
\eqalign{
\d\p^i_+ & =0 ,\cr
\d\p^{\bar i}_+ &=0,\cr
}
}
and
\eqn\aef{
\eqalign{
\d\c^\a_- &= -\bar\ep_- H^\a ,\cr
\d\c^{\bar \a}_- &= -\ep_-  H^{\bar \a} ,\cr
}\qquad
\eqalign{
\d H^\a &= 0,\cr
\d H^{\bar \a} &=0.
}
}
Now we have following 
new action functional 
\eqn\aeg{
\eqalign{
S=& -\bs_+\bbs_+
\left(h_{\a\bar \b}(X^i, X^{\bar i})\c^\a_- \c^{\bar \b}_- \right)
+ {i}\bs_+ \left(\c^\a_- \eufm{S}_\a(X^i)\right) 
 +{i}\bbs_+\left(\c^{\bar \a}_- \eufm{S}_{\bar \a}(X^{\bar i})\right),
}
}
which is the general form of $N_c=(2,0)$ supersymmetric
action functional.  

Note that the above action functional may or may not
have $N_c=(2,2)$ symmetry.  Generically the model does
not have $N_c=(2,2)$ supersymmetry.
Note also that the model has the same supersymmetry
as our toy model  in Sect. $2.1.1$.  Thus the new model
shares the same observables with the toy models, which
are $\hat \a^{p,q}$ obtained by an element $\a^{p,q}=H^{p,q}(X)$
of the cohomology group $H^{p,q}(X)$ after the parity
change $T\!X \rightarrow \widehat{T\! X}$.
The differences with the toy
model are that we have additional Fermi multiplets $(\c_-^\a, H^\a)$
with a different action functional. We call the multiplets
$(\c_-^\a, H^\a)$ Fermi  multiplets. We call $\c_-^\a$
anti-ghosts.
We remark that the action
functional of the toy model may be regarded as zero
by treating the K\"{a}hler form $\varpi$ as an observables.
Now we turn to examine the action functional.

Expanding $S$ we have
\eqn\aegg{
\eqalign{
S
=& h_{\a\bar \b}H^\a H^{\bar \b}
  +{i}\left(\rd_i h_{\a\bar \b} \p^i_- \c^\a_+ 
  -  \eufm{S}_{\bar \b}\right)H^{\bar \b}
  +{i}H^\a\left(\rd_{\bar j} h_{\a\bar \b}\p^{\bar j}_+\c^{\bar \b}_-
  -\eufm{S}_\a\right) 
\cr
&
  -\Fr{\rd \eufm{S}_\a}{\rd X^j} \p^j_+\c^\a_-
  -\Fr{\rd \eufm{S}_{\bar \a}}{\rd X^{\bar j}} \p^{\bar j}_+
    \c^{\bar \a}_-
  +(\rd_{i}\rd_{\bar j} h_{\a\bar \b})\p^i_+\p^{\bar j}_+\c^\a_-\c^{\bar \b}_-.
}
}
After integrating out  the auxiliary fields $H^\a$ and $H^{\bar \b}$
by their algebraic equations of motion
\eqn\aaeh{
\eqalign{
H^\a &
	= -i h^{\a\bar\b}\rd^{}_k h_{\g\bar\b}\p^k_+\c^\g_- 
	+i h^{\a\bar \b}\eufm{S}_{\bar \b},\cr
H^{\bar \a} &
	= +i h^{\a\bar\b}\rd_{\bar k} h_{\a\bar\g}\p^{\bar k}_+\c^{\bar \g}_-
	-ih^{\b\bar \a} \eufm{S}_\b,
}
}
we are left with 
\eqn\aei{
\eqalign{
S^\pr=
h^{\a\bar \b} \eufm{S}_\a \eufm{S}_{\bar \b}
 -\Fr{D \eufm{S}_\a}{D X^j} \p^j_+\c^\a_-
  -\Fr{D \eufm{S}_{\bar \a}}{D X^{\bar j}} \p^{\bar j}_+\c^{\bar \a}_-
  -F_{\a\bar\b  i\bar j}\p^i_+\p^{\bar j}_+\c^\a_-\c^{\bar \b}_-,
}
}
where
\eqn\aek{
F_{\a\bar \b i\bar j} = -\rd_{\bar j}\rd_i h_{\a\bar \b}
+ h^{\g\bar \r}(\rd_i h_{\a\bar \r})(\rd_{\bar j} h_{\g\bar \b})
}
and
\eqn\aej{
\Fr{D \eufm{S}_\a}{D X^j} =\rd_j \eufm{S}_\a 
+ h^{\b\bar \g} (\rd_j h_{\a\bar \g})\eufm{S}_{\b}.
}

\subsubsection{Relations with Hermitian Holomorphic Vector Bundle}

It turns out that we are describing a rank $r$ 
Hermitian holomorphic vector bundle $\E\rightarrow X$
over a K\"{a}hler manifold $X$ with Hermitian structure
$h_{\a\bar\b}$. Here we briefly summarize some properties
of Hermitian holomorphic bundles \Kobayashi.
Consider a rank $r$ complex vector bundle $\E$ over $X$.
Let $\O^{p,q}(X,\E)$ denote the space of $(p,q)$-forms over
$X$ with values in $\E$. A connection (the covariant
derivative)  $d_{\! A}$ can be decomposed as
\eqn\aeea{
d_{\!A} =\Dp  +\Dpp : \O^{p,q}(X,\E)\rightarrow 
\O^{p+1,q}(X,E)\oplus \O^{p, q+1}(X,\E).
}
A connection $d_{\!A}$  endows $\E$ with a structure of a holomorphic
vector bundle if the $(0,2)$-component $F^{0,2}\in \O^2(X, End(\E))$
of its curvature $F$ vanishes, i.e., $\Dpp^2=0$.
A complex vector bundle $\E$ is {\it Hermitian}
if it has a fixed Hermitian structure $h$ which is a
$C^\infty$ field of positive definite Hermitian inner products
in the fibers of $\E$. Given a local frame field $s_U=(s_1,\ldots,s_r)$
of $\E$ over an open subset $U \subset X$ we
set $h_{\a\bar\b}= h(s_\a, s_\b)$ where
$\a,\b=1,\ldots,r$. Gluing them along different coordinate
patches as usual we obtain $h_{\a\bar\b}(z^i, z^{\bar i})$.
A connection $D$ in $(\E,h)$ is called an {\it h-connection}
if $d (h(\xi,\eta)) =h(D\xi, \eta) + h(\xi, D\eta)$ for $\xi,\eta\in
\O^0(\E)$.  The theorem is that given a Hermitian structure
$h$ in a holomorphic vector bundle $\E$, there is a unique
$h$-connection $d_{\! A}$ called {\it Hermitian connection} 
such that $\Dpp = \bar\rd$. 
Finally the curvature two-form of a Hermitian connection
is of type $(1,1)$, thus $F^{2,0}$ also vanishes. The curvature
two-form is given by
the formula 
\eqn\twfg{
F_{\a\bar\b} := F_{\a\bar\b i\bar j} dz^i\wedge dz^{\bar j},
}
where $F_{\a\bar\b i\bar j}$ is defined as \aek.
We note that the K\"{a}hler metric 
$g_{i\bar j}$ on $X$ is a Hermitian structure of $TX$.

We saw that our model describes a  rank $r$ Hermitian holomorphic
vector bundle $\E$ with Hermitian structure $h_{\a\bar\b}(z^i, z^{\bar i})$.
Now $\eufm{S}_\a$ can be identified with a  holomorphic
section of $\E$. In summary a $N_c=(2,0)$ model is associated with
a Hermitian holomorphic vector bundle $(\E,h)$ over a K\"{a}hler
manifold $X$ with holomorphic section. 
Associated with the base manifold $X$ we have
holomorphic multiplets \aed, as in the toy model.
Associated with the fiber space we have Fermi multiplets
\aef.

\subsubsection{The Path Integrals}

Now we examine the path integral of our model
in the various situations.

\lin{Turning Off the Holomorphic Section}

To begin with we consider the case that $\eufm{S}_\a=0$.  
The partition function $Z$ is defined
by
\eqn\aeka{
Z =\int \biggr[
\prod_{k,\bar\k =1}^n\left( d X^k d X^{\bar k} d\p^k_+ d\p^{\bar k}_+
\right)
\prod_{\g,\bar\g=1}^r\left( d \c^\g_- d \c^{\bar\g}_-\right)\biggr] 
exp\left( F_{\a\bar\b  i\bar j}\p^i_+\p^{\bar j}_+\c^\a_-\c^{\bar \b}_-
\right).
}
The bosonic integral is an integral over $X$. 
As before the bosonic integral and
integration over $\p^k_+$ and $\p^{\bar k}_+$
combine into the integration of differential forms on $X$
by replacing $F_{\a\bar\b i\bar j}\p^i_+\p^{\bar j}_+$
with the curvature two-form $F_{\a\bar\b}$ defined in \twfg.
Thus we have
\eqn\aekb{
Z =\int_X \prod_{\g,\bar\g=1}^r \left(d \c^\g_- d \c^{\bar\g}_-\right)
exp\left( F_{\a\bar\b}\c^\a_-\c^{\bar \b}_-
\right).
}
The fermionic
integral of $\chi^\a$ and $\chi^{\bar \b}$
leads to the Paffian of the curvature two-form $F_{\a\bar\b}\in \O^{1,1}(X,End(\E)$.
We immediately see that the integrand is not
a top form on $X$ unless $n=r$. 
For $n=r$
the partition function is the Euler character $\chi(\E)$,
\eqn\ael{
Z=\int_X e(\E)=\chi(\E),
}
otherwise, for  $n\neq r$, the path integral vanishes.
In the case $r < n$ we can insert a set of observables
$\prod \hat\a^{p_\ell, q_\ell}$
with the total ghost number $(n-r,n-r)$ and evaluate
the correlation function
\eqn\aem{
\left<\prod_{\ell=1}^m \hat\a^{p_\ell, q_\ell}\right>
= \int_X e(\E)\wedge \a^{p_1, q_1}\wedge \ldots \wedge\a^{p_m,q_m}
}
The path integral always vanishes for $r > n$.
We see that the vector bundle $\E$ after the parity change
can be viewed as a bundle spanned by anti-ghosts $\chi^\a_-$ over $X$.

\lin{Turning On the Holomorphic Section}

Now we turn on the holomorphic section $\eufm{S}_\a$
of $E\rightarrow X$. 
Applying the fixed point theorem of Witten
we see that the path integral is localized to an
$\bs_+$ and $\bbs_+$ invariant neighborhood
of the vanishing locus $N$ of $\eufm{S}_\a(X^i)$
in $X$, where $\a=1,\ldots,r$ and $i=1,\ldots,n$. 
The condition $\eufm{S}_\a(X^i)=0$ 
implies $\bs_+\eufm(S)_\a = 0$ in the $\bs_+$ invariant neighborhood of 
$N$. We have
\eqn\pzero{
\rd_j \eufm{S}_\a \p^j_+=0.
}
We call a non-trivial solution above a zero-modes of $\p_+$,
which is a degree of freedom tangent to the vanishing locus $N$.
We call a  non-trivial solution of the similar equations
\eqn\czero{
\rd_j \eufm{S}_\a \chi^\a_-=0
}
a zero-mode of $\chi_-$.
For a generic choice of section $\eufm{S}_\a$
the equation $\eufm{S}_\a =0$ cuts out a $(n-r)$
complex dimensional subspace of $X$. Then the
equation \pzero\ implies that we have exactly $(n-r)$ zero-modes
of $\p_+$, while the equation \czero\ implies that
we do not have any zero-modes of $\c_-$, since $n\geq r$.
Assume that the equations $\rd_j \eufm{S}_\a=0$, only for a fixed $\a$
have common roots for all $j=1,\ldots, n$.
Then \pzero\ for the fixed $\a$ do not impose any condition on 
the $\p^j_+$ and we may have $(n-r+1)$ zero-modes of $\p_+$.
Similarly the equations  \czero\ do not impose any condition
on  the fixed component $\c^\a_-$ and we may have one
zero-mode of $\c_-$.  Thus we may draw two conclusions

\def\V{\msbm{V}}
\begin{enumerate} 

\item 
For a generic choice of section we do not have
any zero-modes of anti-ghosts. The vanishing locus
$\eufm{S}^{-1}(0)$ of
the section has the right complex $(n-r)$ dimensions
and the zero-modes of $\p_+$ span the tangent space
of $\eufm{S}^{-1}(0)$.

\item
For a non-generic choice of section we may have
anti-ghost zero-modes. The vanishing locus
$\eufm{S}^{-1}(0)$ of
the section have dimension higher than the right one.
In any cases we have
\eqn\tindex{
n -r = \#(\tilde\p_+) - \#(\tilde\chi_-)
}
where $\#(\tilde{fermi})$ denotes the number of fermionic zero-modes.
We call the above the {\it formal} or {\it virtual} complex dimension of 
$\eufm{S}^{-1}(0)$.
The space of anti-ghost zero-modes span a vector bundle
$\V$ over $\eufm{S}^{-1}(0)$ called the anti-ghost bundle.
The fiber dimension of $\V$ may jump when $\eufm{S}^{-1}(0)$
develops singularities.

We also see that our action functional $S^\pr$ \aei\
restricted to the $\bs_+$ and $\bbs_+$ invariant neighborhood
$\CC$ of the fixed point locus is given by
\eqn\aewi{
\eqalign{
S^\pr|_\CC=
  -F_{\a^\pr\bar\b^\pr  i^\pr\bar j^\pr}\tilde\p^{i^\pr}_+
  \tilde\p^{\bar j^\pr}_+\tilde\c^{\a^\pr}_-\tilde\c^{\bar \b^\pr}_-,
}
}
where it is understood all the fermions 
$(\p^i_+,\p^{\bar i}_+,\c^\a_-,\c^{\bar\a}_-$) are replaced by
their zero-modes $(\tilde\p^{i^\pr}_+,\tilde\p^{\bar i^\pr}_+,\tilde\c^{\a^\pr}_-,
\c^{\bar\a^\pr}_-$) and the curvature above is the curvature
of the anti-ghost bundle $\V$ over $\eufm{S}^{-1}(0)$.

\end{enumerate}

Now we examine the path integral.
For $n=r$ and with a generic section 
the vanishing locus $\eufm{S}^{-1}(0)$ is
zero-dimensional and the path integral counts
the number of zeros of the section. For $n=r$
and with a non-generic section the zeros of the section
can be a positive dimensional submanifold $\eufm{S}^{-1}(0) \subset X$
of $X$.
The path integral reduces to an integral over $\eufm{S}^{-1}(0)$ and over
anti-ghost zero-modes.
Note that the rank of the anti-ghost bundle $\V$ over $\eufm{S}^{-1}(0)$ 
is the same as the complex dimension of $\eufm{S}^{-1}(0)$.
The path integral becomes $\chi(\V)$
\eqn\aeo{
Z=\int_{\eufm{S}^{-1}(0)} e(\V)=\chi(\V),
}
which in turn can be identified with $\chi(\E)$.

Now  we consider the case  $r < n$.
The partition function still evaluates the Euler class $e(\V)$
of the anti-ghost bundle $\V$ over $\eufm{S}^{-1}(0)$. 
Since, by the formula \tindex,
the rank of $\V$ is smaller than the complex dimension
of $\eufm{S}^{-1}(0)$. Thus the Euler class $e(\V)$
is not a top form and the partition function vanishes.
To get a non-trivial result we should  insert a set of observables
and evaluate the expectation value
\eqn\aep{
\left<\prod_{\ell=1}^m \hat\a^{p_\ell, q_\ell}\right>
= \int_{\eufm{S}^{-1}(0)} e(\V)\wedge \a^{p_1, q_1}\wedge \ldots \wedge\a^{p_m,q_m},
}
where
\eqn\aeq{
\sum_{\ell=1}^m p_\ell =\sum_{\ell=1}^m q_\ell = n-r,
}
and otherwise the path integral vanishes.
If there are no anti-ghost zero-modes 
we have $e(\V)=1$ and the above correlation function reduces
to the intersection number of homology cycles
Poincar\'e dual to $\a^{p_{\ell}, q_{\ell}}$ in $\eufm{S}^{-1}(0)$.
The selection rule above can be understood in more physical terms.
The path integral measure contains a ghost number anomaly
due to the fermionic zero-modes. The net ghost number
violation of the path integral measure is
$(n-r, n-r)$, which follows from the formula \tindex\
and   the ghost numbers of the fermions;
\eqn\jhgrs{
\eqalign{
\p^i_+ : (1,0),\cr
\p^{\bar i}_+:(0,1),\cr
}\qquad
\eqalign{
\c^\a_-:(-1,0),\cr
\c^{\bar \a}_-:(0,-1).
}
}
To cancel the ghost number anomaly we have to
insert observables according to the selection rule \aeq\
to soak up the fermion zero-modes in the path
integral measure.

\lin{Specializing to $N_c=(2,2)$ Model}

Finally we consider a special case of $N_c=(2,0)$
model which actually has $N_c=(2,2)$ supersymmetry.
We have the following properties
 
\begin{enumerate}

\item For a generic choice of holomorphic potential $\CW(X^i)$
we do not have any anti-ghost zero-modes. The critical set
$V_i^{-1}(0)$ where $V_i =\rd_i \CW(X^j)$ consists of 
a collection of non-degenerate points. The partition function
is the number of such points.

\item For a non-generic $\CW(X^i)$ we may have anti-ghost
zero-modes. The critical set $V_i^{-1}(0)$ may be a higher
dimensional subvariety of $X$. The net ghost number
violation in the path integral measure is always zero. 
Thus the rank of the anti-ghost
bundle $\V$ is exactly the same as the complex dimension
of $V_i^{-1}(0)$. Thus the partition function is well-defined
and computes the Euler characteristic $\chi(\V)$ of $\V$.
We can identify $\V$ with the tangent bundle of  $V_i^{-1}(0)$.
Thus the partition function is the Euler characteristic
of $V_i^{-1}(0)$. This, in turn, can be identified with
the Euler characteristic of $X$. 

\end{enumerate}

\newsec{Equivariant Cohomological Field Theory}

In the previous chapter we developed standard models 
of cohomological field theories associated with
a K\"{a}hler manifold $X$, tangent bundle $TX$ and Hermitian 
holomorphic vector bundle $\E$ over $X$. 
In this chapter  we generalize those modes to
the cases when there is a certain group $\CG$ action. 
This generalization is relevant since most of field theory 
has a certain gauge symmetry. 
The models in the previous chapter are obviously empty
if the target  space $X$ is linear. On the other hand
models in this chapter have rich structures both
for linear\foot{The relation between the previous
section and the present section is best compared with
that of non-linear sigma-models and linear gauged
sigma models in two-dimensions.} 
and non-linear target spaces. This also
allows us to consider more general classes of target spaces
like the space of a certain set of matrices, the space of
a certain set of fields on a manifold, etc.

The central tool will be the notion of  
equivariant cohomology and symplectic quotients.
The only practical difference between the models
in the previous chapter and their equivariant generalizations
are that the later models further localize the path integrals
to the vanishing locus of $\CG$-momentum map, modulo
the $\CG$ symmetry. If the $\CG$ acts freely on such  locus
we recover the standard models in the previous chapters
now associated with the symplectic quotients.
The momentum map is a generalization of the familiar
angular momentum associated with a group of rotations
in the classical mechanics.

\subsec{Equivariant Toy Model}

We return to our toy model in Sect.~$2.1.1$, where
we considered a $n$-dimensional K\"{a}hler manifold
$(X,\varpi)$ with K\"{a}hler form $\varpi$ as the target space.
Now we assume that there is a group $\CG$ action 
\eqn\ala{
\CG\times X\rightarrow X,
}
preserving the complex and K\"{a}hler structures.
We consider the toy model  with
the action functional $S$ in \aad. 
The action functional
is invariant under $\CG$ thus the path integral
is degenerated. We want to remove the gauge degree
of freedom as follows (compare with \aae\ )
\eqn\alb{
\eqalign{
Z &= \Fr{1}{vol(\CG)}\int_X [\CD X \CD\bar X\CD\p\CD\bar\p] \;e^{-S}
\cr
&= \Fr{1}{\#(\CG)}\int_{X/\CG} 
[\CD X \CD\bar X\CD\p\CD\bar\p]^\pr \;e^{-S}
\cr
&= \Fr{1}{\#(\CG)}\int_{X} [\CD X \CD\bar X\CD\p\CD\bar\p\CD (ghosts)]
\;e^{-S -S_{gf} -S_{gh}},
}
}
where $\#(\CG)$ denotes the number of central elements of $\CG$,
$S_{gf}$ and $S_{gh}$ denote the gauge fixing and ghost terms.
The above procedure is the well-known  
Faddeev-Popov-BRST quantization
on which I do not want to review here.\foot{ I only want to
remark that it involves the Lie algebra cohomology with the parity
change.}

A general problem with the path integral above is
that the quotient space $X/\CG$ rarely has good
topology and geometry. This means that it is
difficult make sense out of our (even for finite dimensional)
path integral.  Furthermore 
the geometrical meaning of the $\bs$ and $\bbs$
supercharges on the quotient space is not  quite obvious.
This problem can be avoided by considering
equivariant cohomology. For general references see \MQ\AB\BGV.

\subsubsection{Extending Our Toy Model}

A nice route to introduce the equivariant cohomology is
a simple generalization of our toy model in Sect.~$2.1.1$.
Now we assume that there is a group $\CG$ action $\CG\times X\rightarrow X$
on our target space $X$
preserving the complex and K\"{a}hler structures.
Our goal is to extend our target space and 
supercharges $\bs$ and $\bbs$ by introducing extra fields
such that 

\begin{enumerate}

\item
If  $\CG$ acts freely on $X$ the degrees of freedom
due to the extra  fields disappear,

\item 
the supercharges become $\rd$ and $\bar\rd$ 
operators, after the parity change, on the 
{\it $\CG$-invariant subspace}.

\end{enumerate}

To implant the above idea  we need the notion of Lie derivative.
Consider a manifold with $\CG$ action. 
Let $Lie(\CG)$ be the Lie algebra of $\CG$.
We will always assume that we have a bi-invariant
inner product $<\;\;,\;\;>$ on $Lie(\CG)$ such that
we can identify $Lie(\CG)$ with its dual $Lie(\CG)^*$.
Let $X^I$ be
the local coordinate fields on $X$. The $\CG$ action induces
a vector $V^I_a T^a$ such that an infinitesimal
$\CG$ action is represent by 
\eqn\ald{
X^I \rightarrow X^I + \e^a V^I_a.
} 
We denote by $j_a$ the interior derivative
with respect to the vector $V_a$, i.e.,
\eqn\ale{
\eqalign{
j_a :\O^r(X)&\rightarrow \O^{r-1}(X),\cr
(j_a \a)_{I_2I_3\cdots I_r} &= r V^{I_1}\a_{I_1I_2\cdots I_r}.
}
}
Let $\CL_a$ be the Lie derivative with
respect to the vector field $V_a$;
\eqn\alf{
 \CL_a = dj_a + j_a d
} 
Then the infinitesimal
$\CG$ action on $\a \in \O^*(X)$ is given by
$
\a \rightarrow \a + \e^a\CL_a \a.
$
Thus a differential form $\a$ is $\CG$-invariant
if $\CL_a \a =0$. We note an obvious relation
$\e^a\CL_a X^I = \e^a V_a^I$.

Now we extend our target space $X$ by 
introducing a $Lie(\CG)$-valued scalar $\phi=\phi^a T_a$
and modify the commutation relation \aab\
as\foot{
where $\phi^a$ is $\e^a$ in \ald\
incarnated  as a field.}
\eqn\alg{
\bs^2=0,\qquad \{\bs,\bbs\}=-i\phi^a\CL_a,\qquad
\bbs^2=0.
}
Thus $\{\bs,\bbs\}=0$ on the $\CG$-invariant subspace
of $X$ and the supercharges are related with
the $\rd$ and $\bar\rd$ operators on the invariant
subspace as in the case of our previous toy model.
The ghost numbers of $\phi$ should be assigned 
$(1,1)$ to match the ghost numbers in the anti-commutation
relations above. The above defines $\CG$-equivariant Dolbeault
cohomology \Park\HyunPa.

By the new anti-commutation relations \alg\
the supersymmetry transformation laws \aad\ should be modified
as follows
\eqn\alh{
\eqalign{
\bs X^i &= i\p^i,\cr
\bbs X^i&=0,\cr
\bs X^{\bar i}&=0,\cr
\bbs X^{\bar i}&=i\p^{\bar i},\cr
}\qquad
\eqalign{
\bs \p^i &= 0,\cr
\bbs \p^i&=-\phi^a \CL_a X^i,\cr
\bs \p^{\bar i}&=-\phi^a\CL_a X^{\bar i},\cr
\bbs \p^{\bar i}&=0.\cr
}\qquad
\eqalign{
\bs  \phi =0,\cr
\bbs \phi=0,
}
}
where we obtained the  conditions $\bs \phi=\bbs\phi=0$ 
by demanding the algebra to be closed. 
Assume that we have a model with an action functional
which is invariant under the supersymmetries generated by 
the above new supercharges. Then
we can apply the fixed point theorem of Witten and 
we have the following fixed point
equation, deduced from the above
\eqn\alkk{
\phi^a \CL_a X^I=0.
}
This equation tells us that $\phi^a=0$ if $\CG$ act freely on $X$
while $\phi^a$ can be non-zero on a fixed point of the $\CG$ action.
Thus we achieved our initial two goals.

Now we consider a supersymmetric action functional $S$.
Compare with the non-equivariant case in Sect.~$2.1.1$,
an action functional should be invariant under $\CG$
in addition to $\bs S=\bbs S=0$. These conditions
imply that one can also apply the Poincar\'e lemma
since the new supercharges are also nilpotent if they
are acting on $\CG$ invariant  quantities.
Thus $S$  can be written by the same form as the previous toy model
\eqn\ali{
S = i\bs\bbs \CK(X^i, X^{\bar i}),
}
where $\CK$ should be $\CG$ invariant. \foot{
Actually $\CK$ only needs to satisfy a weaker
condition that it should invariant under gauge transformations
connected to the identity.
}
Applying the transformation laws \alh\
we have
\eqn\alj{
S =-i \bigl<\phi, \m\bigr> -i g_{i\bar j} \p^i\p^{\bar j},
}
where
\eqn\momentmap{
\m_a = i \Fr{\rd \CK}{\rd X^{\bar i}}\left(\CL_a X^{\bar i}\right).
}
Later we shall see  that $\m =\m_a T^a$ is the equivariant $\CG$
momentum map on $X$. 
Maintaining all the supersymmetry we consider the following
more general action functional $S(\zeta)$
\eqn\alk{
\eqalign{
S(\zeta) &=i\bs\bbs \CK  +i \phi^a \zeta_a\cr 
&= -i<\phi, \m -\zeta> -i g_{i\bar j}\p^i\p^{\bar j},
}
}
where $\zeta$ belongs to the center of $\CG$.
We call the additional term a FI coupling.

Now we consider the partition function  for the new action.
We have
\eqn\alm{
\eqalign{
Z(\zeta) &=\Fr{1}{vol(\CG)} 
\int [\CD\phi\CD X \CD\bar X\CD\p\CD\bar\p] \;e^{-S(\zeta)}\cr
 &=\Fr{1}{vol(\CG)} \int_X \d(\m -\zeta)\prod_{k,\bar k} dX^\k dX^{\bar k} 
d\p^k d\p^{\bar k}\cdot e^{-ig_{i\bar j}\p^i\p^{\bar j}}
\cr
&=\Fr{ 1}{\# \CG} \int_{\m^{-1}(\zeta)/\CG} \Fr{\tilde \varpi^r}{r!}\cr
&= \Fr{ 1}{\# \CG} vol(\CN_\zeta),
}
} 
where 
\eqn\aln{
\CN_\zeta = \m^{-1}(\zeta)/\CG.
}
In the above we assumed that $\CG$ acts freely on the locus 
$\m^{-1}(\zeta)\subset X$. Thus we could simply integrate
$\phi$ out, which gives rise to the delta function supported on 
$\m^{-1}(\zeta)$. Then  the quotient space $\CN_\zeta$
is smooth. Our action functional $S_\zeta$ reduces to
the K\"{a}hler form on the subspace $\m^{-1}(\zeta)$.
Since it is $\CG$ invariant it becomes, after the
parity change, the K\"{a}hler from $\tilde\varpi$ on 
the quotient space $\CN_\zeta$. What we showed is the
symplectic reduction theorem of Marsden and Weinstein \MarW.

We call our extended toy model the equivariant toy model.
We note that the equivariant toy model makes perfect sense
even if we start from a flat K\"{a}hler manifold $X$ as our
initial target space. We call the space $\CN_\zeta$ {\it the effective
target space}, which can be a very complicated non-linear
space even if our initial target space $X$ is flat.

Before examining further properties of our model, we turn
to a review of the equivariant cohomology and
momentum map.
We refer for details on
the equivariant cohomology and relation with
momentum maps to a beautiful exposition of Atiyah and Bott \AB.
The idea is to replace $X$ by a bigger space 
$X\times E\CG$
such that the extended space has a nice quotient
$$X_\CG=(X\times_\CG E\CG)$$ 
which is equivalent to the original
quotient $M/\CG$ when it has a nice quotient.\foot{
The additional space $E\CG$ is a fixed universal
$\CG$-bundle over the classifying space $B\CG$.
The homotopy quotient $X_\CG$ forms
a fiber bundle $\pi: X_\CG \rightarrow B\CG$
with fiber $X$. Then we have the following diagram
\eqn\alc{
\matrix{
E\CG &\leftarrow & E\CG\times X & \rightarrow X \cr
\downarrow & & \downarrow &\downarrow\cr
B\CG &\leftarrow& E\CG\times_\CG X &\rightarrow X/\CG
}
}
}
The $\CG$-equivariant cohomology $H^*_\CG(X)$ 
of $X$
is defined as the ordinary cohomology $H^*(X_\CG)$
of $X_\CG$.
For instance the $\CG$-equivariant cohomology of $M$
is the ordinary cohomology of $X/\CG$ if $\CG$
acts freely on $X$.\foot{Note, however, $\CG$-equivariant
cohomology of a point is $H^*(B\CG)$ which is highly
non-trivial.}

We will briefly review a convenient 
model of  equivariant cohomology due to Cartan, 
of which variants will be used in this thesis.  A crucial reference
on the Cartan model for us is Witten's paper \tdYM.  The path
integral of the equivariant toy model  reproduces
a K\"{a}hler version of  Witten's non-Abelian equivariant integration 
formula.

\subsubsection{Equivariant Cohomology and Momentum Map}

Consider a manifold $X$ with $\CG$ action. 
Let $Lie(\CG)$ be the Lie algebra of $\CG$.
We will always assume that we have a bi-invariant
inner product $<\;\;,\;\;>$ on $Lie(\CG)$ such that
we can identify $Lie(\CG)$ with its dual $Lie(\CG)^*$.

Let $Fun^*(Lie(\CG))$ denote the algebra of polynomial
functions on $Lie(\CG)$ so that an $m^{th}$ 
order homogeneous 
polynomial is considered to be of degree $2m$.
The  equivariant differential forms  $\O^*_\CG(X)$ 
on $X$ are
represented by
\eqn\ama{
\O^*_\CG(X):=\left(\O^*(X) \otimes Fun^*(Lie(\CG))\right)^\CG,
}
where ${}^\CG$ denote the $\CG$-invariant part.
The degree of such a form is the sum of degrees
of  $\O^*(X)$ and  $Fun^*(Lie(G))$. 
One endows $\O^*_\CG(X)$ with the equivariant differential
operator $d_\CG$
\eqn\amb{
 d_\CG=d -  i\phi^a j_{a}, \qquad  d_\CG^2 = -i\phi^a\CL_a,
}
where $j_a^2=0$ and  $\phi=\phi^a T^a\in Lie(\CG)$. 
That is,  $d_\CG^2=0$
modulo an infinitesimal gauge transformation generated by $\phi^a$.
Thus on the space $\O^*_\CG(X)$ we have\foot{
An element in $\O^*_\CG(X)$ is annihilated by
$L_a = \CL_a + f_{ab}{}^c\phi^b\Fr{\rd}{\rd\phi^c}$,
where $f_{ab}{}^c = -f_{ba}{}^c$ are the structure constants
of $\CG$. Then,
it is also annihilated by $\phi^a\CL_a$ since
$\phi^a L_a = \phi^a\CL_a$.} 
$$\hat d_\CG^2=0.$$
The  $\CG$-equivariant de Rham cohomology on $X$
is the cohomology of the complex  $(\O^*_\CG(X), d_\CG)$. 
The equivariant cohomology
of $X$ is the ordinary cohomology of the quotient space if
the group acts freely, otherwise it is something else.
For example $H^*_\CG(pt)$ is $Fun^*(Lie(\CG))$.

\lin{The Symplectic Case}

Now we consider a symplectic manifold $X$ with
symplectic form $\varpi$. Assume that we have a $\CG$ action
on $X$. Under an infinitesimal $\CG$ action 
$X^I \rightarrow X^I + \e^a V^I_a$ the symplectic
form transforms as $\varpi \rightarrow \varpi + \e^a\CL_a\varpi$.
Thus we have a vector field $V^I_a$ which is an infinitesimal
symplectic transformation whenever $\CL_a\varpi=0$.
Since $d\varpi=0$ we have $d (j_a\varpi)=0$, thus at least
locally we can write
\eqn\amc{
j_a \varpi = d\m_a.  
}
The $\m = \m_a T^a: X \rightarrow Lie(\CG)^*$ is called the $\CG$-momentum
map.\foot{Note that we identified  $Lie(\CG)$
with its dual $Lie(\CG)^*$.} The obstruction for global
existence of $\m_a$ is $H^1(X)$.
The momentum map is a generalization of the familiar classical
mechanical notion that $X$ is a classical phase
space and $\CG$ is a group of rotation and
$\m$ is the angular momentum.
The momentum map is equivariant
if $\m(g(x)) = (ad\; g)^*(\m(x))$. Then
$\CG$ preserve the subspace $\m^{-1}(\zeta)$
when $\zeta$ is a centeral element. Then the reduced
phase space or the symplectic quotient
is defined by
\eqn\amd{
\CN_\zeta = (X\cap\m^{-1}(\zeta))/\CG
}
The quotient space is a smooth symplectic manifold
if $\zeta$ is a regular value. The symplectic form
$\widetilde \varpi$ on $N_\zeta$ is obtained
from $\varpi$ by restriction and reduction \MarW.

The equivariant cohomology and the momentum map
are closely related \AB. Note that the symplectic form
$\varpi$ is not equivariantly closed, $d_\CG\varpi \neq 0$.
We have a unique form, due to the degree, of 
equivariant extension $\varpi_\CG$ of $\varpi$
\eqn\ame{
\varpi_\CG = \varpi  +  i(\phi, \m)
}
The condition $d_\CG \varpi_\CG=0$
reduces to using $d_\CG \phi=0$
\eqn\amf{
<\phi, d\m - j\varpi>=0.
}
Thus $\varpi_\CG$ is equivariantly closed iff $\m$ is
the momentum map \amc.
Note that $\varpi_\CG$ is $\CG$ invariant, $L_a \varpi_\CG=0$,
iff the momentum
map $\m$ is equivariant.

\lin{The K\"{a}hler Case}

Now we specialize to the case that $X$ is a K\"{a}hler
manifold with K\"{a}hler form $\varpi$ and with $\CG$ action,
which  preserve the complex structure and the K\"{a}hler
form. The vector field $V^I$ induced by the $\CG$ action
is decomposed into $V^I = V^i + V^{\bar i}$. Thus 
one can introduce interior derivatives $\iota_a$
and $\bar\iota_a$ by contracting with $V^i_a$ and
$V^{\bar i}_a$, respectively, such that $j_a = \iota_a +\bar\iota_a$;
\eqn\amg{
\eqalign{
\iota_a :\O^{p,q}(X) \rightarrow \O^{p-1,q}(X),\cr
\bar\iota_a :\O^{p,q}(X) \rightarrow \O^{p,q-1}(X).\cr
}
}
{}From the relation $j_a^2=0$ we have
\eqn\amh{
\iota_a^2=0,\qquad \{\iota_a ,\bar\iota_a\} =0,
\qquad \bar\iota_a^2=0.
}
It follows that 
\eqn\ami{
\CL_a = \rd \iota_a +\iota_a\rd + \bar\rd \bar\iota_a +\bar\iota_a\bar\rd.
}
We also decompose $Fun^*(Lie(\CG))$ such
that an $m^{th}$ order homogeneous 
polynomial in $Fun(Lie(\CG))$ is considered to be of 
degree $(m,m)$.
Then  equivariant differential forms  $\O^{*,*}_\CG(X)$ 
on $X$ are
represented by
\eqn\amj{
\O^{*,*}_\CG(X):=\left(\O^{*,*}(X) \otimes Fun^*(Lie(\CG))\right)^\CG
}
Similarly we decompose $d_G$ into 
\eqn\amk{
 d_\CG=\rd_G + \bar\rd_G: \O^0_\CG(M)
= \O^{1,0}_\CG(M) \oplus \O^{0,1}_\CG(M).
}
where
\eqn\aml{
\eqalign{
\rd_\CG=\rd - i\phi^a \bar\iota_a,\cr
\bar\rd_\CG= \bar\rd -i\phi^a\iota_a.
}
}
Remark that $\phi$ is assigned to degree $(1,1)$.
The anti-commutation relations between $\rd_\CG$ and $\bar\rd_\CG$
are
\eqn\amm{
\rd_\CG^2=0,\qquad \{\rd_\CG,\bar\rd_\CG\}=-i\phi^a\CL_a,\qquad
\bar\rd_\CG^2=0.
}
This defines equivariant Dolbeault cohomology on a K\"{a}hler
manifold. Comparing with the anti-commutation relations \alg\
we can identify our supercharges $\bs$ and $\bbs$ with
$\rd_\CG$ and $\bar\rd_\CG$ after the parity change \aao.
Thus
\eqn\amn{
\eqalign{
\bs &= i\p^i \Fr{\rd}{\rd X^i} 
	-\phi^a V_a^{\bar i}\Fr{\rd}{\rd \p^{\bar i}},\cr
\bbs &= i\p^{\bar i} \Fr{\rd}{\rd X^{\bar i}} 
-\phi^a V_a^{i}\Fr{\rd}{\rd \p^{i}}.
}
}

Now we examine the relation between the momentum map
and equivariant Dolbeault cohomology.
For the K\"{a}hler case the relation \amc\ becomes, by matching
form degrees
\eqn\amq{
\iota_a\varpi = \bar\rd \m_a.
}
Since the K\"{a}hler form $\varpi$ can be written locally
in terms of a K\"{a}hler potential $f$,
\eqn\amr{
\varpi=i\rd\bar\rd f,
}
we have
\eqn\ams{
i\iota_a (\rd\bar\rd f)= \bar\rd \m_a.
}
Using the relations $\{\iota_a,\bar\rd\}=\{\rd,\bar\rd\}=0$
we deduce that
\eqn\amt{
\m_a=i\iota_{a}(\rd  f)
}
up to a constant. 
Combining all together we find an important identity
\eqn\amu{
i\rd_\CG\bar\rd_{\CG} f = \varpi + i<\phi,\m >,
}
which
we obtained earlier in \ali\ and \alj.
Thus minus the action functional, $-S$, is 
a $\CG$-equivariant K\"{a}hler form after the parity
change. Note that the momentum map derived above
is equivariant if the K\"{a}hler potential is $\CG$
invariant. Thus we showed all the assertions made in
Sect. $3.1.1$.

\subsubsection{The Path Integrals and Non-Abelian Localization Theorem}

We now return to the equivariant toy model.
We return to the partition function $Z$ \alm\ and
ask what will happen as we vary the FI term $\zeta$.

We have a classical theorem; the image of a proper momentum
map of a compact group is a convex polytope divided
by walls \DH\Atiyah\Kirwan. As we vary $\zeta$
the symplectic quotient $\CN_\zeta$ may undergo birational
transformations if the path of $\zeta$ crosses a wall, otherwise
diffeomorphic.
For the non-proper case the symplectic quotient does not
exist. This does  not imply that the partition function
is empty.
Recall that space of all bosonic fields is a copy of $X$
and the space of all $\phi$. The correct picture is
that the path integral is localized to $\CN_\zeta \subset X/\CG$
for regular values of $\zeta$.
The full equation for the localization is
\eqn\ana{
\eqalign{
\m -\zeta =0,\cr
\phi^a \CL_a(X^i)=0.\cr
}
}
We call the non-trivial solutions $\phi_0$ 
of above equations
the zero-modes of $\phi$. It is clear that we have zero-modes
of $\phi$ whenever the $\CG$ action has fixed points
in $\m^{-1}(\zeta)$, thus when $\zeta$ lies on
a non-regular value $\zeta_0$.
Clearly the path integral degenerates at   such a value since
the path integral measure contains zero-modes of $\phi$. 
Let $\zeta_+< \zeta_0< \zeta_-$ 
we have
\eqn\anb{
Z(\zeta_+) \neq Z(\zeta_-)
}
due to topology change. At $\zeta_0$ the partition function
should be singular.

It is clear how to resolve the singularity of the path integral.
We have to regularize.
We consider a more general action functional
$S(\zeta,\e)$,
\eqn\anc{
\eqalign{
S(\zeta,\e)&=S(\zeta) +\Fr{\e}{2}<\phi, \phi>\cr 
&= -i<\phi,\m-\zeta > +\Fr{\e}{2}(\phi,\phi)
-i g_{i\bar j}\p^i\p^{\bar j}.
}
}
Note that the additional term is invariant under
$\CG$ as well as all the supersymmetry.
The additional $\e$ dependent term changes the
fixed point equations of the supersymmetry
since the  $\phi$ equation of motion is now
\eqn\anddd{
i(\m -\zeta) = \e \phi.
} 
Consequently, for $\e\neq 0$, the path integral
is localized to the locus of the following
equations
\eqn\ane{
{\left(\Fr{\rd \m_a}{\rd X}\right)}(\m^a -\zeta^a )=0.
}
Now we also have contributions from {\it higher critical points}.
Thus, we have two branches; (i) $\m^a -\zeta^\a=0$, 
(ii) $\Fr{\rd }{\rd X}\m =0$.
Clearly  the  quotient $\CN_\zeta$ space develops
singularities when branches (i) and (ii) intersect.
In such a case the integrand of path integral contains
the Gaussian measure
\eqn\anf{
e^{ - \e \sum_\ell |\phi_{0,\ell}|^2}
}
for the space of 
zero-modes  $\phi_{0,\ell}$ of $\phi$.
Thus the path integral is non-singular.
Consequently the politically correct version of
the model is defined by the action functional
$S(\zeta,\e)$.

Now we consider the correlation functions.
A supersymmetric observable should be 
$\CG$-invariant as well as invariant under $\bs$ and $\bbs$.
Such an observable should be constructed
from an equivariantly closed differential form.
An equivariant differential form $\CO^{p,q}$ of total degree
$(p,q)$ can be expanded as
\eqn\ang{
\CO^{p,q} = \a^{p,q}_0 + \phi^a \a^{p-1,q-1}_a
+ \phi^a \phi^b  \a^{p-2,q-2}_{ab} +\cdots ,
}
where $\a^{p,q}_*\in \O^{p,q}(X)$.
Let $\hat\CO^{p,q}$  be the parity change
of $\CO^{p,q}$, thus carrying ghost number $(p,q)$.
We have
\eqn\anh{
\bbs \hat \CO^{p,q} = \widehat {\bar\rd_\CG \CO}^{p,q}.
}
Thus $\hat\CO^{p,q}$ is an $\bbs$-invariant observable
if $\CO^{p,q}$ is an $\bar\rd_\CG$-closed equivariant
differential form.

The correlation function of observables or the expectation
value is defined by
\eqn\aam{
\left<\prod_{m=1}^r\hat\CO^{p_m, q_m}\right>
= \int [\CD\phi\CD X \CD\bar X\CD\p \CD\bar\p] 
\prod_{m=1}^r\hat\CO^{p_m, q_m}\cdot 
e^{-S}.
}
For the present model
one can show that
\eqn\aan{
\eqalign{
\left<\prod_{m=1}^r\hat\CO^{p_m, q_m}\right>
= &
\Fr{1}{vol(\CG)}\int_\CG \Fr{d\phi_1 d\phi_2\ldots d\phi_s}{(2\pi)^s}
\cr
\times 
\int_X \CO^{p_1,q_2}\wedge &\ldots\wedge \CO^{p_r,q_r} \cdot 
\exp\left(\varpi +i(\phi, \m -\zeta) -\Fr{\e}{2}(\phi, \phi)\right),
}
}
where $s =dim(\CG)$.
Applying the fixed point theorem for the global supersymmetry
we see that the above  integral can be written as
a sum of contribution of the
critical points \ane\ of $I=<\m, \m>$.
This is the non-Abelian localization theorem of Witten \tdYM,
generalizing the more familiar abelian Duistermaat-Heckman (DH) integration formula \DH. 
In the end  our equivariant toy model turns out to be
very non-trivial.

\subsec{The Equivariant $N_c=(2,2)$ Model}

\def\mapr{\!\smash{
	    \mathop{\longrightarrow}\limits^{\bs_+}}\!}
\def\mapl{\!\smash{
	    \mathop{\longleftarrow}\limits^{\bs_-}}\!}
\def\mapbr{\!\smash{
	    \mathop{\longrightarrow}\limits^{\bbs_+}}\!}
\def\mapbl{\!\smash{
	    \mathop{\longleftarrow}\limits^{\bbs_-}}\!}
\def\mapd{\Big\downarrow
 	 \rlap{$\vcenter{\hbox{$\scriptstyle \bbs_-$}}$}}
\def\mapu{\Big\uparrow
	  \rlap{$\vcenter{\hbox{$\scriptstyle\bbs_+$}}$}}
\def\ne{\nearrow}
\def\se{\searrow}
\def\nw{\nwarrow}
\def\sw{\swarrow}

In this section
we develop the equivariant generalization
of the $N_c=(2,2)$ model in Sect.~$2.2$.
We assume the same group $\CG$ acting on $X$ as in the previous
section. This naturally extend to the tangent space $TX$.
Recall that the partition function of our toy model 
is the symplectic volume of the target space, while
the partition function of the equivariant toy model is
the symplectic volume of the symplectic quotient $\CN_\zeta$,
for generic values of $\zeta$, of $X$ by $\CG$.
Similarly, the partition function the equivariant version of 
$N_c=(2,2)$ model, without  holomorphic potential $\CW$,
will be the Euler characteristic $\chi(T\CN_\zeta)$ 
of  the 
symplectic quotient $\CN_\zeta$, for  generic value of $\zeta$, 
of $X$ by $\CG$.
After turning on $\CW$, the path integral reduces
to the symplectic quotient $\CM_\zeta$ of the critical
subset $X_{crit}\subset X$ of the potential $\CW$ by $\CG$.

We consider the same "type" of supercharges
carrying the same ghost numbers $(p,q)$;
\eqn\aoa{
\eqalign{
\bs_+:(+1,0),\cr
\bs_-:(-1,0),\cr
}\qquad
\eqalign{
\bbs_+:(0, +1),\cr
\bbs_-:(0,-1).\cr
}}
Now we postulate the
supercharges to satisfy the following
anti-commutation relations
\eqn\aob{
\eqalign{
\{\bs_+,\bs_+\}=0,\cr
\{\bs_+,\bs_-\}=0,\cr
\{\bs_-,\bs_-\}=0,\cr
}\qquad
\eqalign{
\{\bs_+,\bbs_+\}&=-i\phi^a_{++}\CL_a,\cr
\{\bs_+,\bbs_-\}&=-i\s^a\CL_a,\cr
\{\bbs_+ ,\bs_-\}&= -i\bar\s^a\CL_a,\cr
\{\bs_-,\bbs_-\}&=-i\phi^a_{--}\CL_a,
}\qquad
\eqalign{
\{\bbs_+,\bbs_+\}=0,\cr
\{\bbs_+,\bbs_-\}=0,\cr
\{\bbs_-,\bbs_-\}=0,\cr
}
}
which are equivariant generalizations of the
commutation relations \aba\ and \abac\ for the $N_c=(2,2)$
model. For the $\CG$-invariant subspace 
the equivariant supercharges
are the same as the non-equivariant ones. 
Here, in total, we introduced four 
bosonic fields $\phi_{\pm\pm}$, $\s$ and $\bar\s$
taking values in $Lie(\CG)$. They carry the
following ghost numbers
\eqn\aoc{
\eqalign{
\phi_{++}:(+1,+1),\cr
\phi_{--}:(-1,-1),\cr
}\qquad
\eqalign{
\s:(+1,-1),\cr
\bar\s:(-1,+1).
}
}
The anti-commutation
relations above define balanced  $\CG$-equivariant
Dolbeault cohomology \DPS. 
This is the K\"{a}hler version of the balanced equivariant 
cohomology  \DM.\foot{ 
In our approach a balanced cohomological field theory \BT\VW\DM\
is a $N_c=(1,1)$ supersymmetric sigma-model in $(0+0)$ 
dimensions, whose target space can be a general Riemannian
space. }

We should remark that the above algebra can
be obtained by  dimensional reduction of
the $N=1$ supersymmetry algebra of four-dimensional
super-Yang-Mill theory and, equivalently, the algebra
of $N_{ws}=(2,2)$ super-Yang-Mills theory in
two-dimensions. Thus we may introduce other
quantum numbers, as in two-dimensions,
the left and right $U(1)$ $\CR$-charges $(J_L,J_R)$
as follows 
\eqn\rcharge{\eqalign{
&\bs_+ : (+1,0),\qquad \bbs_+ : (-1,0),\cr
&\bs_- : (0,+1),\qquad \bbs_- : (0,-1).
}
}
The analogy with the two-dimensional $N_{ws}=(2,2)$
space-time supersymmetric gauge theory, equivalently
the linear gauged sigma-model \GLSM\Wittengr\
will be very useful. Indeed it is a trivial step
to obtain a $N_{ws}=(2,2)$ model, and vice versa, just by replacing
$\phi^a_{\pm\pm}\CL_a$ by the left and right
moving covariant derivatives $D_{\pm\pm}$ everywhere.
Then the indices $\pm$ are identified with the left and
right spinor indices in two-dimensions. For example
requiring the ghost number symmetry is equivalent
to requiring the two-dimensional Lorentz symmetry.

\subsubsection{The Basic Structure}

Now we examine the basic structure of the model.

\lin{The $N_c=(2,2)$ Multiplets}

\begin{itemize}

\item Chiral multiplets

We have the same chiral multiplets introduced in
the non-equivariant $N_c=(2,2)$ model,
\eqn\aoddd{
\bbs_\pm X^i=0.
}
We have
\eqn\chiralm{
\def\normalbaselines{\baselineskip20pt
\lineskip3pt \lineskiplimit3pt}
\matrix{
\p^i_- &\mapl & X^i &\mapr & \p^i_+\cr
             & 
 	 \rlap{\lower.3ex\hbox{$\scriptstyle s_{\!+}$}}\searrow
 & &\swarrow\!\!\!\rlap{\lower.3ex\hbox{$\scriptstyle s_{\!-}$}} 
& \cr
&&H^i&&
}.
}
We denote their anti-chiral partners $(X^{\bar i}, \p^{\bar i}_\pm,
H^{\bar i})$,  which are  their  Hermitian conjugates.

\item Gauge multiplet

The internal consistency of the anti-commutation relations \aob\
determines uniquely the following multiplet
\eqn\vectorm{
\def\normalbaselines{\baselineskip20pt
\lineskip3pt \lineskiplimit3pt}
\matrix{
\bar\s     & \mapr & \eta_+ & \mapl & \phi_{++}     \cr
\mapd      &       & \mapd  &       & \mapd      \cr
\bar\eta_- & \mapr & D      & \mapl & \bar\eta_+ \cr
\mapu      &       & \mapu  &       & \mapu      \cr
\phi_{--}     & \mapr & \eta_- & \mapl & \s         \cr
}
}
where $D$ is real auxiliary field. All the fields above take
values in $Lie(\CG)$.
We call the above multiplet  a $N_c=(2,2)$
{\it gauge multiplet} since it originated from the
$\CG$ action on $X$. 
Remark that $\s$ is {\it twisted-chiral}; i.e.,
\eqn\aoeef{
\bs_+\s =\bbs_-\s =0.
}

\item The ghost numbers

The   ghost numbers $(p,q)$ of the fields in the gauge multiplet
are determined from the assignments \aoa\ and
the commutation relations \aob. We set the
ghost number of $X^i$ to $(0,0)$.
For the bosonic
fields we have
\eqn\aof{
\matrix{
    &\phi &\bar\phi & \s & \bar\s& D & X^i & H^i \cr
p & +1& -1 & +1 & -1 & 0&0&0\cr
q&  +1 & -1 & -1& +1 &0&0&0\cr
}
}

\item The $\CR$-charges 

The   $\CR$-charges  $(J_L,J_R)$ of the fields in the gauge multiplet
are also determined from the assignments \rcharge\ and
the commutation relations \aob.  We set the
 $\CR$-charges
of $X^i$ to $(0,0)$. For the bosonic
fields we have
\eqn\aog{
\matrix{
    & \phi_{++} & \phi_{--} & \s & \bar\s & D & X^i & H^i 
\cr
J_L & 0      & 0      & +1 & -1     & 0 &0 &1
\cr
J_R & 0      & 0      & -1 & +1     & 0 &0 & 1
}
}

\end{itemize}

\lin{The Supersymmetry Transformation Laws}

The explicit transformation
laws for the fields in the  $N_c=(2,2)$ gauge multiplet
are uniquely determined by the internal consistency
\eqn\vector{
\eqalign{
\d \phi  &= i\bar\ep_+\eta_+ + i\ep_+\bar\eta_+,\cr
\d \bar\phi&= i\bar\ep_-\eta_- + i\ep_-\bar\eta_-,\cr
\d \s    &= -i\bar\ep_+\eta_- -i\ep_-\bar\eta_+,\cr
\d\bar\s &= -i\bar\ep_-\eta_+ -i\ep_+\bar\eta_-,\cr
\d\eta_{+}
     &=+i\ep_+ D
	-\Fr{1}{2}\ep_+[\s,\bar\s] 
	-\Fr{1}{2}\ep_+[\phi_{++},\phi_{--}]
	-\ep_-[\phi_{\!++},\bar\s]
,\cr
\d\bar\eta_{+}
     &= -i\bar\ep_+D
	+\Fr{1}{2}\bar\ep_+[\s,\bar\s] 
  	-\Fr{1}{2}\bar\ep_+ [\phi_{++},\phi_{--}]
	-\bar\ep_-[\phi_{\!++},\s]
,\cr
\d\eta_{-} 
	&=+i\ep_-D
	+\Fr{1}{2}\ep_-[\s,\bar\s]
	+\Fr{1}{2}\ep_-[\phi_{++},\phi_{--}]
	-\ep_+[\phi_{\!--},\s]
,\cr
\d\bar\eta_{-} 
 	 &=-i\bar\ep_-D
	-\Fr{1}{2}\bar\ep_-[\s,\bar\s] 
  	+\Fr{1}{2}\bar\ep_-[\phi_{++},\phi_{--}]
   	-\bar\ep_+ [\phi_{\!--},\bar\s]
,\cr
\d D
&=
+\Fr{1}{2}\bar\ep_-[\phi_{\!++},\eta_{-}]
+\Fr{1}{2}\bar\ep_-[\s,\eta_+]
+\Fr{1}{2}\bar\ep_+[\phi_{\!--},\eta_{+}]
+\Fr{1}{2}\bar\ep_+[\bar\s,\eta_-]
\cr
&\phantom{=}
-\Fr{1}{2}\ep_-[\phi_{\!++},\bar\eta_-]
-\Fr{1}{2}\ep_-[\bar\s,\bar\eta_+]
-\Fr{1}{2}\ep_+[\phi_{\!--},\bar\eta_{+}]
-\Fr{1}{2}\ep_+[\s,\bar\eta_-],
}
}
where $D$ is an auxiliary field and
the commutators are for $Lie(\CG)$.

The transformation laws for
chiral multiplets are also uniquely determined from the
conditions $\bbs_\pm X^i =0$.
\eqn\chiral{
\eqalign{
 \d X^i
	=& i\bar\ep_+\p^i_- +i\bar\ep_-\p^i_+ ,
\cr
\d\p^i_+ =&
	+\bar\ep_+ H^i
	-\ep_-\phi^a_{\!++}\CL_a(X^i)
	-\ep_+\s^a\CL_a(X^i),
\cr
\d\p^i_- =&
	-\bar\ep_- H^i
	-\ep_+\phi^a_{\!--}\CL_a(X^i)
	-\ep_- \bar\s^a \CL_a(X^i),
\cr
\d H^i =&
	+i\ep_-\phi^a_{\!++}\CL_a(\p^i_-) 
	+i\ep_-\eta^a_+\CL_a(X^i)
	-i\ep_- \bar\s^a\CL_a(\p^i_+) 
\cr
&	-i\ep_+\phi^a_{\!--}\CL_a(\p^i_+)
	-i\ep_+\eta^a_-\CL_a(X^i)
	+i\ep_+\s^a\CL_a(\p^i_-),
}
}
where $H^i$ are auxiliary fields as in the non-equivariant
$N_c=(2,2)$ model.
The details of the transformation laws above depend on the
ways the group $\CG$ acts on $X^i$.
One may have several different chiral multiplets.
Their transformation laws are also determined as above
once the complex structure and the group action are given
for the bosonic fields.

\lin{The Fixed Point Equations}

One can never over emphasize the importance
of the fixed point theorem of Witten. We have seen
many times that the existence of global supersymmetry
determine the theories almost uniquely. Such uniqueness
becomes stronger as many global supercharges we have.

{}From the above supersymmetry transformation laws we
see that the simultaneous fixed point equations for
all the $N_c=(2,2)$ are given by
\eqn\ttfixed{
\eqalign{
H^i=0,\cr
D=0,\cr
\w^a_m\CL_a(X^i)=0,\cr
[\w_m,\w_n]=0,\cr
}
}
where $\w_m$, $m=1,\ldots,4$ denote the four independent
real $Lie(\CG)$-valued scalar components of 
$\phi_{\pm\pm}$, $\s$ and its Hermitian
conjugate $\bar\s$. The action functional, 
in many respects,  just gives the detailed form of
the values of the auxiliary fields $D$ and $H^i$.
The path integral is localized to the solution space of
the above set of equations modulo  the $\CG$-action. 
The third
equation implies that $\w_m$ are identically zero if
$\CG$ act freely on the subset  $H^{-1}(0)\cap D^{-1}(0) \subset X$.
In such a case the path integral reduces to an integral
over the quotient space 
\eqn\ttlocus{
\left(H^{-1}(0)\cap D^{-1}(0)\right) /\CG.
}
We call this {\it the effective target space}.
The $N_c=(2,2)$ supersymmetry further implies, as we shall
see shortly, that
the above space is a K\"{a}hler manifold.

If one is interested in evaluating correlation
functions of observables invariant only under the supersymmetry
generated by $\bs_+$ and $\bbs_+$, the path integral
is localized to the locus of the following equations
\eqn\tzfixed{
\eqalign{
H^i=0,\cr
D-\Fr{i}{2}[\s,\bar\s]=0,\cr
\bar\s^a\CL_a(X^i)=0,\cr
}
}
and
\eqn\tzfixedb{\eqalign{
\phi^a_{++}\CL_a(X^i)=0,\cr
[\phi_{++},\phi_{--}]=0,\cr
[\phi_{++},\bar\s]=0.\cr
}
}

\subsection{Action Functional and Partition Function}

We define the general action functional $S$
by demanding $N_c=(2,2)$ supersymmetry,
the $\CG$-symmetry and the ghost number
symmetry.  We may, however, not require the $U(1)_\CR$
symmetry in general.
Then $S$ should have the following
form\foot{The total "K\"{a}hler" potential 
$K(X^i, X^{\bar i}) - \bigl<\s,\bar\s\bigr>$
can be generalized to an arbitrary $\CG$-invariant
real functional $\tilde \CK(X^i, X^{\bar i}; \s,\bar\s)$.
Then we may obtain a model
whose effective target space is non-K\"{a}hler but has torsion
and generally a dilaton.
}
\eqn\ttaction{
\eqalign{
S= 
&\bs_+\bs_-\bbs_+\bbs_-\CK\bigl(X^i, X^{\bar i})  
+\bs_+\bs_- \!\CW\bigl(X^i\bigr)
+\bbs_+\bbs_-\bar\CW\bigl(X^{\bar i}\bigr)
\cr 
&
-\bs_+\bs_-\bbs_+\bbs_-\bigl<\s,\bar\s\bigr>
+ \bbs_+\bs_- <t, \s>
+ \bs_+\bbs_- \tr<\bar\s,\bar t>,
}
}
where all potentials $\CK(X^i, X^{\bar i})$, $\CW(X^i)$ and
its Hermitian conjugate $\bar\CW(X^{\bar i})$ are $\CG$-invariant
and\foot{The theta term plays no roles in the $(0+0)$-dimension
we are considering here.}
\eqn\aoh{
t = \Fr{\th}{2\pi} - i\zeta
}
belongs to the center of $Lie(\CG)$.
The first line of the action functional \ttaction\ has the same form 
as the non-equivariant
$N_c=(2,2)$ action functional.
We remark that the above action functional can be a quite
strange object if $\CK(X^i, X^{\bar i})$ is non-linear as well as
if  $X^i$ are certain matrices.

Expanding the action functional above we have
the following terms depending on the auxiliary fields
\eqn\aoi{
S = \bigl<D,D\bigr> - i\bigl<D,  \m - \zeta\bigr> 
+\bigl< g_{i\bar j}H^i, H^{\bar j}\bigr> 
-{i} \bigl<H^i ,\rd_i \CW\bigr>
- {i} \bigl<H^{\bar i}, \rd_{\bar i} \bar\CW\bigr>
+\ldots,
}
where $\m$ is the $\CG$-momentum map on the target
space\foot{A better terminology is to regard
$X$ as the space of all $X^i$'s.} 
$X$ as defined earlier in \momentmap,
$g_{i\bar j} :=\rd_i\bar\rd_j \CK$ and $\rd_i \CW=\rd \CW/\rd X^i$.
We integrate out the auxiliary fields $D$, $H^i$ and $H^{\bar i}$ 
by imposing the following algebraic equations of motion
\eqn\aoj{
\eqalign{
D &= \Fr{i}{2} (\m-\zeta),\cr
H^i&= ig^{i\bar j}\Fr{\rd \bar{\CW}}{\rd X^{\bar j}}.
}
}
{}From our general discussion earlier, we see that
the path integral is localized to the space of solutions
of the following equations
\eqn\aok{
\eqalign{
\m - \zeta=0,\cr
\Fr{\rd \CW}{\rd X^i}=0,\cr
}
}
modulo the $\CG$-symmetry. In other words
the effective target space \ttlocus\ is the symplectic
quotient at  level $\zeta$ of the critical set 
$H_i^{-1}(0)\subset X$ of the holomorphic potential
\eqn\aol{
\CM_\zeta := \left(H_i^{-1}(0)\cap \m^{-1}(\zeta)\right)/\CG.
}
Equivalently $\CM_\zeta$ is the restriction of $\CN_\zeta$,
the symplectic quotient of $X$ by $\CG$, to the critical subset.
Those are compatible since $H^i$ is $\CG$-equivariant as $\CW$ and
$S$  are $\CG$-invariant. Thus $\CM_\zeta$ is a K\"{a}hler
manifold, provided that $\zeta$ is generic.
Note that the space of all bosonic fields is much bigger
than $X$ due to the additional  affine space of four real scalars
$\w^{m}$, $m=1,\ldots, 4$. The path integral is localized, 
in addition to \aok, to the space of solutions of
\eqn\aokk{
[\w^m,\w^n]=0,\qquad \w^a_m\CL_a(X^i)=0,
}
modulo the gauge symmetry. As the basic principle of
the equivariant cohomology $\w_m=0$ if $\CG$ acts freely
while, otherwise, there is something else.

Now we assume that $\CM_\zeta$ is smooth. Then our model
is equivalent to the non-equivariant $N_c=(2,2)$ model
with target space $\CM_\zeta$. Thus the partition
function is the Euler characteristic of the effective target space;
\eqn\tttheorem{
Z=\chi(T\CM_\zeta)=\chi(\CM_\zeta).
}
A beautiful fact is that our initial target space 
$X$ may be infinite dimensional
with an infinite dimensional group $\CG$ acting on it, while the
final target space $\CM_\zeta$ can be finite dimensional.

\subsubsection{The Geometry of Effective Target Space}

It is obvious that
the group action preserves the condition $H_i=0$ and the subvariety 
$H_i^{-1}(0)\subset X$ inherits the complex and K\"{a}hler structures 
by restriction. The quotient space $\CM_\zeta$ 
inherits the K\"{a}hler structure from $H_i^{-1}(0)$
by the restrictions and the reduction. 

If $\zeta$ takes on a generic value, the group $\CG$ acts freely and
$\CM_\zeta$ is a smooth K\"{a}hler manifold. For such a case
the model can be identified with the 
non-linear non-equivariant $N_c=(2,2)$ model  in
Sect. $2.2.2$ with target space $\CM_{\zeta}$. 
This property is equivalent to the property of equivariant
cohomology that the equivariant cohomology is
the ordinary cohomology of the quotient space if
it is smooth.

For non-generic $\zeta$ the quotient space develops singularities
or even may not exist at all.  For such cases however one always has 
some extra degrees of freedom not described by the moduli space, 
due to the extension of $X/\CG$ to $X_\CG$. Those extra degrees
of freedom are represented by the solutions of \aokk\
modulo gauge symmetry. The first equation in \aokk\
show that no such a solution exists if the $\CG$ action
act freely, without fixed points, on $X$.
If there are solutions they span an affine space,\foot{We will
relate those degrees of freedom, in certain cases, with the degrees
transverse to the D-brane world volume in the bulk.}
which looks
like a symmetric products of $\msbm{R}^4$.

The beautiful relation between the symplectic
and geometrical invariant theory (GIT) quotients also is an important
part of the story \Mumford\Newstead\Kirwan. 
The essential point is that the condition $H_i=0$
is preserved by the complexified group action $\CG^\msbm{C}$, while
the condition $D=0$ is only preserved by the real group action. 
Thus we may consider a complex quotient $H_i^{-1}(0)/\CG^\msbm{C}$
and try to compare with the real quotient $(H_i^{-1}(0)\cap D^{-1}(0))/\CG$.
In general there can be $\CG^\msbm{C}$-orbits in $H_i^{-1}(0)$
which contain several $\CG$ orbits in $H_i^{-1}(0)\cap D^{-1}(0)$.
Thus we need to consider a suitable subset in $H_i^{-1}(0)$
for which a $\CG^\msbm{C}$-orbit contains exactly one solution
of the equation $D=0$. Then the real equation $D=0$
can be identified with the gauge fixing condition of the complex
gauge symmetry of the complex equations $H_i =0$.

The complex gauge group in general does not act freely on 
the submanifold $H_i\inv(0)$, so that taking the quotient directly 
would lead to unwanted singularities. One first removes such
obvious bad points $B$. However there are subsets in $(H_i^{-1}(0)-B)$
which can be arbitrarly close to $B$ by $\CG^\msbm{C}$ action.
One call a point in $H^{-1}_i(0)$ semi-stable if the closure of
its $\CG^\msbm{C}$ orbit does not contain $B$.
Let $H^{-1}_i(0)_{ss}$ be the semi-stable subset of $H^{-1}_i(0)$.
Now the beautiful fact is that the complex quotient
$H^{-1}_i(0)_{ss}/\CG^\msbm{C}$ contains the
symplectic quotient $\CM_0$ as open subset.
A stable orbit is a semi-stable orbit if the points of the orbit
have at most finite stabilizers under the real $\CG$ action.
Then the various symplectic quotients $\CM_\zeta$ can
be identified with the quotient space $H^{-1}(0)_s/\CG^\msbm{C}$
in dense open subset.  Thus we have
\eqn\gitth{
H^{-1}_i(0)_{ss}/\CG^\msbm{C}\supset \CM_\zeta \supset
H^{-1}(0)_s/\CG^\msbm{C}.
}
The first relation implies that we have a natural
compactification of $\CM_\zeta$ by taking the closure
in  $H^{-1}_i(0)_{ss}/\CG^\msbm{C}$.
The second relation implies that the various symplectic
quotients $\CM_\zeta$ are birational with each others.

\subsec{Generalization to Equivariant  $N_c=(2,0)$ Model}

Now we consider the equivariant extension of
the $N_c=(2,0)$ model introduced in Sect.~$2.3$ or,
equivalently, the generalization of the equivariant $N_{ws}=(2,2)$
model in the previous section. We consider the same group $\CG$
acting on $X$ as before but now we allow the $\CG$ action to extend
to a Hermitian holomorphic 
vector bundle $\E \rightarrow X$ preserving the
Hermitian structure. We have two supercharges
$\bs_+$ and $\bbs_+$, isomorphic to the differentials
of $\CG$-equivariant Dolbeault cohomology 
as in the equivariant toy model in
Sect.~$3.1$;
\eqn\ara{
\bs_+^2=0,\qquad \{\bs_+,\bbs_+\} = -i\phi^a_{++} \CL_a,\qquad
\bbs_+^2=0.
}

Comparing with the non-equivariant counterpart, the equivariant
$N_c=(2,0)$ model has essentially one addition structure
that the path integral is further localized to the vanishing
locus $\m^{-1}(\zeta)$ of $\CG$-moment map.
If $\CG$ acts freely on  $\m^{-1}(\zeta)$ the model reduce
to a standard $N_c=(2,0)$ model associated with
the symplectic quotients. The observables of the model
are given by $\CG$-equivariant closed differential forms,
after the parity changes,
as our equivariant toy model. If $\CG$ acts freely on $\m^{-1}(\zeta)$ 
those observables become ordinary closed differential form
on the symplectic quotient. Comparing with our equivariant
toy model the additional structure is that the path integral
is further localized to the locus of vanishing holomorphic
sections on $\E$. We will use such property to define
a more general hybrid $N_c=(2,0)$ model.
Following the discussion in Sect.~$3.2$ the model
is related with $N_{ws}=(2,0)$ world-sheet gauged sigma-model
in $(1+1)$ dimensions by dimensional reduction \GLSM.

\subsubsection{Basic Structures}

We may follow  exactly the same route as we followed to arrive 
at the non-equivariant $N_c=(2,0)$ model from the 
non-equivariant $N_c=(2,2)$ models.

First we write the $N_c=(2,2)$ action functional $S$ \ttaction\  in a form
such that only the $\bs_+$ and $\bbs_+$ are manifest - compare with \aea -
\eqn\arb{
\eqalign{
S(\zeta)=& -\bs_+\bbs_+
\biggl(
\left<\phi_{--}, \m(X^i, X^{\bar i}) -\zeta\right>
 - \bigl<\eta_-,\bar\eta_-\bigr>
+\left<g_{i\bar j}(X^k, X^{\bar k})\p^i_-,\p^{\bar j}_- \right>
\biggr)
\cr
&
+ i\bs_+\!\left<\p^i_-, V_i(X^j)\right> 
+i\bbs_+\!\left<\p^{\bar i}_-, V_{\bar i}(X^{\bar j})\right>,
}
}
where $V_i =\rd \CW/\rd X^i$.
Similarly we disconnect the diagram \chiralm\  by removing the
link $\bs_-$,
\eqn\arbb{
\def\normalbaselines{\baselineskip20pt
\lineskip3pt \lineskiplimit3pt}
\matrix{
\p^i_- & & X^i &\mapr & \p^i_+\cr
             & 
 	 \rlap{\lower.3ex\hbox{$\scriptstyle s_{\!+}$}}\searrow
 & & 
& \cr
&&H^i&&
}.
}
Now we  regard the above as two independent sets
of multiplets. Then we rename various fields as
follows,  exactly the same as earlier \aec
\eqn\arc{
\eqalign{
\p^i_- \rightarrow \chi^\a_-,\cr
\p^{\bar i}_- \rightarrow \chi^{\bar \a}_-,\cr
}\qquad
\eqalign{
H^i \rightarrow H^\a,\cr
H^{\bar i} \rightarrow H^{\bar \a},\cr
}\qquad
\eqalign{
V_i \rightarrow \eufm{S}_\a(X^j),\cr
V_{\bar i} \rightarrow \eufm{S}_{\bar \a}(X^{\bar j}),\cr
}\qquad
 g_{i\bar j} \rightarrow h_{\a\bar\b}(X^i, X^{\bar i}),
}
where the new indices run as $\a,\b=1,\ldots,r$ and we
maintain the Hermiticity of $h_{\a\bar\b}$.
The $N_c=(2,0)$ multiplets $(X^i, \p^i_+)$ are holomorphic,
i.e., $\bbs_+ X^i=0$. We call the multiplets $(\c^\a_-, H^\a)$
Fermi  multiplets.
We also disconnect the diagram \vectorm\ for the $N_c=(2,2)$
gauge multiplet by removing the links $\bs_-$ and $\bbs_-$,
\eqn\ard{
\def\normalbaselines{\baselineskip20pt
\lineskip3pt \lineskiplimit3pt}
\matrix{
\bar\s     & \mapr & \eta_+ &  & \phi_{++}     \cr
    &       &  &       &      \cr
\bar\eta_- & \mapr & D      &  & \bar\eta_+ \cr
\mapu      &       & \mapu  &       & \mapu      \cr
\phi_{--}     & \mapr & \eta_- &  & \s         \cr
}.
}
Note that $\bar\s$ is holomorphic, i.e., $\bbs_+\bar\s =0$.
Thus the $N_c=(2,0)$ multiplet $(\bar\s, \eta_+)$ is another
holomorphic multiplet, while their Hermitian conjugates $(\s,\bar\eta_+)$
form  an anti-holomorphic multiplet. 
We may simply remove them, or keep them 
as they are still valued in $Lie(\CG)$, or 
just regard them as another holomorphic multiplet 
supplementing the multiplets $(X^i, \p^i_+)$.\foot{It is our convention that
all the holomorphic multiplets are collectively denoted as 
$(X^i,\p^i_+)$ where each multiplet may transform differently
under $\CG$ and other global symmetries. We also denote $X$
as the space of all $X^i$'s.} 
We call the multiplet $(\phi_{--},\eta_-,\bar\eta_-,D)$ $N_c=(2,0)$
gauge multiplet taking values in $Lie(\CG)$.

Now we consider the transformation laws for the
$\bs_+$ and $\bbs_+$ supersymmetry.
For the holomorphic multiplets $(X^i, \p^i_+)$, i.e., $\bbs_+ X^i=0$,
and their conjugates we have
\eqn\are{
\eqalign{
\bs_+ X^i&= i\p^i_+ ,\cr
\bbs_+ X^{i}&= 0 ,\cr
\bs_+ X^{\bar i} &=0,\cr
\bbs_+ X^{\bar i} &=i\p^{\bar i}_+,\cr
}\qquad
\eqalign{
\bs_+\p^i_+ &=0  ,\cr
\bbs_+\p^i_+ &= \phi^a_{++}\CL_a X^i  ,\cr
\bs_+ \p^{\bar i}_+ &= \phi^a_{++}\CL_a X^{\bar i},\cr
\bbs_+\p^{\bar i}_+ &=0,\cr
}\qquad
\eqalign{
\bs_+\phi_{++}=0,\cr
\bbs_+\phi_{++} =0,\cr
}
}
which are, of course,  the same  as \alh.
The transformation laws for Fermi multiplets $(\c^\a_-, H^\a)$
and their conjugates are given by
\eqn\uselater{
\eqalign{
\bs_+\c^\a_- &= - H^\a  ,\cr
\bbs_+\c^\a_- &=  \eufm{J}^\a(X^i),\cr
\bs_+\c^{\bar \a}_- &= \eufm{J}^{\bar  \a}(X^{\bar i}) ,\cr
\bbs_+\c^{\bar \a}_- &= - H^{\bar a},\cr
}\qquad
\eqalign{
\bs_+ H^\a &= 0
,\cr
\bbs_+ H^\a &= -i\phi^a_{++}\CL_a \c^\a_- 
+i\p^{i}_+ \rd_i \eufm{J}^\a(X^j)
,\cr
\bs_+ H^{\bar \a} &=
-i\phi^a_{++}\CL_a \c^{\bar \a}_- 
+i\p^{\bar i}_+ \rd_{\bar i} \eufm{J}^{\bar \a}(X^{\bar j})
,\cr
\bbs_+ H^{\bar \a} &=0
,\cr
}
}
where $\rd_{\bar i} \eufm{J}^\a(X^j)=0$. Note that $\bbs_+ \c^\a_-\neq 0$
but rather equals $\eufm{J}^\a(X^i)$, while the above transformation laws
are consistent, since $\bbs_+^2\c^\a_- =\bbs_+\eufm{J}^\a(X^i)=0$, 
with the commutation relations \ara.
Finally the transformation laws for the $N_c=(2,0)$
gauge multiplet  $(\phi_{--},\eta_-,\bar\eta_-)$ 
are given by
\eqn\arf{
\eqalign{
\bs_+ \phi_{--} = i\eta_-,\cr
\bbs_+\phi_{--}=i\bar\eta_-,\cr
}
\qquad
\eqalign{
\bs_+ \eta_-&=0,\cr
\bbs_+\eta_- &=+i D + \Fr{1}{2}[\phi_{++},\phi_{--}],\cr
\bs_+\bar\eta_-&=-iD +\Fr{1}{2}[\phi_{++},\phi_{--}],\cr
\bbs_+\bar\eta_-&=0.\cr
}
}

The general $N_c=(2,0)$ action functional, with the vanishing
ghost number, is given by the following form\foot{The repeated indices 
are summed over unless otherwise stated.}
\eqn\tzaction{
\eqalign{
S(\zeta)=& -\bs_+\bbs_+\biggl(\left<\phi_{--}, \m-\zeta\right>  
-\bigl<\eta_-,\bar\eta_-\bigr>
+\left<h_{\a\bar \b}\c^\a_-,\c^{\bar \a}_- \right>
\biggr)
\cr
&
+ i\bs_+\!\left<\c^\a_-,\eufm{S}^\a(X^i)\right> 
+i\bbs_+\!\left<\c^{\bar \a}_-, \eufm{S}^{\bar \a}(X^{\bar i})\right>.
}
}
Here $h_{\a\bar\b}(X^i, X^{\bar i})$ is a Hermitian structure on 
a Hermitian vector
bundle $\E$ over $X$, $\eufm{S}^\a(X^i)$ a holomorphic section and
$\m(X^i, X^{\bar i})$ is the $\CG$-momentum map on $X$.
Note that $N_c=(2,0)$ symmetry of the above action
functional is not obvious due to the second line in \tzaction.
For example the $\bbs_+$ supersymmetry of the term 
$\bs_+\left<\chi^\a_-, \eufm{S}_\a\right>$ 
is not obvious if $\eufm{J}(X^i)\neq 0$ due to the
transformation law $\bbs_+\chi^\a_-=\eufm{J}(X^i)$.
The condition that the action functional $S(\zeta)$ has
$N_c=(2,0)$ is
\eqn\arh{
\bbs_+\left<\chi^\a_-, \eufm{S}^\a(X^i) \right> 
= \left<\eufm{J}^\a(X^i), \eufm{S}^\a(X^i) \right>=0.
}

Let us summarize the basic structure of an equivariant
$N_c=(2,0)$ model.

\begin{enumerate}

\item A complex K\"{a}hler target space $X$ with a $\CG$ symmetry
as an isometry. These data determine holomorphic multiplets
and gauge multiplets as well as their transformation laws
and $\CG$-equivariant momentum map $\m:X\rightarrow Lie(\CG)^*$.

\item A Hermitian holomorphic vector bundle $\E\rightarrow X$
over the target space $X$ with the $\CG$ action preserving
the Hermitian structure.
We may have up to two $\CG$-equivariant holomorphic sections 
$\eufm{S}$ and $\eufm{\eufm{J}}$
orthogonal with each others by a natural non-degenerated $\CG$
invariant parings.
Those sections determine Fermi multiplets and their
transformation laws.

\end{enumerate}

Given the data above, we have an unique family of   
equivariant $N_c=(2,0)$ 
models parameterized by the FI term $\zeta$.

\subsubsection{The Path Integrals}

Expanding the action functional $S$ \tzaction\ we have
the following terms depending on the auxiliary fields
$D$, $H^\a$ and $H^{\bar \a}$,
\eqn\arj{
S = \bigl<D,D\bigr> - \bigl<D,  \m - \zeta\bigr> 
+\bigl< h_{\a\bar \b}H^\a, H^{\bar \b}\bigr> 
-{i} \bigl<H^\a ,\eufm{S}^\a\bigr>
- {i} \bigl<H^{\bar \a}, \eufm{S}^{\bar\a}\bigr>
+\ldots.
}
We integrate the auxiliary fields out
by imposing the following algebraic equations of motion,
\eqn\ari{
\eqalign{
D &= \Fr{1}{2} (\m-\zeta),\cr
H_\a&= ih_{\a\bar \b}\eufm{S}^{\bar\b}.
}
}
{}From our general discussion earlier, we see that
the bosonic part of the path integral reduces to an integral
over the space of solutions
of the following equations,
\eqn\arj{
\eqalign{
\eufm{J}^\a(X^i) =0,\cr
\eufm{S}_\a(X^i)=0,\cr
\m - \zeta=0,\cr
}
}
and
\eqn\ark{
\eqalign{
\phi^a_{++}\CL_a X^i=0,\cr
[\phi_{++},\phi_{--}]=0,\cr
}
}
modulo $\CG$-symmetry.

Now we examine the properties of the path integral
in some detail by applying the fixed
point theorem of Witten. For simplicity assume that the space $X$
and the Hermitian holomorphic bundle $E$ are flat.
We also turn off the section $\eufm{J}^\a$, keeping $\eufm{S}$
only. Then the fixed point locus of the $\bs_+$ and $\bbs_+$
supersymmetry is  the symplectic
quotient $\CM_\zeta$ of $\eufm{S}_{\a}^{-1}(0)\subset X$
by $\CG$;
\eqn\arka{
\CM_{\zeta} = \left(\m^{-1}(\zeta) \cap \eufm{S}_{\a}^{-1}(0)\right)/\CG.
}
We have the same set of observables as in the equivariant
toy model, given by $\bs_+$ and $\bbs_+$ closed
$\CG$-equivariant differential forms $\widehat \CO^{r,s}$
wite ghost numbers $(r,s)$.

The explicit expression of the action functional is
\eqn\arac{
\eqalign{
S^\pr =&
D^2 + \sum |H_\a|^2
-\Fr{1}{4}[\phi_{++},\phi_{--}]^2
-i [\phi_{++},\eta_{-}]_a \bar\eta^a_-
\cr
&
-\bar\eta^a_- \rd_i\m_a \p^i_+
-\eta^a_-\rd_{\bar i}\m_a \p^{\bar i}_+
+\c^\a_- \rd_i \eufm{S}_\a \p^i_+
+\c^{\bar\a}_- \rd_i \eufm{S}_{\bar\a} \p^{\bar i}_+
\cr
&
-i h_{\a\bar\b}\phi^a_{++}\CL_a \c^\a_- \c^{\bar\b}_-
+i\phi^a_{--}\left(\phi^b_{++}\rd_{\bar i}\m_a  V^{\bar i}_b
+ \rd_{i}\rd_{\bar j}\m_a  \p^i_{+}\p^{\bar j}_+\right),
}
}
where $V^{\bar i}_b = \CL_a X^{\bar i}$.
In doing the path integral one replaces all fields yb their 
zero-modes. The zero-modes of the fermions are solutions of the
following equations 
\eqn\urkj{
\eqalign{
\rd_{\bar i} \m_a \p^{\bar i}_+=0,\cr
\rd_{\bar i}\eufm{S}_{\bar\a}\p^{\bar i}_+=0,\cr
}\qquad
\eqalign{
\eta_-^a\rd_{\bar i}\m_a=0,\cr
\c^{\bar\a}\rd_{\bar i}\eufm{S}_{\bar\a}=0.\cr
}
}
The above equations implies that 
the net ghost number violation $\triangle$ in the path integral
measure due to fermionic zero-modes of $(\p^{\bar i}_+,
\c^{\bar\a}_-,\eta_-)$  always equals 
\eqn\urkjs{
\triangle =n-r - dim\,\CG.
}
We call $\triangle$
the virtual complex dimension of $\CM_{\zeta}$.
{}From the equations $\rd_{\bar i} \m_a \p^{\bar i}_+=0$
we have the following integrability condition
\eqn\urkl{
\phi^b_{++}\rd_{\bar i}\m_a  V^{\bar i}_b
+ \rd_{i}\rd_{\bar j}\m_a  \p^i_{+}\p^{\bar j}_+=0,
}
which is also the $\phi^a_{--}$ equation of motion.
This implies that one can simply replace $\phi^a_{++}$
with the solutions of the above. Such an argument
can not be justified if there are zero-modes of $\phi^a_{++}$,
which are given by the non-trivial solutions of \ark, for instance
$\phi^b_{++}V^{\bar i}_b=0$.

Here we specialize to the case that  $\CG$ acts freely,
thus there are no zero-modes
of $\eta^a_-$ and $\phi^a_{\pm\pm}$. 
Then the only non-trivial term in teh action functional $S^\pr$ 
in the $\bs_+$ and $\bbs_+$ invariant  neighborhood $\CC$  of 
the fixed point locus is
\eqn\urkm{
S^\pr|_{\CC} = -i h_{\a\bar\b}\phi^a_{++}\CL_a \c^\a_- \c^{\bar\b}_-|_\CC.
}
Using \urkl\ we can solve
$\phi^a_{++}$ in terms of the zero-modes $(u^{\bar i^\pr}, 
\tilde \p^{\bar i^\pr}_+,\tilde \chi_-^{\bar\a^\pr})$
of $(X^{\bar i},\p^{\bar i}_+,\c^{\bar\a}_-)$
\eqn\urkn{
<\phi_{++}^a(u^{i^\pr},u^{\bar i^\pr})> = -\left(\rd_{\bar \ell^\pr}\m_b 
V^{\bar \ell^\pr}_a\right)^{-1}  \rd_{i^\pr}\rd_{\bar j^\pr}\m_b  
\tilde\p^{i^\pr}_{+}\tilde\p^{\bar j^\pr}_+
}
where the primed indices above are understood to label independent
zero-modes - $\bar i^\pr = 1,\ldots,n^\pr$, $\bar\a^\pr =1,\ldots,r^\pr$,
with the condition 
\eqn\urko{
\triangle=n^\pr - r^\pr = n-r - dim\,\CG.
}
Then we may write
\eqn\urkp{
S^\pr|_{\CC} = -\CF(u^{\ell^\pr}, u^{\bar \ell^\pr})_{i^\pr \bar j^\pr \a^\pr
\bar\b^\pr}  
\tilde\p^{i^\pr}_{+}\tilde\p^{\bar j^\pr}_+
\tilde \chi_-^{\a^\pr}\tilde \chi_-^{\bar\b^\pr},
}
where $\CF_{i^\pr \bar j^\pr \a^\pr\bar\b^\pr}  
\tilde\p^{i^\pr}_{+}\tilde\p^{\bar j^\pr}_+$ can be interpreted
as the curvature two-form of the anti-ghost bundle $\V$ over
$\CM_{\zeta}$.
Consequently the path integral reduces to
\eqn\urkq{
\eqalign{
\left<\prod _{m=1}^k\widehat \CO^{r_m, s_m}\right>\!
= \!\int_{\CM_\zeta} 
&\prod_{\g^\pr = 1}^{r^\pr}
d\tilde \chi_-^{\g^\pr}d\tilde \chi_-^{\bar\g^\pr}
\prod_{\ell^\pr=1}^{n^\pr}
du^{\ell^\pr} du^{\bar \ell^\pr}
d\tilde\p^{\ell^\pr}_{+}\tilde\p^{\bar \ell^\pr}_+
\cr
&\times
\exp\left(\CF_{i^\pr \bar j^\pr \a^\pr
\bar\b^\pr}  
\tilde\p^{i^\pr}_{+}\tilde\p^{\bar j^\pr}_+
\tilde \chi_-^{\a^\pr}\tilde \chi_-^{\bar\b^\pr}\right)
\prod \widetilde \CO^{r_m, s_m},
}
}
where $\widehat{\widetilde \CO}$ denote the expression of an
observable $\widehat \CO$ in terms of zero-modes
and $<\phi_{++}>$. The necessary condition for
a non-vanishing correlation function is
\eqn\urkr{
\sum_{m=1}^k (r_m,s_m) = (\triangle,\triangle).
}

Let us first assume that the section is generic
and $\CG$ acts freely on $\eufm{S}_{\a}^{-1}(0)\subset X$.
Then $\CM_{\zeta}$ is a smooth non-linear K\"{a}hler manifold
with complex dimensions
\eqn\arkb{
dim_\msbm{C} \CM_{\zeta} =\triangle= n-r - dim\,\CG.
}
The above counting goes as follows.
Since $\eufm{S}_\a$, $\a=1,\ldots,r$, are generic
they are all independent and transverse. Thus the condition
$\eufm{S}_\a =0$ cuts out a complex $(n-r)$ smooth submanifold
inside the complex $n$-dimensional ambient space $X$.
On the subspace we further impose $dim \CG$ real equations
$\m_a -\zeta =0$ and take the quotient by the free $\CG$ action.
Now we do not have zero-modes of $\chi_-$ and
the path integral becomes
\eqn\urks{
\eqalign{
\left<\prod_{m=1}^k \widehat \CO^{r_m, s_m}\right>\!
&= \!\int_{\CM_\zeta} 
\prod_{\ell^\pr=1}^{\triangle}
du^{\ell^\pr} du^{\bar \ell^\pr}
d\tilde\p^{\ell^\pr}_{+}\tilde\p^{\bar \ell^\pr}_+
\prod_m \widehat{\widetilde\CO}^{r_m, s_m}
\cr
&= \int_{\CM_\zeta} 
\widetilde \CO^{r_1, s_1}\wedge\ldots\wedge 
\widetilde \CO^{r_k, s_k}.
}
}
A non-generic situation arises when 
$\eufm{S}_{\a^\pr}$, $\a^\pr=1,\ldots,r^\pr$, are linearly
dependent to the remaining sections. Then the complex
dimension of $\CM_{\zeta}$  is given by $n^\pr =\triangle + r^\pr$.
The resulting space is smooth if the linearly independent components
of the section are transverse.
We have $r^\pr$ $\chi_-$ zero-modes
which span the anti-ghost bundle $\V$ over $\CM_{\zeta}$.
The path integral becomes
\eqn\urkt{
\eqalign{
\left<\prod _{m=1}^k\widehat \CO^{r_m, s_m}\right>\!
= \int_{\CM_\zeta} e(\V)\wedge
\widetilde \CO^{r_1, s_1}\wedge\ldots\wedge 
\widetilde \CO^{r_k, s_k}.
}
}

A beautiful fact about this is that
$(X, \E,\CG)$ can be all infinite dimensional while
the space $\CM_{\zeta}$ can be a finite dimensional
space. In particular $X$ can be a certain function space
defined by the space of all fields of a certain gauge field
theory on a manifold $M$. Then the integral we are dealing
with is a genuine path integral of a non-trivial quantum
field theory on $M$, while the path integral eventually
reduces to an ordinary integral on a smooth finite
dimensional space $\CM_{\zeta}$.
The above is a key principle underlying cohomological
field theory \TFT\Wittengwzw. In principle the
above path integral formalism is well-defined
regardless of the properties the moduli
space $\CM_{\zeta}$.

Finally we remark that a proper mathematical interpretation
of our formalism may be a certain equivariant version
of Fulton and MacPherson's intersection theory \Fulton.

\newsec{Generalizations}

In this section we consider three geneneralizations of
an equivariant $N_c=(2,0)$ model.

\subsection{Deformation to Holomorphic $N_c=(2,0)$ Model}

In this subsection we introduce  hybrid $\CN_c=(2,0)$ model 
of the equivariant $N_c=(2,0)$ mode and the equivariant toy model
in Sect.~$2.2$.  The resulting hybrid model will have
much better behavior than the original 
model when the effective target space $\CM_\zeta$
has singularities.
To motivate such a model we first compare
the two models.

First of all both the models have the same supersymmetry
generated by $\bs_+$ and $\bbs_+$, which are the differentials
of equivariant Dolbeault cohomology after the parity change.
Secondly both the models share the same holomorphic multiplets
$(X^i, \p^i_+)$ and their Hermitian conjugates, which are anti-holomorphic
multiplets $(X^{\bar i}, \p^{\bar i}_+)$. Thus they share 
the same observables, given by equivariantly closed differential
forms on $X$, the space of all $X^i$, after the parity change.

A difference is that the equivariant $N_c=(2,0)$ model
has the additional Fermi multiplets $(\chi^\a_-, H^\a)$
and their Hermitian conjugates $(\chi^{\bar\a}_-, H^{\bar\a})$.
The roles of the Fermi multiplets are to restrict  the (path) integral
over $X$ to the subspace defined by  
$\eufm{J}_\a^{-1}(0)\cap \eufm{S}_\a^{-1}(0)\subset X$.
For convenience we denote this subspace by $X^{1,1}\subset X$.
We saw that the path integral of the $N_c=(2,0)$
model is localized to the symplectic quotients 
$\CM_\zeta= (X^{1,1}\cap \m^{-1}(\zeta)$
of $X^{1,1}$ by $\CG$. 
Now we consider an equivariant
toy model whose initial target space is $X^{1,1}$. 
Then, its path integral is also localized
to the same space $\CM_\zeta$, provided that we set $\e=0$ in the
action functional $S(\zeta,\e)$ defined by \anc.
We also note that the partition function of the above equivariant
toy model is the expectation value of 
$\exp\left(\widehat\varpi_\CG\right)$ evaluated by the $N_c=(2,0)$
model, where
\eqn\asa{
\widehat\varpi_\CG :=S(\zeta,0)= i\left<\phi_{++}, \m -\zeta\right> 
+ i g_{i\bar j}\p^i_+\p^{\bar i}_+.
}
The first term above is irrelevant
as the path integral of  the $N_c=(2,0)$ is localized the to
the locus $\m-\zeta =0$, while the second term above becomes
the K\"{a}hler from $\tilde\varpi$ on $\CM_\zeta$.
We note that it is the $N_c=(2,0)$ gauge multiplet
$(\phi_{--},\eta_-,\bar\eta_-,D)$, which is  responsible 
for such a localization.
One the other hand, the above is the action functional 
of the equivariant toy model on $X^{1,1}$ and the integration
over $\phi$ localizes the path integral by a delta function
supported on $\CM_\zeta$ in $X^{1,1}$. Note
that $\CM_\zeta = \CN_{\zeta}|_{X^{1,1}}$ is the restriction
of $\CN_\zeta$ - the symplectic quotient of $X$ by $\CG$ -
to $X^{1,1}$.

The above discussion 
motivates us to define a new $N_c=(2,0)$
model with
the following action functional $S_h(\zeta,0)$, modifying the
original $N_c=(2,0)$ action functional $S$ in \tzaction
\eqn\asb{
\eqalign{
S_h(\zeta, 0)=& -i\bs_+\bbs_+
\left<h_{\a\bar \b}\c^\a_-,\c^{\bar \a}_- \right>
+ i\bs_+\!\left<\c^\a_-,\eufm{S}_\a\right> 
+i\bbs_+\!\left<\c^{\bar \a}_-, \eufm{S}_{\bar \a}\right>
\cr
&
-i\left<\phi_{++}, \m -\zeta\right> 
- i g_{i\bar j}\p^i_+\p^{\bar i}_+,
}
}
where we removed the $N_c=(2,0)$ gauge multiplet
$(\phi_{--},\eta_-,\bar\eta_-,D)$ 
and added the action functional $S(\zeta,0)$ of the
equivariant toy model. According to the previous discussion
we see that the partition function defined by the new action
$S_h(\zeta,0)$ is equivalent to the expectation value
of $\exp(\widehat \varpi_\CG)$, evaluated by the
original $N_c=(2,0)$ action functional $S$ \tzaction.

Now we define more general action functional $S_h(\zeta,\e)$
by
\eqn\htzaction{
\eqalign{
S_h(\zeta,\e) :=
&-\bs_+\bbs_+
\left<h_{\a\bar \b}\c^\a_-,\c^{\bar \a}_- \right>
+ i\bs_+\!\left<\c^\a_-,\eufm{S}_\a\right> 
+i\bbs_+\!\left<\c^{\bar \a}_-, \eufm{S}_{\bar \a}\right>
\cr
&
-i\left<\phi_{++}, \m -\zeta\right> 
- i g_{i\bar j}\p^i_+\p^{\bar i}_+
+\Fr{\e}{2}\left<\phi_{++},\phi_{++}\right>.
}
}
We call the $N_c=(2,0)$ model with the above action
functional $S_h(\zeta,\e)$ a holomorphic $N_c=(2,0)$
model, see \HYM\ for the first example. 
Now we immediately see that the path integral
of the holomorphic $N_c=(2,0)$ model is governed
by Witten's non-Abelian localization principle \tdYM. 
The first line of the above action functional localizes
the path integral to $X^{1,1}$. Then, following the
discussions in Sect.~$2.2.3$, the path integral
can be written as the sum of contributions of the
critical points $I=\left<\m-\zeta,\m-\zeta\right>$ in $X^{1,1}$.
Also from the discussions in Sect.~$2.2.3$ the
$\e$-dependent term regularizes the path integral
when $\CM_\zeta$ develops singularities.

\subsubsection{The Mapping Between the Two Models}

Now we will give more wider viewpoints which contain
the original and holomorphic $N_c=(2,0)$ models
as two special limits, following the original method of Witten
\tdYM. Witten considered the case without the Fermi multiplets
but for  general manifolds.  The Fermi multiplets will be purely 
spectators, and the specialization to a K\"ahler case will simplify
the procedure.

Consider the following one-parameter family of 
$N_c=(2,0)$ supersymmetric action functional $S(\zeta)_\l$,
\eqn\tzgaction{
\eqalign{
S(\zeta)_\l 
 :=& S(\zeta) +\Fr{\l}{2}\bs_+\bbs_+\bigl<\phi_{--},\phi_{--}\bigr>
\cr
=&
-\bs_+\bbs_+\left<h_{\a\bar \b}\c^\a_-,\c^{\bar \a}_- \right>
+ i\bs_+\!\left<\c^\a_-,\eufm{S}_\a\right> 
+i\bbs_+\!\left<\c^{\bar \a}_-, \eufm{S}_{\bar \a}\right>
\cr
&-\bs_+\bbs_+\biggl(\bigl<\phi_{--}, \m -\zeta -\Fr{ \l}{2}\phi_{--}\bigr>
-\left<\eta_-,\bar\eta_-\right>\biggr).
}
}
If we set $\l=0$ 
we have the original $N_c=(2,0)$ model.
For $\l\neq 0$ we can integrate out the $N_c=(2,0)$
gauge multiplet, and  we are left with
\eqn\tzgac{
\eqalign{
S^\pr(\zeta)_\l 
 =&
-\bs_+\bbs_+\left<h_{\a\bar \b}\c^\a_-,\c^{\bar \a}_- \right>
+ i\bs_+\!\left<\c^\a_-,\eufm{S}_\a\right> 
+i\bbs_+\!\left<\c^{\bar \a}_-, \eufm{S}_{\bar \a}\right>
\cr
&+\Fr{1}{2\l}\bs_+\bbs_+\left<\m -\zeta ,\m-\zeta\right>
+\CO(1/\l^2).
}
}
Since the additional $\l$-dependent term is closed by
$\bs_+$ and $\bbs_+$, 
the path integral does not depend on $\l$ as
long as $\l\neq 0$. The models with $\l=0$ and $\l\neq 0$
can be different since new fixed points can flow from the infinity
$\l\rightarrow \infty$ in the field space \tdYM.

If we take the limit $\l\rightarrow 0$, while $\l\neq 0$,
we see that the dominant contributions to the
path integral come from the critical points of
$I=\bigl<\m-\zeta,\m -\zeta\bigr>$.
Now we add $\bs_+$ and $\bbs_+$-closed observables, 
$-\widehat\varpi +\Fr{\e}{2}\bigl <\phi_{++},\phi_{++}\bigr>$,
to the above action functional,
\eqn\tzgac{
\eqalign{
S^\pr(\zeta,\e)_\l 
 =&
-\bs_+\bbs_+\left<h_{\a\bar \b}\c^\a_-,\c^{\bar \a}_- \right>
+ i\bs_+\!\left<\c^\a_-,\eufm{S}_\a\right> 
+i\bbs_+\!\left<\c^{\bar \a}_-, \eufm{S}_{\bar \a}\right>
\cr
&
-i\left<\phi_{++}, \m -\zeta\right> 
- i \left<g_{i\bar j}\p^i_+,\p^{\bar i}_+\right>
+\Fr{\e}{2}\left<\phi_{++},\phi_{++}\right>
\cr
&+\Fr{1}{2\l}\bs_+\bbs_+\left<\m -\zeta ,\m-\zeta\right>
+\CO(1/\l^2).
}
}
In the above the path integral should be independent of $\l\neq 0$.
Consequently we see that the partition function of the above action
functional can still be written as a sum of contributions from
the critical points of $I$. Finally we may take the limit 
$\l\rightarrow \infty$
to remove all the $\l$-dependent terms and obtain the action functional
$S_h(\zeta,\e)$ \htzaction\ of the holomorphic $N_c=(2,0)$ model.
Thus we showed that the partition function of the holomorphic
$N_c=(2,0)$ model can be written as a sum of contributions
from the critical points of $I=\bigl<\m-\zeta,\m-\zeta\bigr>$.

\subsec{Flows from $N_c=(2,2)$ to
$N_c=(2,0)$ Models}

Consider an equivariant $N_c=(2,0)$ model as described in
Sect. $2.4$. with $\eufm{J}^\a=0$.  
Such a model was classified by
a $\CG$-equivariant Hermitian holomorphic bundle $E\rightarrow X$
with holomorphic section $\eufm{S}_A$. In a generic situation
the model is equivalent to a non-linear $N_c=(2,0)$
model which target space $\CM_\zeta$ is 
$\CM_\zeta =(X\cap \eufm{S}^{-1}(0)\cap \m^{-1}(\zeta))/\CG$.
In this section we define a canonical
embedding of such a model to $N_c=(2,2)$ model
based on the tangent space $TE$ of the total space
of bundle $E\rightarrow X$. Then we study the mapping
from the $N_c=(2,2)$ to the $N_c=(2,0)$ model.
We will see that the above circle of ideas leads
us to find a $N_c=(2,0)$ model which is
"equivalent" to the original $N_c=(2,0)$ model.
{}From the viewpoint of the original $N_c=(2,0)$ model
there is no a priori reason of such a "equivalence"
to a completely different model.
For simplicity we restricted to the linear model.

\def\mapr{\!\smash{
	    \mathop{\longrightarrow}\limits^{\bs_+}}\!}
\def\mapl{\!\smash{
	    \mathop{\longleftarrow}\limits^{\bs_-}}\!}
\def\mapbr{\!\smash{
	    \mathop{\longrightarrow}\limits^{\bbs_+}}\!}
\def\mapbl{\!\smash{
	    \mathop{\longleftarrow}\limits^{\bbs_-}}\!}
\def\mapd{\Big\downarrow
 	 \rlap{$\vcenter{\hbox{$\scriptstyle \bbs_-$}}$}}
\def\mapu{\Big\uparrow
	  \rlap{$\vcenter{\hbox{$\scriptstyle\bbs_+$}}$}}
\def\ne{\nearrow}
\def\se{\searrow}
\def\nw{\nwarrow}
\def\sw{\swarrow}

\subsubsection{Embedding of a $N_c=(2,0)$
Model to $N_c=(2,2)$ Model.}

The basic idea behind extension to $\CN_c=(2,2)$
model is that one can regard the total space
of holomorphic bundle $E\rightarrow X$
as the target space of a $\CN_c=(2,2)$
model. Then we have to supply local holomorphic
coordinates fields for fiber space of the bundle  $E \rightarrow X$.
Thus we introduce adjoint-valued bosonic  spectral 
fields $B^\a$ and its superpartner $\c^\a_+$.
Now the former equivariant holomorphic section $\eufm{S}^\a(X^i)$
of bundle $E\rightarrow X$ corresponds
to holomorphic vector on the target space $E$
but being supported only on $X$. Thus the
$\CG$-equivariant holomorphic vector  $\eufm{S}^\a(X^i)$ should be 
extended over the whole space $E$.
Furthermore  $\CN_c=(2,2)$ supersymmetry
requires that such holomorphic vector should be gradient  vector 
of a non-degenerated $\CG$-invariant holomorphic function 
$\CW$, i.e, $\bbs_+\CW=0$,  
of  the target space $\msbm{E}$.

Now demanding $N_c=(2,2)$ supersymmetry will take
care of everything.
Recall that the  $N_c=(2,0)$ model has a $Lie(\CG)$-valued 
gauge multiplet associated with the group action of $\CG$.
We add a $Lie(\CG)$-valued holomorphic multiplet
$
\bar\s\; \mapr\; \eta_+,
$
to form a  $N_c=(2,2)$ gauge multiplet
\eqn\aha{
\def\normalbaselines{\baselineskip20pt
\lineskip3pt \lineskiplimit3pt}
\matrix{
\bar\s     & \mapr & \eta_+ & \mapl & v_{++}     \cr
\mapd      &       & \mapd  &       & \mapd      \cr
\bar\eta_- & \mapr & D      & \mapl & \bar\eta_+ \cr
\mapu      &       & \mapu  &       & \mapu      \cr
v_{--}     & \mapr & \eta_- & \mapl & \s         \cr
}.
}
We had holomorphic multiplets 
$
(X^i\;\mapr\; \p^i_+)
$, $i=1,\ldots,n$,
associated with the base space $X$ of $\E\rightarrow X$.
By adding new Fermi multiplets $(\p^i_-\;\mapr H^i)$,
we extend them to $N_c=(2,2)$ chiral multiplets;
\eqn\ahc{
\def\normalbaselines{\baselineskip20pt
\lineskip3pt \lineskiplimit3pt}
\matrix{
\p^i_- &\mapl & X^i &\mapr & \p^i_+\cr
             & 
 	 \rlap{\lower.3ex\hbox{$\scriptstyle s_{\!+}$}}\searrow
 & &\swarrow\!\!\!\rlap{\lower.3ex\hbox{$\scriptstyle s_{\!-}$}} 
& \cr
&&H^i&&
}.
}
We also had Fermi multiplets $(\c^\a_-\;\mapr\; H^\a)$, 
$\a=1,\ldots,r$, associated with
the fibre of $\E\rightarrow X$. 
By adding new holomorphic multiplets $(B^\a\;\mapr \chi^\a_+)$,
we extend them to  $N_c=(2,2)$ chiral multiplets;
\eqn\ahd{
\def\normalbaselines{\baselineskip20pt
\lineskip3pt \lineskiplimit3pt}
\matrix{
\c^\a_- &\mapl & B^\a &\mapr & \c^\a_+\cr
             & 
 	 \rlap{\lower.3ex\hbox{$\scriptstyle s_{\!+}$}}\searrow
 & &\swarrow\!\!\!\rlap{\lower.3ex\hbox{$\scriptstyle s_{\!-}$}} 
& \cr
&&H^\a&&
}.
}

Now we consider the following $N_c=(2,2)$ supersymmetric action
functional
\eqn\ahe{
\eqalign{
S=& \bs_+\bbs_+\bs_-\bbs_-\biggl(\sum_{i=1}^n\bigr<X^i, X^{\bar i}\bigr>
+ \sum_{\a=1}^r\bigr<B^\a, B^{\bar \a}\bigr> - <\s,\bar\s>
\biggr)
\cr
&
+ \bs_+\bs_-\CW\left(X^{i}, B^{\a}\right)
+ \bbs_+\bbs_-\bar\CW\left(X^{\bar i}, B^{\bar \a}\right)
\cr
&
+  \bs_+ \bbs_-<t,\bar\s>
+  \bbs_+\bs_- <\s,\bar t>.
}
}
To relate the above model with the initial $N_c=(2,0)$ model
we assume the following conditions
\eqn\ahf{
\eqalign{
\Fr{\rd \CW}{\rd B^\a}&= \eufm{S}_\a(X^i),\cr
}
}
where $\eufm{S}_\a$ is the
holomorphic section of $\E$. This condition implies
that $\CW(X^i, B^\a)$ is linear in $B^\a$. We will utilize
this property later.
It is useful to rewrite the action functional $S$ \ahe\ such that
only the $N_c=(2,0)$ symmetry is manifest
\eqn\ahg{
\eqalign{
S=& -i\bs_+\bbs_+\biggl(\bigr<\phi_{--}, \m_X + \m_F -\zeta\bigr>
+\sum\bigr<\p_-^i,\p_-^{\bar i}\bigr>
+ \sum\bigl< \c^\a_-,\c^{\bar \a}_-\bigr>
- \bigr<\eta_-,\bar\eta_-\bigr>
\biggr)
\cr
&
+ i\bs_+\biggl(\bigl< \p^i_-,G_i\bigr> 
+ \bigl<\c^\a_-,\eufm{S}_\a\bigr>\biggr)
+ i\bbs_+\biggr(\bigl<\p^{\bar i}_-, G_{\bar i}\bigr> 
+ \bigl<\c^{\bar\a}_-,\eufm{S}_{\bar\a}\bigl>\biggr)
}
}
where $\m_X$ and $\m_F$ are the momentum maps
on $X$ and the fibre of $\E$, respectively, while
\eqn\ahh{
G_i (X^j, B^\a):= \Fr{\rd}{\rd X^i}\CW(X^j, B^\a).
}
Note that $G_i$ is linear in $B^\a$ since $\CW$ is linear
in $B^\a$.

Applying the fixed point theorem we see 
that the path integral is localized to the
solution space of the following equations, modulo the group
action of $\CG$
\eqn\ahi{
\eqalign{
\eufm{S}_\a(X^i)=0,\cr
G_i(X^j, B^\a)=0,\cr
\m_X(X^i, X^{\bar i}) + \m_F(B^\a,B^{\bar \a}) -\zeta =0,\cr
}
}
and
\eqn\ahj{
\eqalign{
\w^a_m \CL_a(X^i)=0,\cr
\w^a_m \CL_a(B^\a)=0,\cr
[\w_m, \w_n]=0,\cr
[\w_m,\bar \w_m]=0.
}
}
This model, for a generic value of $\zeta$ implying $\w_m=0$ as
usual,  reduce to the non-linear $N_c=(2,2)$ model whose
target space $\eufm{M}_\zeta$ is the space of all solutions of 
the equations \ahi\
modulo $\CG$-symmetry.

\subsubsection{Perturbation to a $N_c=(2,0)$ Model}

Now we want to perturb the $N_c=(2,2)$ model above
to a $N_c=(2,0)$ model by breaking the $N_c=(0,2)$
supersymmetry generated by $\bs_-$ and $\bbs_-$.
This can be done by giving bare "mass" to 
all the newly introduced multiplets given by
\eqn\ahk{
(\bar\s, \eta_+),\qquad (\p^i_-, H^i),\qquad (B^\a, \chi^\a_+)
}
and their conjugates.
Then the model flows to the original $N_c=(2,0)$
model if we take the bare "mass" to infinity.
Such bare mass terms will have special geometrical
meaning.

Note that there is a natural $U(1)=S^1$ group acting 
on $B^\a$, while leaving fixed the $X^i$,
such that the momentum map $\m_F$ remains invariant.
This  $S^1$-action is given by
\eqn\ahl{
 S^1:(X^i, B^\a)\rightarrow (X^i,\xi B^\a),
}
where $\xi\bar\xi=1$. Note that the above $S^1$-action
does not change the first and the last equations of \ahi.
The LHS of the second equation of \ahi\ will be multiplied by
$\zeta$, which does not alter the solution space of the
equation. Thus the $S^1$-action is a symmetry of
the effective target space $\eufm{M}_\zeta$.

It is important to note that the above $U(1)$ needs {\it not}
be a symmetry of our $N_{ws}=(2,2)$ model.
To be such a symmetry, the $S^1$ action \ahl\ should be
extended to all the superpartners. That is, $\p^i_\pm$ and $H^i$ should
be invariant under $U(1)$ while $\chi^\a_{\pm}$ and $H^\a$
should carry the $U(1)$-charge $1$. 
We, however, demand
that the above $U(1)$ is compatible with the $N_c=(2,0)$
supersymmetry generated by $\bs_+$ and $\bbs_+$ supercharges.
{}From the expression \ahg\ of $S$ with  manifest $N_c=(2,0)$
symmetry we see that the $\p^i_-$ should carry $U(1)$ charge $-1$,
since $G_i(X^j, B^\a)$ is linear in $B^\a$.
Then, by examining the supersymmetry transformation laws
for the supercharges $\bs_+$ and $\bbs_+$, we see
that the $S^1$-symmetry \ahl\ should be extended to
all the $N_c=(2,0)$ multiplets in \ahk\ as follows
\eqn\ahn{
\eqalign{
\qquad\qquad 
&S^1:(B^\a, \chi^\a_+)\rightarrow \xi(B^\a, \chi^\a_+)\cr
&S^1:(\p^i_-, H^i) \rightarrow  \bar\xi(\p^i_-, H^i),\cr
&S^1:(\bar\s, \eta_+) \rightarrow \bar\xi (\bar\s, \eta_+). 
}
}
That is,  we give $U(1)$-charges to the fields in \ahk\ while
all the other fields remain neutral. Clearly this can't be
done while maintaining  the full $N_c=(2,2)$ supersymmetry.

Recall that  the $N_c=(2,0)$ 
supercharges $\bs_+$ and $\bbs_+$ satisfy
now familiar anti-commutation relations
\eqn\aho{
\bs_+^2=0, \qquad\{\bs_+,\bbs_+\} = -i\phi^a_{++}\CL_a,
\qquad \bbs_+^2=0,
}
defining the $\CG$-equivariant Dolbeault cohomology.
Since we have an additional $S^1$ acting on our system
it is natural to extend the above to $\CG\times S^1$-equivariant
cohomology. Then the new supercharges, still to be
denoted $\bs_+$ and $\bbs_+$, satisfy the following
anti-commutation relations
\eqn\ahp{
\bs_+^2=0, \qquad\{\bs_+,\bbs_+\} = -i\phi^a_{++}\CL_a -i m \CL_{S^1},
\qquad \bbs_+^2=0,
}
where we introduced a parameter $m$ taking values in $Lie(S^1)$.
The supersymmetry transformation laws should be modified
accordingly.

Finally we define the following $N_c=(2,0)$ supersymmetric
action functional
\eqn\haq{
S(m,\bar m) = S^\pr 
+ \bar m\bs_+\bbs_+\biggl(\sum_{\a=1}^r \bigl<B^\a,B^{\bar\a}\bigr> - 
\bigl<\s,\bar\s\bigr>\biggr),
}
where $S^\pr$ is defined by the same formula as the action functional
in \ahg\  but with the modified supersymmetry.
The new action functional $S(m,\bar m)$, compared to
the $N_c=(2,2)$ symmetric action $S$, is
\eqn\ahr{
\eqalign{
S(m,\bar m) 
&= S + m\bar m \sum_\a \bigl<B^\a,B^{\bar\a}\bigr>
 -i\bar m \sum_\a\bigl<\c^\a_+,\c^{\bar\a}_+\bigr>
+m\bar m\bigl<\s,\bar\s\bigr>-i\bar m \bigl<\bar\eta_+,\eta_+\bigr>
\cr
&
-i m\bigl<\phi_{--},\m_F-[\s,\bar\s]\bigr> 
+i\bar m\bigl<\phi_{++}, \m_F-[\s,\bar\s]\bigr>
+ im \sum_i \bigl<\p^i_-,\p^{\bar i}_-\bigr>,
}
}
containing the desired mass terms.
We note that the mass terms contain
the {\it Hamiltonian} $H_{S_1}$ of the $S^1$ symmetry
on the space of all $B^\a$ and $\s$;
\eqn\ahs{
H_{S^1} =i\sum_\a \bigl<B^\a,B^{\bar\a}\bigr> + i \bigl<\s,\bar\s\bigr>.
}
This fact will play a crucial role later.

Now we examine the equation for fixed points. Since we only have
$\bs_+$ and $\bbs_+$ supersymmetry the path integral
is localized to the fixed point locus of those symmetries, modulo
the $\CG$ symmetry.
We have
\eqn\aht{
\eqalign{
\bar\s^a\CL_a(X^i)=0,\cr
\bar\s^a\CL_a(B^\a)=0,\cr
\eufm{S}_\a(X^i)=0,\cr
G_i(X^j, B^\a)=0,\cr
\m_X(X^i, X^{\bar i}) + \m_F(B^\a,B^{\bar \a}) -[\s,\bar\s]-\zeta =0,\cr
}
}
and
\eqn\ahr{
\eqalign{
[\phi,\bar\phi]=0,\cr
\phi^a \CL_a(X^i)=0,\cr
\phi^a\CL_a(B^\a) + m B^\a=0,\cr
[\phi,\bar\s] - m\s=0.\cr
}
}
The set of equations in \aht\ 
cut out a subspace of the space of all $X^i$, $B^\a$ and $\s$.
After modding out the $\CG$-symmetry we get the effective
target space $\tilde\eufm{M}_\zeta$ of our $N_c=(2,0)$ model. 
Following the previous general discussions we expect that
$\tilde\eufm{M}_{\zeta}$ is a K\"{a}hler manifold at least for the generic
case. The set of equations in \ahr\ represent
gauge degrees of freedom. In particular those equations
implies the path integral
is localized to the fixed point of $S^1$ action on 
$\tilde\eufm{M}_{\zeta}$.

We always have trivial fixed points, namely
$B^\a=\s=0$. We call such fixed points branch (i).
In branch (i) the path integral is localized to the solution
space of the following equations, modulo $\CG$-symmetry,
\eqn\ahra{
\eqalign{
\phi^a \CL_a(X^i)=0,\cr
\m_X(X^i, X^{\bar i}) -\zeta =0.\cr
}
}
which are exactly the generic fixed point equations for the
original $N_c=(2,0)$ model. 
There are other fixed points with $B^\a,\bar\s \neq 0$ when the $S^1$-action
can be undone by the $\CG$ action. The last two equations in \ahr\
exactly stand for such property. We call such fixed
points branch (ii).

The above localization principle can  also be obtained
from a different viewpoint. We consider a limit $|m|\rightarrow \infty$.
Then the dominant contributions to the path integral come
from the set of critical points of the Hamiltonian $H_{S^1}$
defined by \ahs. It is well-known that the critical points of 
the Hamiltonian 
of a $S^1$ action are exactly the same as the fixed points
of the $S^1$ action. One may evaluate the partition function
in such a limit and set $|m|=0$ afterwards, to get the partition
function of the $N_c=(2,2)$ model.

Now we assume that everything is generic, so that we do not
have any zero-modes of anti-ghosts, $\chi^\a_-,\p^i_-$, 
as well as any zero-modes of the $N_c=(2,0)$ gauge
multiplets. Then the partition function of the action functional
$S(m,\bar m)$ in \ahr\ reduces  to the following
integral
\eqn\ahu{
Z = \int_{\tilde\eufm{M}_\zeta} exp 
\left(i mH_{S^1}
 +i\sum_\a\bigl<\c^\a_+,\c^{\bar\a}_+\bigr>
+i \bigl<\bar\eta_+,\eta_+\bigr>\right),
}
where we regard $m$ and $\bar m$ as independent numbers
and scaled away the overall $\bar m$.
The above resembles the DH integration formula
on $\tilde\eufm{M}_\zeta$. We see, however, that 
there is a missing term
since the fermionic terms above correspond to
the K\"{a}hler form only on the subspace of $\tilde\eufm{M}_{\zeta}$
given by $X^i=0$. We can provide the missing term
by evaluating the correlation function of 
$\exp(i\bigl<\phi_{++}, \m_X\bigr> +i \sum \bigl<\p^i_+,\p^{\bar i}_+\bigr>
)$, where the exponent is the $\CG$-equivariant K\"{a}hler 
form $\widehat\varpi_{\CG}^X$ on $X$. Note that it is an observable
of the original $N_c=(2,0)$ model  we started from.
Assuming the same generic situation as above, 
the correlation function reduces to the following integral
\eqn\ahu{
\left<e^{\widehat\varpi_{\CG}^X}\right>= \int_{\tilde\eufm{M}_\zeta} exp 
\left(i mH_{S^1} + \tilde\varpi\right),
}
where $\widetilde\varpi$ denote the K\"{a}hler form on 
$\tilde\eufm{M}_{\zeta}$.
Now we have exactly the DH integration formula \DH.
The integral can be written as the sum of contributions from
the fixed points of the $S^1$-action. 

We saw that we have
two branches. In branch (i) the fixed point locus is
the effective target space $\CM_\zeta$ of 
the original $N_c=(2,0)$
model. The Hamiltonian $H_{S_1}$ in this branch is
simply zero. Thus we are evaluating the symplectic volume
of $\CM_\zeta$. This is a correlation function of the
original $N_c=(2,0)$ model. In branch (ii) the value $H^f_{S^1}$ 
of $H_{S^1}$ at a fixed point is non-zero. So the integral for each
fixed point is weighted by a phase factor $\exp(i m H^{f}_{S^1})$.
For both branches the integral is weighted by a one loop determinant
coming from the transverse degrees of freedom. We note that
such a determinant contains factors of $m$ with certain weights
depending on the particular fixed points.
After evaluating the DH integral we can set $m=0$.
Then we may obtain many relations by imposing that
the poles should be cancelled order by order 
between the two different branches,
since the limit $m\rightarrow 0$ should be smooth in the path integral of 
the massive $N_c=(2,0)$ model.  The partition function of the $N_c=(2,2)$
model is given by a sum of terms with order zero in $m$.
One can also obtain the symplectic volume of $\CM_{\zeta}$
in terms of a sum of contributions coming from branch (ii).

In the real situation life is more complicated since it is
difficult to achieve the generic conditions and the
space $\tilde\eufm{M}_\zeta$ may be non-compact. 
Its is in principle
possible to elaborate on the above procedure and perform
the integral. Even if we can't do such an integral due to
technicalities we can at least see that the essential
information on the correlation function of the original
$N_c=(2,0)$ model is contained in the fixed points
which belongs to branch (ii).

\section*{Acknowledgement}
\baselineskip 10pt
{\small
This article is partly based on the introductory chapters
of my PhD thesis and lectures given at KIAS.
It is my great pleasure to thanks  my adviser Herman Verlinde 
for various suggestions and encouragement. 
I am grateful to KIAS for hospitality and a financial
support during my visit.
I am  grateful to Robbert  Dijkgraaf, Christiaan Hofman,
Seungjoon Hyun and Bernd Schroers for fruitful collaborations
on the related subjects. 
I am also grateful to Christiaan Hofman for correcting
English and an error after a patient proof-reading.
This work is supported by a pionier fund of NWO and
DOE grant \# DE-FG02-92ER40699.
}
\vfill\eject

\end{document}